\documentclass[acmsmall]{acmart}

\AtBeginDocument{%
  }

\setcopyright{acmcopyright}
\copyrightyear{2024}
\acmYear{2024}
\acmDOI{XXXXXXX.XXXXXXX}

\usepackage{cleveref}

\usepackage{tikz}
\newcommand*\emptycirc[1][1ex]{\tikz\draw[thick] (0,0) circle (#1);} 
\newcommand*\halfcirc[1][1ex]{%
  \begin{tikzpicture}
  \draw[fill] (0,0)-- (90:#1) arc (90:270:#1) -- cycle ;
  \draw[thick] (0,0) circle (#1);
  \end{tikzpicture}}
\newcommand*\fullcirc[1][1ex]{\tikz\fill (0,0) circle (#1);}

\newcommand*\emptycirctxt[1][0.6ex]{\tikz\draw[thick] (0,0) circle (#1);}
\newcommand*\halfcirctxt[1][0.6ex]{%
  \begin{tikzpicture}
  \draw[fill] (0,0)-- (90:#1) arc (90:270:#1) -- cycle ;
  \draw[thick] (0,0) circle (#1);
  \end{tikzpicture}}
\newcommand*\fullcirctxt[1][0.6ex]{\tikz\fill (0,0) circle (#1);}

\usepackage{adjustbox}
\usepackage{array}
\newcolumntype{R}[2]{%
    >{\adjustbox{angle=#1,lap=\width-(#2)}\bgroup}%
    l%
    <{\egroup}%
}
\newcommand*\rot{\multicolumn{1}{R{45}{1em}}}

\usepackage{lscape}

\begin{document}

\title{Collaborative Cybersecurity Using Blockchain: A Survey}

\author{Loïc Miller}
\email{loic.miller@imt-atlantique.fr}
\orcid{0000-0003-4717-641X}
\affiliation{%
  \institution{IMT Atlantique, Cyber CNI Chair}
  \city{Rennes}
  \country{France}
}

\author{Marc-Oliver Pahl}
\email{marc-oliver.pahl@imt-atlantique.fr}
\affiliation{%
  \institution{IMT Atlantique, Cyber CNI Chair}
  \city{Rennes}
  \country{France}
}

\renewcommand{\shortauthors}{Miller and Pahl}

\begin{abstract}
Collaborative cybersecurity relies on organizations sharing information to boost security, but trust management is a key concern.
Decentralized solutions like distributed ledgers, particularly blockchain, are crucial for eliminating single points of failure.
However, the existing literature on blockchain-based collaborative cybersecurity is limited, lacking comprehensive insights.

This paper addresses this gap by surveying blockchain's role in collaborative cybersecurity from 2016 to 2023.
It explores various applications, trends, and the evolution of blockchain technology, focusing on access control, data validation policies, underlying tech, and consensus mechanisms.

A key finding is the fragmentation of the field with no dominant research group or venue.
Many recent projects poorly select consensus protocols for their blockchain. To aid researchers and practitioners, this paper offers guidelines for choosing the right blockchain for specific purposes and highlights open research areas and lessons learned from past blockchain applications in collaborative cybersecurity, encouraging further exploration in this field.
\end{abstract}

\ccsdesc[500]{General and reference~Surveys and overviews}
\ccsdesc[500]{Security and privacy~Distributed systems security}

\keywords{security, survey, distributed ledgers, blockchain, collaborative cybersecurity, literature review}


\maketitle

\section{Introduction}
\label{introduction}

Collaborative cybersecurity systems rely on sharing information among potentially non-trusting organizations.
This type of system is designed to enable organizations to work together and share information such as threat intelligence, in order to collectively improve their overall security and defend against cyber attacks better.
Collaborative cybersecurity systems can help organizations to more effectively prevent, identify, monitor, and respond to potential threats.
They also provide a more comprehensive view of the cyber threat landscape.
By working together and sharing information, organizations can improve their ability to protect against and respond to ongoing cyber attacks.

One of the main concerns in those systems is how to manage trust.
The traditional way to solve this issue is to give a single node the sensitive resources to protect.
This centralized approach is cost-effective because only one node requires protection, but this node becomes a single point of failure in the system.
If this node becomes compromised, the entire system is at risk.
Decentralized approaches provide an alternative, more costly but eliminating the single point of failure.
Distributed ledgers are one of those alternatives.

Among distributed ledgers, the most used underlying structure is the blockchain.
Blockchain is a distributed ledger in which users append data to existing records.
It enables the storage of information without the need to trust a single third party.

Even though the concept of blockchain predates its creation, the most famous use of blockchain is Bitcoin~\cite{nakamoto2008bitcoin}.
However, blockchain has also been used for other applications.
For example, blockchain has been used to solve traceability and non-repudiation issues in supply chains~\cite{etemadi2020blockchain}, as well as security and privacy issues for health records~\cite{akarca2019blockchain}.


Blockchain has received a lot of attention from both academia and industry.
Researchers have focused on the study of blockchain pertaining to specific domains or applications.
However, there are few surveys focusing on lessons learned in general, concentrating instead on a specific application.
Additionally, there are no surveys focusing on the use of blockchain for collaborative cybersecurity purposes.
To draw conclusions about blockchain and its use in this context, this paper analyzes several studies.

Our literature review provides an overview of the use of distributed ledgers for collaborative cybersecurity, regardless of the application or the domain.
Furthermore, we identify general trends and analyze what blockchains are used for.

The contributions of this paper are as follows:
\begin{itemize}
  \item A cross-domain literature review of 71 papers between 2016 and 2023 analyzing how blockchain is used in collaborative cybersecurity systems.
  \item The identification of key applications, and examination of trends in blockchain in those systems.
  \item The provision of guidance to researchers and practitioners in their choice of blockchain.
  \item The identification of open research issues and lessons learned to encourage further works in the area.
\end{itemize}



\Cref{related} gives an overview on existing surveys, confirming our methodology and showing the need for an additional survey.
\Cref{methodology} describes the methodology employed to realize this study.
We give our research questions and describe the methods used in this structured literature review.
\Cref{background} gives background information on blockchain.
\Cref{survey} gives a detailed survey on recent research developments in the domain to complement past use cases and motivate future ones.
We focus on the evolution of the research topic, relevant venues, and active groups.
\Cref{lessons} then gives lessons learned and open research questions.
\Cref{conclusion} concludes the paper.

\section{Related Work}\label{related}



We identified the related work during the structured literature review process.
Some surveys were added during the recommendations phase~\cite{dawit2020suitability,lavaur2022evolution}, the others were identified via a structured search~\cite{lin2017survey,hasanova2019survey,taylor2020systematic,wang2020survey,dibaei2021investigating,kumar2021survey,abdel2022privacy,mathew2022integration,vasilomanolakis2015taxonomy,meng2018intrusion,baiod2021blockchain,krichen2022blockchain,hakak2021recent,gimenez2021achieving,vance2019cybersecurity,belchior2022survey,liu2021blockchain,zhuang2020blockchain,leng2020blockchain,etemadi2020blockchain,zheng2018blockchain,wylde2022cybersecurity}.
We also considered relevant studies from research groups and conferences known to work in this area.
%
%
We argue that our set of surveys is comprehensive because it was identified structurally during the literature review process.


Many works exist on cybersecurity of the blockchain.
They are only of marginal relevance for this paper.
Therefore, we point the interested reader to the literature for further information~\cite{lin2017survey,hasanova2019survey,taylor2020systematic}.

Table~\ref{tab:surveys} lists all surveys we considered in our literature review.
Besides the year, reference and number of citations, we characterize those studies by two nominal discriminators:
\begin{itemize}
    \item Domain -- General area of interest considered by the survey. A --- signifies the study does not pertain to a domain in particular.
    \item Application -- Application of interest considered by the survey. A --- signifies the study does not pertain to an application in particular.
\end{itemize}

We also differentiate those surveys by four ordinal discriminators.
Each study either completely (\fullcirctxt), partially (\halfcirctxt) or does not fulfill (\emptycirctxt) the discriminator:
\begin{itemize}
    \item Trends analysis -- The study completely fulfills this criteria if it analyzes the evolution of their topic throughout the years in terms of venues, research groups and blockchain, i.e. nature, technology and consensus. The study partially fulfills this criteria if it analyzes at least one but not all of those aspects. The study does not fulfill this criteria if no trend analysis is done.
    \item Area diversity -- The study completely fulfills this criteria if it spans multiple areas. The study does not fulfill this criteria if it focuses on a single area.
    \item Application diversity -- The study completely fulfills this criteria if it spans multiple applications. The study does not fulfill this criteria if it focuses on a single application.
    \item Trait coverage -- The study completely fulfills this criteria if it analyzes for each paper the following characteristics of the blockchain: access control, data validation policy, technology and consensus. The study partially fulfills this criteria if it analyzes at least one characteristic. The study does not fulfill this criteria if it does not analyze those characteristics at all.
\end{itemize}

\begin{table}[ht]
    \caption{Surveys on blockchain and collaborative cybersecurity.}
    \adjustbox{max width=\textwidth}{%
    \centering
    \begin{tabular}{cllllcccccr}
        \toprule
        Year & Reference                                                          & Domain                   & Application                 & \rot{Trends analysis} & \rot{Area diversity} & \rot{Application diversity} & \rot{Trait coverage} & Cited \\
        \midrule
        2015 & Vasilomanolakis~\textit{et al.}~\cite{vasilomanolakis2015taxonomy} & ---                      & Intrusion Detection         & \emptycirc            & \fullcirc            & \emptycirc                  & \emptycirc           & 320   \\
        2018 & Meng~\textit{et al.}~\cite{meng2018intrusion}                      & ---                      & Intrusion Detection         & \emptycirc            & \fullcirc            & \emptycirc                  & \emptycirc           & 440   \\
        2018 & Zheng~\textit{et al.}~\cite{zheng2018blockchain}                   & ---                      & Cybersecurity               & \emptycirc            & \fullcirc            & \emptycirc                  & \emptycirc           & 2806  \\
        2019 & Vance and Vance~\cite{vance2019cybersecurity}                      & Critical Infrastructures & Cybersecurity               & \halfcirc             & \emptycirc           & \emptycirc                  & \emptycirc           & 8     \\
        2020 & Dawit \textit{et al.}~\cite{dawit2020suitability}                  & ---                      & Intrusion Detection         & \emptycirc            & \fullcirc            & \emptycirc                  & \emptycirc           & 11    \\
        2020 & Etemadi~\textit{et al.}~\cite{etemadi2020blockchain}               & Supply chain             & Cybersecurity               & \emptycirc            & \emptycirc           & \emptycirc                  & \emptycirc           & 7     \\
        2020 & Leng~\textit{et al.}~\cite{leng2020blockchain}                     & Industry 4.0             & Cybersecurity               & \emptycirc            & \emptycirc           & \fullcirc                   & \emptycirc           & 101   \\
        2020 & Zhuang~\textit{et al.}~\cite{zhuang2020blockchain}                 & Smart Grid               & Cybersecurity               & \emptycirc            & \emptycirc           & \fullcirc                   & \emptycirc           & 80    \\
        2021 & Baiod~\textit{et al.}~\cite{baiod2021blockchain}                   & ---                      & ---                         & \emptycirc            & \fullcirc            & \fullcirc                   & \emptycirc           & 14    \\
        2021 & Dibaei~\textit{et al.}~\cite{dibaei2021investigating}              & Vehicular Networks       & Cybersecurity               & \emptycirc            & \emptycirc           & \fullcirc                   & \halfcirc            & 41    \\
        2021 & Gimenez-Aguilar \textit{et al.}~\cite{gimenez2021achieving}        & ---                      & Cybersecurity properties    & \emptycirc            & \fullcirc            & \emptycirc                  & \halfcirc            & 20    \\
        2021 & Hakak~\textit{et al.}~\cite{hakak2021recent}                       & ---                      & ---                         & \emptycirc            & \fullcirc            & \fullcirc                   & \halfcirc            & 19    \\
        2021 & Kumar~\textit{et al.}~\cite{kumar2021survey}                       & Vehicular Networks       & Blockchain techniques       & \halfcirc             & \emptycirc           & \fullcirc                   & \emptycirc           & 6     \\ 
        2021 & Liu~\textit{et al.}~\cite{liu2021blockchain}                       & ---                      & Cybersecurity               & \halfcirc             & \fullcirc            & \fullcirc                   & \emptycirc           & 1     \\
        2022 & Abdel-Basset \textit{et al.}~\cite{abdel2022privacy}               & IIoT                     & Intrusion Detection         & \emptycirc            & \emptycirc           & \emptycirc                  & \emptycirc           & 1     \\
        2022 & Belchior~\textit{et al.}~\cite{belchior2022survey}                 & Workflows                & Business Process Management & \emptycirc            & \emptycirc           & \emptycirc                  & \emptycirc           & 1     \\
        2022 & Krichen~\textit{et al.}~\cite{krichen2022blockchain}               & ---                      & ---                         & \emptycirc            & \fullcirc            & \fullcirc                   & \emptycirc           & 4     \\
        2022 & Lavaur~\textit{et al.}~\cite{lavaur2022evolution}                  & ---                      & Intrusion Detection         & \halfcirc             & \fullcirc            & \emptycirc                  & \emptycirc           & 1     \\
        2022 & Mathew \textit{et al.}~\cite{mathew2022integration}                & IIoT                     & Intrusion Detection         & \emptycirc            & \emptycirc           & \emptycirc                  & \emptycirc           & 0     \\
        2022 & Wylde~\textit{et al.}~\cite{wylde2022cybersecurity}                & ---                      & Cybersecurity properties    & \emptycirc            & \fullcirc            & \emptycirc                  & \emptycirc           & 4     \\

        2022 & This work                                                          & Global                   & Global                      & \fullcirc             & \fullcirc            & \fullcirc                   & \fullcirc            & {}    \\
        \bottomrule
    \end{tabular}}
    \label{tab:surveys}
\end{table}

More than a dozen surveys exist on the topics of distributed ledgers and collaborative cybersecurity~\cite{wang2020survey,dibaei2021investigating,kumar2021survey,abdel2022privacy,mathew2022integration,vasilomanolakis2015taxonomy,meng2018intrusion,dawit2020suitability,lavaur2022evolution,baiod2021blockchain,krichen2022blockchain,hakak2021recent,gimenez2021achieving,vance2019cybersecurity,belchior2022survey,liu2021blockchain,zhuang2020blockchain,leng2020blockchain,etemadi2020blockchain,zheng2018blockchain,wylde2022cybersecurity}.
However, the majority of those surveys pertain to a specific application, and/or domain.
Some of the studies are concerned with Vehicular Networks~\cite{wang2020survey,dibaei2021investigating,kumar2021survey}, some with IIoT~\cite{abdel2022privacy,mathew2022integration}, while others only focus solely on Intrusion Detection~\cite{vasilomanolakis2015taxonomy,meng2018intrusion,dawit2020suitability,abdel2022privacy,lavaur2022evolution,mathew2022integration}.

A few surveys~\cite{baiod2021blockchain,krichen2022blockchain,hakak2021recent} open beyond specific applications or areas.
However, those surveys do not identify general trends, and limit themselves to a simple listing of applications and/or issues.


Closest to this study, some surveys extract general information on the topic regardless of area or application.
Baiod~\textit{et al.}~\cite{baiod2021blockchain} consider blockchain technology and its applications across multiple areas.
They give a general overview of blockchain as well as technology and issues surrounding it.
Contrary to us, they do not evaluate trends and identify open issues.
Hakak~\textit{et al.}~\cite{hakak2021recent} explore recent advances in blockchain.
More specifically, they review the different blockchain platforms and compare applications in different areas.
They however do not consider the trends or whether the use of blockchain is justified.
Krichen~\textit{et al.}~\cite{krichen2022blockchain} determine the use of blockchain in several areas like finance, healthcare and governmental services.
They do not look at the underlying blockchain technology and network, whether blockchain use is justified, and do not analyze trends.

A subset of the surveys deal solely with intrusion detection.
Vasilomanolakis~\textit{et al.}~\cite{vasilomanolakis2015taxonomy} investigate collaborative intrusion detection systems.
Some of the considered systems use blockchain.
Meng~\textit{et al.}~\cite{meng2018intrusion} review intrusion detection systems and blockchain.
Dawit~\textit{et al.}~\cite{dawit2020suitability} survey the suitability of blockchain for collaborative intrusion detection systems.
Abdel-Basset~\textit{et al.}~\cite{abdel2022privacy} study both non-blockchain and blockchain solutions to achieve a privacy-preserved intrusion detection system for the Industrial Edge of Things (IEoT).
Lavaur~\textit{et al.}~\cite{lavaur2022evolution} survey federated learning solutions for intrusion detection systems.
Some of those solutions make use of blockchain to enable model sharing between the entities using federated learning.
Mathew~\textit{et al.}~\cite{mathew2022integration} look at blockchain for collaborative intrusion detection systems in the context of Industrial IoT (IIoT).
They identify three blockchain integration approaches and one trust management approach.
Contrary to those works, our study analyzes general trends.

Another subset of the surveys deal with different areas -- cybersecurity~\cite{gimenez2021achieving,liu2021blockchain,ivanov2023security}, vehicular networks~\cite{wang2020survey,dibaei2021investigating,kumar2021survey} and industrial applications~\cite{zhuang2020blockchain,vance2019cybersecurity,leng2020blockchain,etemadi2020blockchain}.
Gimenez-Aguilar~\textit{et al.}~\cite{gimenez2021achieving} investigate ways of achieving cybersecurity with blockchain.
In particular, they look at how researchers and industrial initiatives achieve various cybersecurity properties, e.g.\ authentication, non-repudiation and confidentiality.
They also consider whether the use of blockchain in those papers is justified or not.
Liu~\textit{et al.}~\cite{liu2021blockchain} look at how blockchain is used for cybersecurity, while Ivanov~\textit{et al.}~\cite{ivanov2023security} look specifically at smart contracts.

Wang~\textit{et al.}~\cite{wang2020survey} provide guidelines to further study the application of blockchain in vehicular networks.
Dibaei~\textit{et al.}~\cite{dibaei2021investigating} investigate the different ways blockchain is used to secure vehicular networks.
Kumar~\textit{et al.}~\cite{kumar2021survey} look at blockchain techniques used to secure the Internet of Vehicles.

Zhuang~\textit{et al.}~\cite{zhuang2020blockchain} study how blockchain can be used to secure the Smart Grid.
Vance and vance~\cite{vance2019cybersecurity} look at the use of blockchain to protect critical infrastructures.
Leng~\textit{et al.}~\cite{leng2020blockchain} study how blockchain can be used to overcome challenges in smart manufacturing in industry 4.0.
Etemadi~\textit{et al.}~\cite{etemadi2020blockchain} look at blockchain use in the food supply chain.

Adjacent to collaborative cybersecurity, Belchior~\textit{et al.}~\cite{belchior2022survey} explore the use of blockchain on business process view integration.
Zheng~\textit{et al.}~\cite{zheng2018blockchain} survey blockchain challenges and opportunities.
Wylde~\textit{et al.}~\cite{wylde2022cybersecurity} review issues surrounding cybersecurity, data privacy and blockchain.
\section{Research methodology}\label{methodology}

This section is intentionally detailed to ensure the study's reproducibility and adherence to sound research principles.
It offers a better perspective on the results of the study.
We follow the guidelines of Keele~\textit{et al.}~\cite{keele2007guidelines} as well as Briner and Denyer~\cite{briner2012systematic}.
Following the latter method, we present next the following steps that were taken.
\Cref{methodology:questions} identifies and defines the research questions we address.
\Cref{methodology:slr} shows how we obtain the relevant studies to answer our questions.
\Cref{methodology:id} describes our method to identify the studies meeting our criteria.
\Cref{methodology:syn} explains how we extract and synthetize our findings.

\subsection{Identify and define the research questions}\label{methodology:questions}

The purpose of this paper is to analyze the place of blockchain in collaborative cybersecurity systems.
The main question is thus \textit{``How has blockchain been used in collaborative cybersecurity systems?''}.
To answer this question, we define a set of more refined questions, as follows:
\begin{enumerate}
    \item \textit{RQ1:} What are the main use cases when it comes to collaborative cybersecurity systems using blockchain?
    \item \textit{RQ2:} In what domains has blockchain technology been used to achieve collaborative cybersecurity?
    \item \textit{RQ3:} What type of blockchain has been used to achieve collaborative cybersecurity?
    \item \textit{RQ4:} How has the area of collaborative cybersecurity systems using blockchain evolved over time?
    \item \textit{RQ5:} What lessons and open issues can we deduce from our findings?
\end{enumerate}

Our study targets a diverse range of readers with varying interests.
Specifically, we are focusing on three main groups: collaborative cybersecurity researchers, cybersecurity practitioners, and developers working with blockchain technology.
As such, each research question that we explore in this study will likely appeal to different sets of readers.
Namely, \textit{RQ1} and \textit{RQ2} are relevant for collaborative cybersecurity researchers, interested by an overview of collaborative cybersecurity applications and their domains.
\textit{RQ3} is more interesting for cybersecurity practitioners and blockchain developers, which might gain an insight as to what type of blockchain they need to consider for their specific use case.
Finally, \textit{RQ4} and \textit{RQ5} are of interest to a general audience which can gain a historical perspective of the topic as well as a more complete understanding of the matter.

\subsection{Determine and search the relevant studies}\label{methodology:slr}

To determine and search the relevant studies, we follow the guidelines of Keele~\textit{et al.}~\cite{keele2007guidelines} on structured literature reviews.
The considered set of papers is composed of conference, workshop and journal papers.
Due to the huge amount of papers on one specific application of blockchain in collaborative cybersecurity, the selection method for those papers is slightly different than for the other applications.

We used Google Scholar to retrieve all manuscripts.
The initial set of considered papers is composed of recommendations that were given by coworkers, as well as some papers found using intuitive search to get a first look at the topic.
We then added to this initial set of papers using structured search, with the following query:
\begin{quote}
    (``distributed ledger'' OR ``blockchain'') AND (``collaborative cybersecurity'' OR ``cybersecurity'')
\end{quote}

This query ensures that we capture a set of papers relevant to the matter, even in different forms.
To obtain more surveys related to the topic for our related work, we complete our previous query to obtain the following:
\begin{quote}
    (``distributed ledger'' OR ``blockchain'') AND (``collaborative cybersecurity'' OR ``cybersecurity'') AND (``survey'' OR ``literature review'')
\end{quote}
After this step, a total of 70 papers were retrieved.

\subsection{Identify studies meeting our criteria}\label{methodology:id}

After retrieving this initial set of papers obtained by recommendations, intuitive search and structured search, we carried out a manual review of the papers.
First, the title are abstract of each paper was used to discard those that were not relevant when considering our research questions.
We then carried out a more thorough review of the papers we kept.
After processing each paper, we performed a backward and forward snowball, looking at the papers in the references as well as the papers citing the processed paper.
We used this method to gain access to more relevant papers, which then underwent the same process of filtering and snowballing.

Those steps yielded an additional 65 papers, which after filtering brought our definitive sample to 91 papers.
Out of those 91 papers, 20 are surveys related to our topic which are described in the related work, but not a part of our analysis.
Our analysis was thus performed on a set of 71 papers --- 35 journal papers and 36 conference/workshop papers.

\subsection{Extract and synthesize findings}\label{methodology:syn}

The selected studies were all analyzed to answer the proposed research questions.
To this end, we classified the papers according to different features.
Namely, we characterized each proposal according to their mode of publication, the venue they were published in, the research groups making the research and the countries the authors were from.
We also classified each paper according to its application, as well as its domain.
In the case a paper fit more than one application or domain, it was counted for each one of them.
Finally, we characterized for each paper different aspects of their underlying blockchain.
Specifically, we looked at technology, consensus, access control and permission.
\section{Background}\label{background}

We introduce in this section the foundations of blockchain technology.
\Cref{background:overview} outlines core concepts of distributed ledgers, namely \textit{Nature}, \textit{Technology} and \textit{Consensus}.
To simplify the presentation of some concepts in the analysis, \Cref{background:model} presents a general blockchain model.

\subsection{Blockchain overview}\label{background:overview}

Blockchain technology enables the creation of a decentralized ledger that allows for the addition of data in a sequential manner~\cite{nakamoto2008bitcoin}.
Note that the underlying linked list structure can be changed to other structures, e.g. a graph~\cite{wang2023sok}.
With blockchain, trust is distributed among all participants.
No central party is required.
Therefore, maintaining and updating the ledger requires that all (or a qualified portion of) participants reach a consensus.
The way to reach this consensus depends on the \textit{nature} of the blockchain and their underlying \textit{technology}.

\paragraph*{Nature}

The nature of a blockchain is determined by two key factors: \textit{access control} and \textit{data validation policy}~\cite{gimenez2021achieving}.

Access control is a critical aspect of blockchain technology.
The access control mechanism of a blockchain determines who can access the network, and can be classified into two categories: \textit{public} and \textit{private}.
Public blockchains are open to anyone who wishes to participate, without any restrictions on who can join or access the network.
In contrast, private blockchains have a restricted access control mechanism where only selected members are allowed to join and access the network.

Choosing between a public or private blockchain depends on the specific use case and the requirements of the application.
Public blockchains can offer greater transparency and decentralization, as anyone can view transactions on the network.
They can also lead to a greater number of participants, and therefore potentially increase the value a participant can extract from the ledger.
However, they also have limited privacy and security, which may not be suitable for certain applications.
In direct opposition, private blockchains provide greater privacy and security by limiting access to selected members, but they may lack the transparency and decentralization of a public blockchain.

Data validation is another crucial factor that determines the nature of a blockchain.
The data validation policy determines who has the authority to validate transactions and add blocks to the blockchain.
Blockchains can be categorized into two types based on their data validation policies: \textit{permissionless} and \textit{permissioned}.
In permissionless blockchains, anyone can participate in the validation process and contribute to the consensus mechanism that ensures the integrity of the blockchain.
This means that all nodes on the network are equal, and no specific user or group has special privileges.

In contrast, permissioned blockchains have a restricted set of users who are authorized to validate the blockchain.
These users are typically selected based on their role or authority in the network, and they have a greater responsibility to ensure the accuracy and security of the blockchain.
Permissioned blockchains are commonly used in enterprise applications, where data privacy and security are critical, and a high level of trust among the participants is necessary.
The choice between these two types of blockchains depends on the specific requirements of the application and the level of trust among the participants.

The overall \textit{flavor} of a specific blockchain is thus a combination of those two factors, either public or private, and permissionless or permissioned.
Most public blockchains are permissionless, and most private blockchains are permissioned, but particular use cases may have a need for private permissionless or public permissioned blockchains.

\paragraph*{Technology}

Bitcoin, Ethereum and Hyperledger are the main technologies employed to develop systems based on blockchain.

The first, Bitcoin, is a digital cryptocurrency that was created in 2008 by an unknown person or group of person using the name Satoshi Nakamoto~\cite{nakamoto2008bitcoin}.
Bitcoin allows two parties to send transactions between each other without the involvement of a third party.
It is public and permissionless, meaning that anyone can join and participate in the network without needing permission or authorization.
Transactions are verified by network nodes through cryptography and recorded on the blockchain, which acts as a permanent, transparent record of all transactions.
Despite its early controversies, Bitcoin has become a popular form of currency and store of value, and its underlying blockchain technology has been adopted by numerous industries for its secure and transparent nature.

Ethereum is a blockchain-based cryptocurrency that was launched in 2015.
It is notable for its smart contract functionality, which allows developers to create decentralized applications (dApps) that can run on the Ethereum network.
Smart contracts are self-executing contracts with the terms of agreement between buyer and seller being directly written into lines of code.
The Ethereum Virtual Machine (EVM) is a runtime environment that enables the execution of smart contracts on the Ethereum network.
The main Ethereum network is public, meaning that anyone can participate in the network without needing permission or authorization.
Ethereum has become a popular platform for building decentralized applications, and its blockchain technology is used in various industries such as finance, healthcare, and gaming.

Hyperledger is an open-source collaborative effort that was established at the end of 2015 by the Linux Foundation.
It is a platform that allows developers to create blockchain applications with a focus on scalability, confidentiality, and security.
Hyperledger hosts numerous blockchain technologies, including Iroha, Fabric, Sawtooth, and many more.
Among these projects, Fabric is the most widely used and is designed to be a private permissioned blockchain, which means that access to the network is restricted to authorized participants.
Hyperledger Sawtooth is another popular project and is notable for its modular architecture that allows developers to customize the consensus algorithms that underpin the network.
Hyperledger also provides a framework for building smart contracts, known as Chaincodes, which are written in programming languages such as Go or Java.
The platform is gaining traction in a variety of industries, including finance, healthcare, and supply chain management.

\paragraph*{Consensus}

The different technologies utilized in blockchain networks employ various types of proofs to achieve consensus among the network participants. 

Proof of Work (PoW) is most famously associated with Bitcoin as it was the first and is the most widely known implementation of this consensus algorithm.
The PoW algorithm requires participants, also known as miners, to solve complex mathematical problems to validate transactions and add blocks to the blockchain.
The miner who solves the problem first receives a reward in the form of cryptocurrency.
The PoW algorithm is designed to be resource-intensive, requiring a significant amount of computational power and energy to operate.
This makes it a secure and reliable method of consensus, as it is difficult for malicious actors to manipulate the network.
However, the high energy consumption required by PoW has led to concerns about its environmental impact and sustainability.
Despite these concerns, PoW remains the most widely used consensus algorithms in blockchain technology, particularly in public blockchains such as Bitcoin and Ethereum.

Proof of Stake (PoS) is a consensus algorithm that is increasingly being used in blockchain networks as an alternative to Proof of Work~\cite{vasin2014blackcoin}.
Unlike PoW, where participants validate transactions and add blocks to the blockchain through computational work, PoS operates by having participants, also known as validators, put up a stake in the form of cryptocurrency as collateral.
Validators are then randomly selected to validate transactions and add new blocks to the blockchain based on the size of their stake.
The validators who are selected to add new blocks to the blockchain receive transaction fees as a reward.

PoS is designed to be more energy-efficient than PoW, as it does not require the same level of computational power and energy consumption.
This makes it more sustainable and environmentally friendly.
PoS is also believed to be more secure than PoW as it is harder for malicious actors to acquire a majority stake in the network and manipulate the consensus mechanism~\cite{buterin2014next}.
The PoS consensus algorithm is gaining traction in blockchain networks, with notable implementations including Ethereum's Casper and Cardano's Ouroboros.
Note that Ethereum itself has also switched from PoW to PoS near the end of 2022.
However, there is also criticism that PoS may lead to centralization of power in the hands of a few large stakeholders, potentially compromising the decentralization that is fundamental to blockchain technology.

Delegated Proof of Stake (DPoS) is a consensus algorithm that combines the benefits of both Proof of Work and Proof of Stake.
DPoS is a more democratic version of PoS, where network participants, known as stakeholders, can vote for delegates to validate transactions and create new blocks~\cite{snider2018delegated}.
Delegates are responsible for maintaining the network and receiving rewards for their participation, but they are also subject to being voted out if they fail to perform their duties or act against the interests of the stakeholders.
DPoS is considered more efficient and scalable than both PoW and PoS, and it also reduces the risk of centralization by giving stakeholders the power to choose their delegates.
DPoS is used in several popular blockchain networks, including BitShares, EOS, and TRON.

Proof of Identity (PoI) is a consensus algorithm that verifies the identity of network participants before allowing them to participate in the network~\cite{liu2020blockchain}.
PoI uses a combination of biometric authentication, digital signatures, and other verification methods to ensure that only authorized individuals can access the network.
PoI is considered a more secure and reliable alternative to other consensus algorithms because it eliminates the risk of Sybil attacks, where an attacker can create multiple fake identities to manipulate the network.

Proof of Authority (PoA) is a consensus algorithm that relies on the identity of network participants to validate transactions and create new blocks~\cite{de2018pbft}.
In PoA, validators are known as authorities and are selected based on their reputation, expertise, or other predefined criteria.
Authorities are responsible for maintaining the network and receive rewards for their participation.
Unlike PoW and PoS, PoA does not require computational power or staked cryptocurrency, which makes it more energy-efficient and scalable.
However, it also requires a high degree of trust in the authorities, and the network can become vulnerable to centralization if the authorities collude.
PoA is used in several blockchain networks, including POA Network, Kovan, and Ethereum Classic.


Practical Byzantine Fault Tolerance (PBFT) is another consensus algorithm~\cite{li2020scalable} using a three-phase protocol.
In the first phase, the primary node broadcasts a proposal for a new transaction.
In the second phase, the other nodes validate the proposal and send a response back to the primary node.
In the third phase, the primary node broadcasts the final decision based on the responses received from the other nodes.
PBFT is considered more secure and reliable than other consensus algorithms, but it also requires a higher level of communication and computational power.
PBFT is used in several blockchain networks, including Hyperledger Fabric and Ripple.

Proof of Elapsed Time (PoET) is a consensus algorithm that relies on a trusted execution environment (TEE) to ensure that network participants are selected to create new blocks in a random and fair manner~\cite{chen2017security}.
In PoET, network participants compete to win a random wait time within the TEE, and the participant with the shortest wait time is selected to create the next block.
This process is repeated for each new block, ensuring that all network participants have an equal chance of being selected.
PoET is used in several blockchain networks, including Hyperledger Sawtooth and Quorum.

We point the interested reader to the literature for further information about consensus protocols~\cite{xu2023survey}.

\subsection{Blockchain model}\label{background:model}

\begin{figure}
    \centering
    \includegraphics[width=0.5\linewidth]{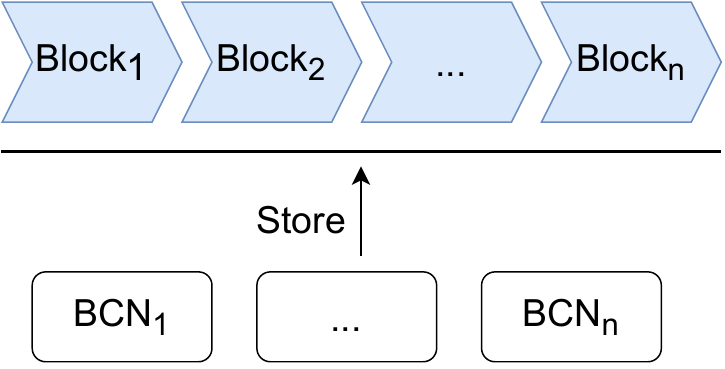}
    \caption{Blockchain model.}\label{fig:bc-model}
\end{figure}

The update and use of a blockchain involve several entities and elements, each with a specific role to play.
We refer here to the model presented by Gimenez-Aguilar~\textit{et al.}~\cite{gimenez2021achieving}.
Figure~\ref{fig:bc-model} represents a general model for blockchain.
The first entity is the set of nodes in charge of keeping the blockchain information itself.
These nodes, known as Blockchain nodes (BCN), cooperate to update blockchain data based on a consensus algorithm.
Consensus is typically reached among a subset of BCNs called miners, which are responsible for validating and adding new blocks to the blockchain.

Apart from BCNs, there is another entity that comes into play -- Blockchain Users (BCUs).
BCUs are those who are willing to insert information into the blockchain.
They might do this for various reasons, such as recording a transaction or recording data that will be shared with others.

Blockchain Observers (BCOs) are individuals who benefit from the information stored in the blockchain without contributing to it by adding new data.
They retrieve data from the blockchain by observing it and do not have the ability to update it.
Even though they do not directly participate in adding data, their retrieval might be recorded in the blockchain through a transaction or an interaction with a smart contract.

Since only BCNs have direct access to the blockchain, BCUs and BCOs must trust that the information they retrieve or exchange is accurate.
However, no single BCN is considered inherently trustworthy, and the blockchain's underlying consensus protocol or access control/permission mechanisms prevent any attempt to alter the information.
To further increase security and prevent mistrust, BCNs rely on incentives such as rewards for successfully adding transactions to the ledger.
These incentives encourage BCNs to act honestly and promote the best interests of the blockchain community.

\section{Blockchain for Collaborative Cybersecurity}\label{survey}

\newcommand{\npapers}{71}
\newcommand{\nconf}{36}
\newcommand{\njournal}{35}

\newcommand{\nintrusion}{38}
\newcommand{\nddos}{15}
\newcommand{\nsharing}{11}

\begin{figure}
    \includegraphics[width=0.6\linewidth]{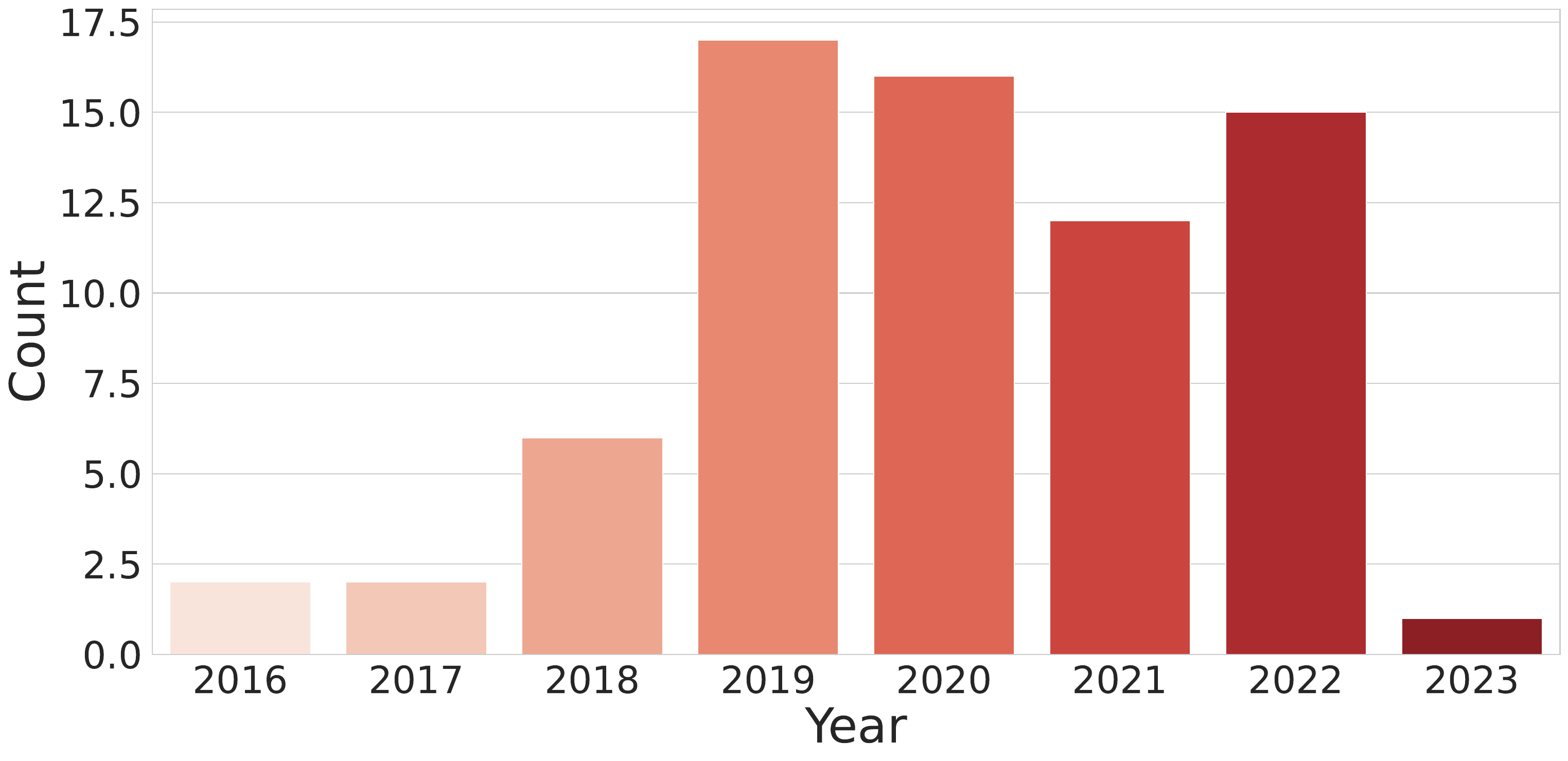}
    \caption{Number of papers per year.}\label{fig:count}
\end{figure}

We analyzed \npapers{} papers from 2016 to 2023.
Regarding the use of blockchain for collaborative cybersecurity, \Cref{fig:count} shows the number of papers published for each year.
Overall, the number of papers has been going up, with the most papers published in the year 2019, closely followed by 2020.
The launch of Ethereum and Hyperledger in 2015 might give a partial explanation as to the lack of papers on the topic before 2016.
Those projects introduce smart contracts and chaincodes respectively, making it easier to implement collaborative cybersecurity systems, which may also explain why we see an increase in the number of papers related to the topic after the launch of these technologies.
Of the \npapers{} papers, \nconf{} were published in conferences, and \njournal{} were published in journals.
The two most popular venues, with five and four papers respectively, are IEEE Access and the IEEE Internet of Things Journal.

To obtain more information about each individual paper on each of the dimensions described below, refer to the tables in \Cref{appendix}.
Specifically, \Cref{tab:papers} contains information on the application, the domain and the number of citations.
\Cref{tab:papers_2} on the other hand contains information about the nature of the blockchain, namely technology, consensus, access control and data validation policy.
A reader looking for a collaborative application using a specific blockchain nature will find great use of those tables.

\subsection{Applications}

Here, we explore the various applications of blockchain for collaborative cybersecurity purposes.
Representing \nintrusion{} out of \npapers{} papers, we find that the majority of papers are on the topic of intrusion detection.
The second most represented application is Distributed Denial of Service (DDoS) defense, with fifteen papers on the subject.
Eleven papers use blockchain in their respective domains to share data securely without a single point of failure, with some doing so out of privacy concerns.

\begin{figure}
    \includegraphics[width=0.8\linewidth]{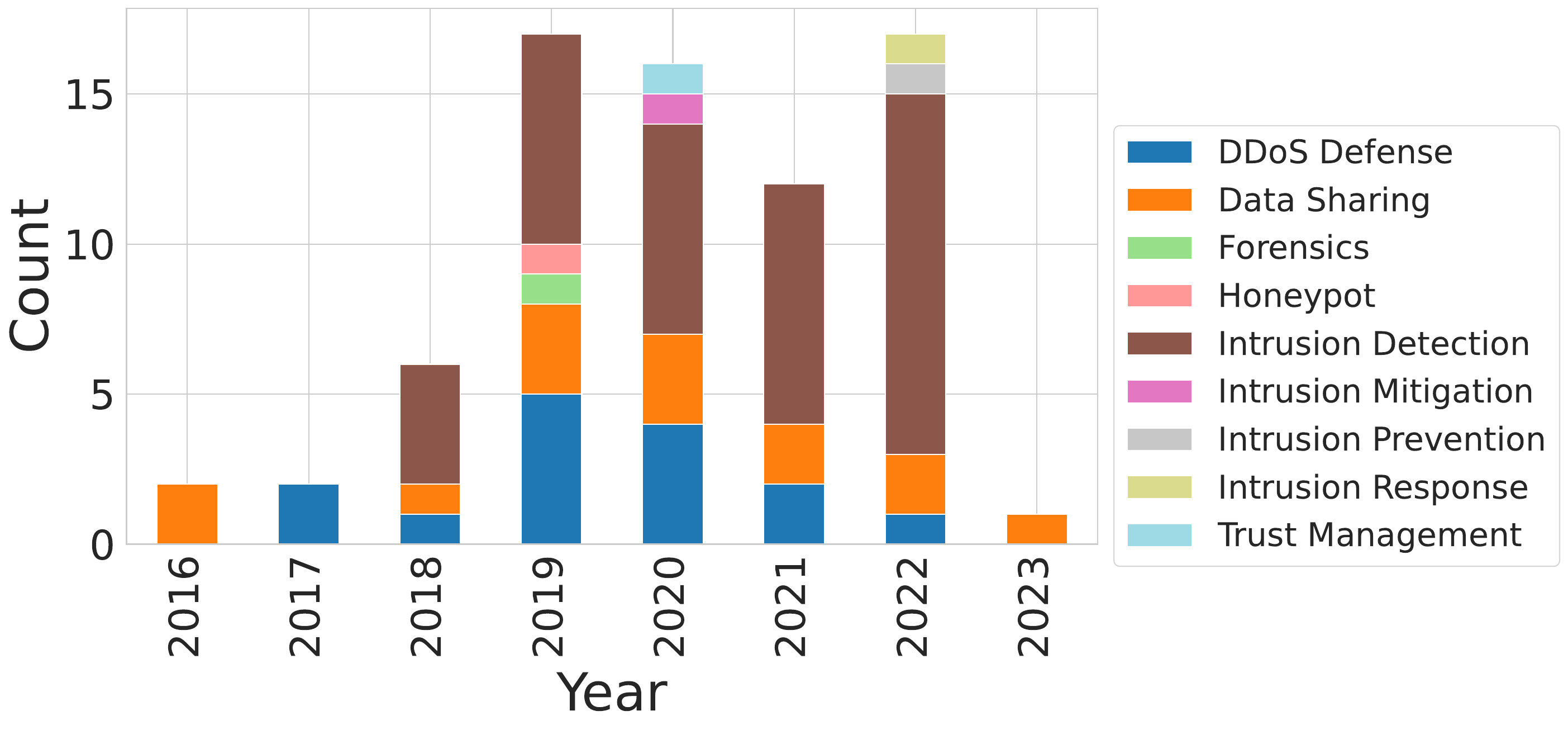}
    \caption{Approaches per application per year.}\label{fig:application}
\end{figure}

The use of blockchain in different applications has evolved over the years.
\Cref{fig:application} details this evolution of application per year.
While the number of papers on DDoS defense has been steadily decreasing, the number of papers on intrusion detection has been increasing.
Some less prevalent applications can also be found in the figure, such as honeypot, forensics or trust management.

In the following paragraphs, we provide a description of each application, as well as an explanation of how blockchain technology is utilized in these applications for collaborative cybersecurity purposes.

\paragraph*{Distributed Denial of Service Defense}
Distributed Denial of Service attacks aim to overwhelm a network with traffic, rendering it unavailable to users.
There are several ways to defend against DDoS attacks, which can be broadly categorized into four categories~\cite{peng2007survey}: attack prevention, attack detection, attack source identification, and attack reaction.
Attack prevention involves taking measures to prevent DDoS attacks from occurring in the first place, such as implementing firewalls and filtering traffic.
Attack detection involves identifying when a DDoS attack is occurring, typically by analyzing traffic patterns or anomalies.
Attack source identification involves identifying the source of the attack, which can be done through various techniques such as tracing the IP address or using honeypots.
Attack reaction involves taking action to mitigate the impact of the attack, such as diverting traffic to other servers or blocking certain traffic.

There are several methods used for mitigating DDoS attacks~\cite{miller2019taxonomy}.
Proactive techniques include:
\begin{itemize}
    \item Designing protocols with reduced amplification factors to minimize the impact of DDoS attacks.
    \item Reducing the number of amplifiers, such as open DNS resolvers and NTP servers, that can be exploited.
    \item Implementing rate limiting, UDP session management, and spoofed packet filtering to control and validate incoming traffic.
\end{itemize}
Reactive techniques involve:
\begin{itemize}
    \item Traffic scrubbing to filter out malicious traffic before it reaches the target.
    \item BGP blackholing to discard malicious traffic~\cite{RFC5635}.
    \item Blockchain signaling.
\end{itemize}

A big advantage of blockchain signaling compared to the other techniques is that DDoS mitigation can be achieved in a collaborative fashion, which we argue leads to better mitigation prospects than individually.
On top of this, blockchain signaling leverages existing infrastructure to minimize costs and system bloat, as the parties taking part in the mitigation do not have to manage and/or create a system to handle trust between collaborating parties.

We identified three main archetypes for blockchain signaling in DDoS mitigation.
We call those archetypes \textit{consortium}, \textit{contractual}, and \textit{aggregator}.

\begin{figure}
    \includegraphics[width=\linewidth]{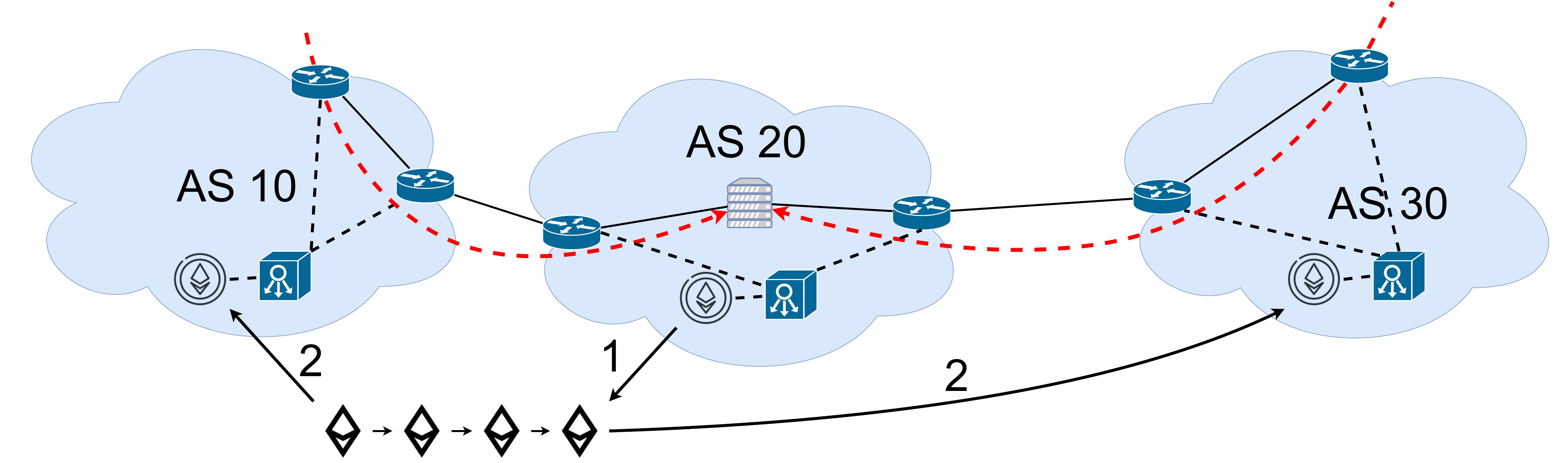}
    \caption{DDoS mitigation using the \textit{consortium} archetype.}\label{fig:consortium}
\end{figure}

In the \textit{consortium} archetype, two or more participants share malicious IP addresses for later mitigation.
An example scenario can be found in \Cref{fig:consortium}.
This example depicts three Autonomous Systems (AS), which can be defined as one or more networks under the control of a single entity (e.g. Hurricane Electric).
The server in AS20 is the victim of a DDoS attack.
To mitigate this attack with the consortium approach, AS 20 will first identify the malicious IP addresses responsible for the attack, and push them to the blockchain.
Then, other ASes belonging to the consortium, here AS 10 and AS 30, will mitigate the attack in any way they see fit.

The creator of the smart contract, here AS 20, is the one managing who is allowed to participate in the consortium.
As such, this archetype makes use of a private blockchain.
There are no requirements on the data validation policy, so the consortium can either be permissioned or permissionless.
Note that all approaches in the literature following this archetype use a private permissioned blockchain.
This approach was originally developed by El Houda~\textit{et al.}~\cite{abou2018chainsecure,abou2019co,abou2020brainchain,abou2020bringing}, but was later improved to include the possibility of sharing whitelists of trusted IP addresses.
El Houda~\textit{et al.} later combined their approach with intra-domain DDoS mitigation in a scheme they name Cochain-SC~\cite{abou2019cochain}.
In the paper they propose multiple intra-domain DDoS mitigation techniques, and coordinate mitigation between multiple SDN domains using blockchain.
Yeh~\textit{et al.}~\cite{yeh2019collaborative} also take the \textit{consortium} approach, but ensure data confidentiality and integrity using Elliptic Curve Cryptography.

A potential problem with this approach is the lack of incentives to mitigate the DDoS attacks targeting other ASes.
Why should an AS spend precious resources to mitigate the attack of another?
Papers using this archetype do not discuss such incentives.
We argue that a potential incentive is that even though an AS might not be the target of an attack, attack traffic may still pass through their AS, taking up bandwidth and resources anyway, so there is an incentive to block the mitigate the attack as soon as possible in terms of hops.
Another incentive may be that an AS will mitigate attacks for others, since it trusts those other ASes will mitigate a potential later attack targeting it.

\begin{figure}
    \includegraphics[width=\linewidth]{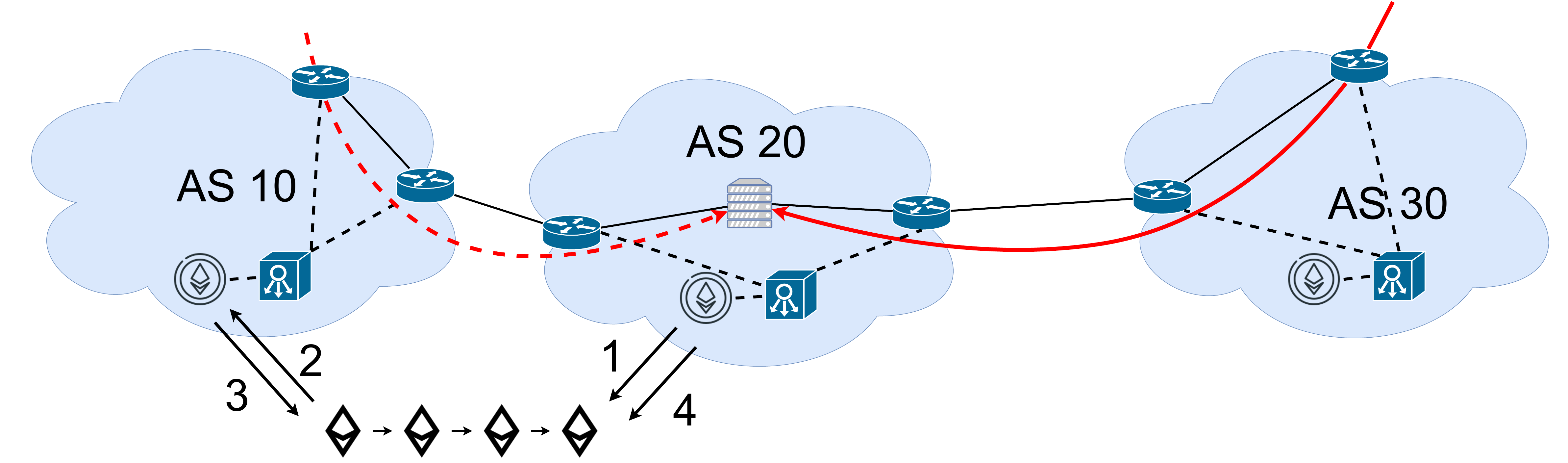}
    \caption{DDoS mitigation using the \textit{contractual} archetype.}\label{fig:contractual}
\end{figure}

The second archetype is \textit{contractual}.
In this archetype, a smart contract is established between two participants which are typically the victim of the attack and the mitigator.
After the establishment of the contract, the victim shares the malicious IP addresses with the mitigator and pays it for the service.
\Cref{fig:contractual} depicts the same example as \Cref{fig:consortium}, but with contractual mitigation.
AS 20 establishes a smart contract between itself and AS 10, which is the mitigator.
After pushing the malicious IP addresses to the blockchain, AS 10 mitigates the attack.
Note that the attack traffic coming from AS 30 is not mitigated, since AS 20 made no contract with it.
Once the attack is mitigated, AS 20 pays AS 10.

Rodrigues~\textit{et al.}~\cite{rodrigues2017blockchain,rodrigues2017enabling} first came up with this approach, naming it ``Blockchain Signaling System (BloSS)''.
They later conducted a performance evaluation of their method~\cite{rodrigues2019evaluating}, and demonstrated it with hardware~\cite{rodrigues2019cooperative}.
Later extensions added a reputation system, with the possibility for the victim to rate the service of the mitigator, and for the mitigator to rate the reputation of the victim~\cite{rodrigues2020sc}.
Such systems include the possibility of refunds if the mitigation was not satisfactory, even though proof of mitigation might be difficult to achieve.
Yeh~\textit{et al.} also contributed using the \textit{contractual} approach~\cite{yeh2020sochain}.
They point out the lack of incentive mechanisms in previous works, as well as the lack of discussion of incentive mechanisms in general.
In this contribution, they propose a decentralized DDoS data exchange platform called SOChain, which uses a \textit{DDoS\_coin} token to generate incentive to mitigate attacks.
Chiu~\textit{et al.} equally propose to mitigate DDoS attacks using FolketID, a system designed to reward participants mitigating DDoS attacks with cryptocurrency~\cite{chiu2022folketid}.

\begin{figure}
    \includegraphics[width=\linewidth]{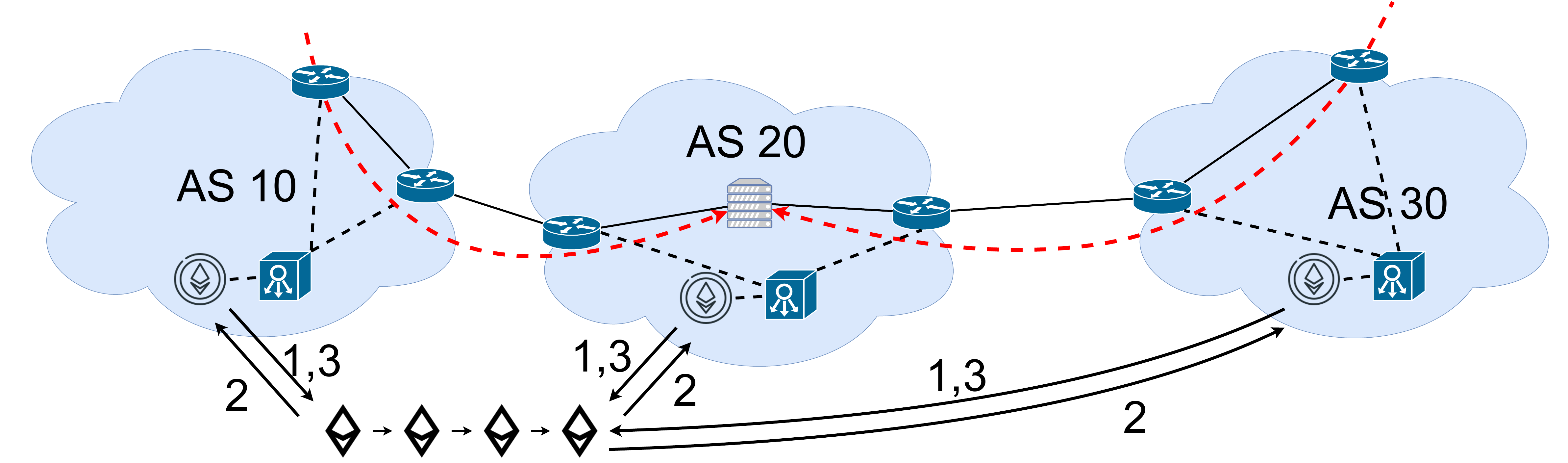}
    \caption{DDoS mitigation using the \textit{aggregator} archetype.}\label{fig:aggregator}
\end{figure}

The third and final archetype, \textit{aggregator}, sees the victim sharing traffic features among two or more participants.
The participants then collectively aggregate and process a defense digest, which they use to mitigate the DDoS attack.
\Cref{fig:aggregator} shows the same example as before, but with aggregator mitigation.
AS 10, 20 and 30 share traffic features of the attack with each other using the blockchain.
By aggregating this shared data, each AS processes a defense digest which is then pushed to the blockchain.
This defense digest is used to mitigate the attack.

Guo~\textit{et al.}~\cite{guo2022distributed} introduce this archetype.
These systems are designed to process large volumes of data quickly, which can be useful in identifying and mitigating DDoS attacks.
By analyzing traffic patterns in real-time using MapReduce, these systems can quickly identify and respond to DDoS attacks, reducing the impact on target servers.
As a side-note, they also complement their approach with a DDoS attack tracing scheme using data in the blockchain~\cite{guo2021ldbt}.

\paragraph*{Intrusion Detection Systems}
Blockchain has also been used a lot in Intrusion Detection Systems (IDS).
Intrusion Detection Systems come in many forms.

HIDS (Host-based Intrusion Detection Systems) and NIDS (Network-based Intrusion Detection Systems) are two distinct strategies for detecting intrusion attempts.
HIDS operate on individual host devices such as servers and workstations, monitoring system logs, processes, and file system activity to detect intrusions and other irregular behavior.
This method enables the identification and prevention of attacks originating from within the host, as well as those that are externally based.
In contrast, NIDS focus on monitoring network traffic to identify suspicious patterns and signs of intrusion.
By analyzing data packets and communication between devices, NIDS can detect anomalies and malicious activities that traverse the network infrastructure.
While both systems are essential for comprehensive security, HIDS excel at detecting and mitigating attacks on specific devices, whereas NIDS provide a broader, network-wide perspective for identifying threats and potential vulnerabilities.

There is another aspect that can be distinguished regarding the detection of attacks, which relates to the method used to identify them.
There are two main ways schemes, Signature-based intrusion detection systems (SIDS) and anomaly-based intrusion detection systems (AIDS), the latter sometimes being called \textit{misuse detection}.
SIDS primarily rely on a database of known attack patterns called \textit{signatures} to detect intrusions.
This approach is effective in identifying and mitigating known threats; however, it is limited in its ability to detect novel or emerging attacks, as it requires constant updates to stay relevant.
On the other hand, AIDS leverage machine learning algorithms to establish a baseline of normal network behavior, and subsequently, identify any deviations from this established norm.
This allows for the detection of unknown or zero-day attacks, as well as subtle anomalies that may otherwise go undetected.
While anomaly-based systems offer greater adaptability and the potential for identifying novel threats, they may suffer from higher false-positive rates due to the inherent challenge of differentiating between benign irregularities and malicious activity.

Lastly, Collaborative Intrusion Detection Systems (CIDS) offer an advanced approach to network security by leveraging the collective intelligence of multiple IDS nodes.
By sharing threat information and detection patterns, CIDS can enhance the overall accuracy and efficiency of intrusion detection.
This collaborative effort enables the system to adapt more quickly to emerging threats, minimizing the likelihood of false positives and negatives.
Furthermore, CIDS can effectively counteract sophisticated attacks that exploit the limitations of isolated IDS deployments.

On the topic of AIDS and CIDS, blockchain is sometimes used in support of federated learning to enhance the security and privacy of the system.
Federated learning is a decentralized machine learning approach where multiple devices or nodes collaboratively train a global model.
Blockchain is favored to fill in this role, because it can be used to store and verify the integrity of the model and training data, as well as to ensure that the participating nodes are trustworthy.

When looking at the literature, there are basically two ways to use blockchain for Intrusion Detection Systems.
Participants can either share the data, or can share the models.
Sharing models is particularly useful when data privacy is a concern.

\begin{figure}
    \centering
    \includegraphics[width=0.6\linewidth]{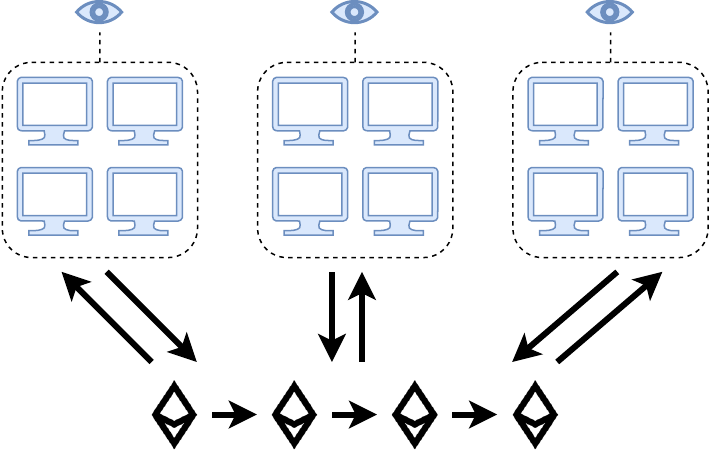}
    \caption{Intrusion detection systems sharing data or models to train a global collaborative model.}\label{fig:ids-data}
\end{figure}

\Cref{fig:ids-data} represents intrusion detection systems collaborating.
What they share using the blockchain is either the data used for the detection of the attacks, or their models directly.

Several papers share data to achieve their CIDS.
Tug~\textit{et al.} propose to use the blockchain to share the signatures used in a SIDS, in a scheme they name CBSigIDS~\cite{tug2018cbsigids}.
In their paper, they show blockchain enhances the effectiveness and robustness of Signature-based IDSs under adversarial scenarios.
In a later study, Tug~\textit{et al.} evaluate the performance of CBSigIDS under different scenarios~\cite{li2019designing}.
Kolokotronis\textit{et al.} propose a packet-filtering mechanism based on CIDS and blockchain~\cite{kolokotronis2019blockchain}.
They propose a consensus protocol which combines Proof of Work and Proof of Stake.
They later propose their CIDS for the Internet of Medical Things~\cite{kolokotronis2022intelligent}.
Alexopoulos~\textit{et al.} propose a generic architecture to incorporate blockchain in CIDS, and discuss about design decisions for such systems~\cite{alexopoulos2018towards}.
After discussing design decisions, they present TRIDEnT~\cite{alexopoulos2019trident,alexopoulos2020trident}, a collaborative platform aiming to incentivize parties to exchange data.
TRIDEnT enables entities in potentially competitive positions to selectively share, market, and obtain security notifications.
Alexopoulos~\textit{et al.} then introduce a game-theoretic model for alert sharing, and demonstrate that collaboration will inevitably happen.
Yenugunti and Yau present an approach using a trust score as a reputation system, with the goal of sharing logs~\cite{yenugunti2020blockchain}.
Fan~\textit{et al.} build a prototype using smart contracts and IPFS~\cite{fan2020blockchain}.
They use this scheme to efficiently share and distribute detection signatures in real-time.
Kumar and Singh propose utilizing the cloud as a support to a blockchain-based CIDS~\cite{kumar2020distributed}.
Alkadi~\textit{et al.} propose a framework to enable CIDS for protecting IoT and Cloud networks~\cite{alkadi2020deep}.
Liu~\textit{et al.} work on adding an SGX component to their IDS~\cite{liu2021sedid}, while Saveetha and Maragatham use deep learning~\cite{saveetha2022design}.
Alevizos~\textit{et al.} build their CIDS within a zero-trust architecture~\cite{alevizos2022blockchain}.
Abdel-Basset~\textit{et al.} use a CIDS for the Industrial Edge of Things, and test their proof of concept using two public datasets~\cite{abdel2022privacy}.

Some papers share their models to achieve a collaborative IDS.
Preuveneers~\textit{et al.} propose a federated learning AIDS scheme, where they consider an adversary trying to poison one of the local machine learning models with malicious data~\cite{preuveneers2018chained}.
They use a blockchain to support the machine learning process, providing them with full transparency and a way to hold the adversary accountable.
Golomb~\textit{et al.} propose CIoTA, an AIDS using the blockchain to perform distributed and collaborative anomaly detection for devices with limited resources in the IoT~\cite{golomb1803ciota}.
Some papers use blockchain to achieve a Snort-based IDS~\cite{ujjan2019snort,gurung2022cids}.
In a more specific context, Li~\textit{et al.} propose a challenge-based CIDS using blockchain to enhance the robustness of said challenges~\cite{li2019towards}.
In another paper, they defend against advanced insider attacks and evaluate their approach in both a simulated and real network environment~\cite{meng2020enhancing}.
They evaluate their approach with an alarm aggregation system.
Li~\textit{et al.} then later use their challenge-based CIDS as a test case to describe a general framework for CIDS in SDN~\cite{li2020framework,li2021toward,li2022blockcsdn}.
Rathore~\textit{et al.} propose a CIDS to share Signature-based IDS models between parties in the context of IoT networks~\cite{rathore2019blockseciotnet}.
Santin~\textit{et al.} use random forests as well as a secure execution environment to achieve their CIDS~\cite{santin2022framework}.

Some data-sharing CIDSs concentrate on a domain in particular.
Demertzis proposes a CIDS for smart cities~\cite{demertzis2021blockchained}.
A few papers concentrate on power systems~\cite{pahlevan2021secure,ramanan2021blockchain}.
Aljuhani focuses on the Internet of Medical Things~\cite{aljuhani2022ids}.
Likewise, some model-sharing CIDSs concentrate on a specific domain.
For example, Zhang~\textit{et al.} use a blockchain-based CIDS in the context of the Industrial IoT~\cite{zhang2020blockchain}.
Here, they use their system to detect device failures.
Adhikari and Davis on the other hand propose the use of CIDS for aviation cybersecurity~\cite{adhikari2020application}.
Liu~\textit{et al.} develop their CIDS in the context of vehicular edge computing~\cite{liu2021blockchain}.
Closely related, He~\textit{et al.} give their approach for unmanned aerial vehicles~\cite{he2022cgan}.
Like Aljuhani~\cite{aljuhani2022ids}, Ashraf~\textit{et al.} consider the Internet of Medical Things, but use a model-sharing approach~\cite{ashraf2022fidchain}.

Lastly, several papers study the suitability of blockchain for Intrusion Detection Systems~\cite{dawit2020suitability,diro2021comprehensive}.
In relation, Winanto~\textit{et al.} design a consensus algorithm specifically for collaborative SIDS~\cite{winanto2021designing}.
At the frontier of IDS and DDoS defense, Kumar~\textit{et al.} propose to use a CIDS to detect DDoS attacks in IoT networks~\cite{kumar2022distributed}.
Blockchain offers several benefits for intrusion detection systems, such as full transparency and traceability.
This allows participants to oversee data being shared.
Participants can also track who shares what data, meaning a reputation system can be created to remove bad actors.
Unrelated to Intrusion Detection, blockchain has also been used in other federated learning applications~\cite{wang2022infedge,singh2023fusionfedblock,tang2022blockchain}.

Overall, the archetypes of blockchain usage for intrusion detection systems are quite simple.
What varies greatly in the papers using blockchain to do intrusion detection is the context they are applied in, and, when applicable, the machine learning models.

\begin{figure}
    \centering
    \includegraphics[width=0.6\linewidth]{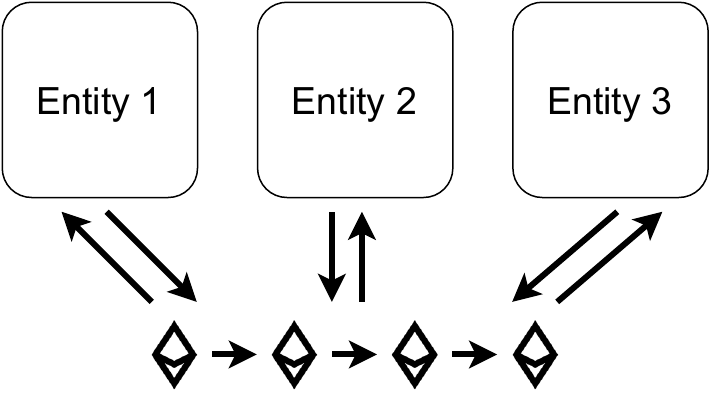}
    \caption{Entities using blockchain as a decentralized record management system.}\label{fig:data-sharing}
\end{figure}

\paragraph*{Secure Data Sharing}
Data sharing is often considered the primary and most common use case for blockchain technology, serving as a decentralized record management system.
Compared to the two previous use cases which also securely share data, here we focus on the more general case.
Blockchain enables secure data sharing by providing robust authentication, confidentiality, and accountability measures.
By utilizing blockchain for data sharing, organizations can enhance their cybersecurity efforts while maintaining full control over their sensitive information.
\Cref{fig:data-sharing} depicts different entities using blockchain as a decentralized record management system.

Azaria~\textit{et al.} propose to use blockchain to handle the access to medical records~\cite{azaria2016medrec}.
This enables patients to easily access their medical data, as well as an immutable log of access.
The paper paves the way to medical data economics, giving researchers access to the data, and patients the choice the release it.
Akarca~\textit{et al.} also consider medical records, detailing the kind of blockchain to use~\cite{akarca2019blockchain}.

Christidis and Devetsikiotis on the other hand explore the use of blockchain for IoT, mostly for process automation~\cite{christidis2016blockchains}.
Abdulkader~\textit{et al.} propose a scheme involving edge computing to scale the blockchain for the IoT~\cite{abdulkader2019lightweight}.
Neureither~\textit{et al.} leverage the blockchain to manage a trust relation graph for trust relationships in the IoT~\cite{neureither2020legiot}.
Close to the IoT, Ossamah considers the use of blockchain as a means to enhance drone cybersecurity~\cite{ossamah2020blockchain}.
Lee and Kwon use distributed watchdogs enhanced with blockchain to securely manage software updates in the industrial IoT~\cite{lee2021distributed}.

Blockchain can also be useful for supply chain management, where parties need to collaborate and share information about products and deliveries.
Mylrea and Gourisetti for example examine how blockchain can help meet compliance standards and cybersecurity for the power grid~\cite{mylrea2018blockchain}.
Gürcan~\textit{et al.} on the other hand apply blockchain to energy performance contracts, used for energy conservation measurements~\cite{gurcan2018industrial}.
They store predictive models and data in the blockchain, to ensure transparency for auditing.
Likewise, Cali~\textit{et al.} leverage blockchain for renewable energy certificates~\cite{cali2021cybersecurity}.
Demirkan~\textit{et al.} look at the impact of blockchain for business and accounting, namely in terms of cybersecurity and auditing~\cite{demirkan2020blockchain}.

Closer to cybersecurity, Brotsis~\textit{et al.} consider the use of blockchain to store forensic evidence metadata~\cite{brotsis2019blockchain}.
Shi~\textit{et al.} make a distributed honeypot, which is a decoy system designed to attract and trap malicious actors~\cite{shi2019dynamic}.
In this paper, Shi~\textit{et al.} use the blockchain to store port access data.
Hu~\textit{et al.} on the other hand use blockchain to collect user behavioral data, with the purpose of computational trust~\cite{hu2020contract}.

\subsection{Blockchain}\label{survey:bc}

In this section, we take a look at blockchains used in collaborative cybersecurity systems.
In particular, we explore underlying technologies, consensus methods, access control and data validation policies.
We find that, surprisingly, a lot of papers in the literature do not use the best possible blockchain for their use cases.

\paragraph*{Technology}

Surprisingly, a lot of the papers do not specify any particular technology, even when they implemented a proof of concept.
Twenty-one papers out of \npapers{} seem to think of and use blockchain more as a conceptual object than as an actual technology.
One reason possibly explaining this is that the paper might not need a specific technology, but rather a general-purpose blockchain, and that as such, the technology can be any kind of blockchain.

Ethereum is the most used technology with 27 papers.
This is not surprising considering the frequent use of smart contracts in the implementations.
Ten papers use custom-made blockchains, and six papers use Hyperledger Fabric.
The use of Hyperledger is to be expected considering the availability of chaincodes for implementation purposes.
Papers using custom-made blockchains are for the most part more concerned with developing a proof of concept than an actual implementation on a prominent platform.

\begin{figure}
    \includegraphics[width=0.8\linewidth]{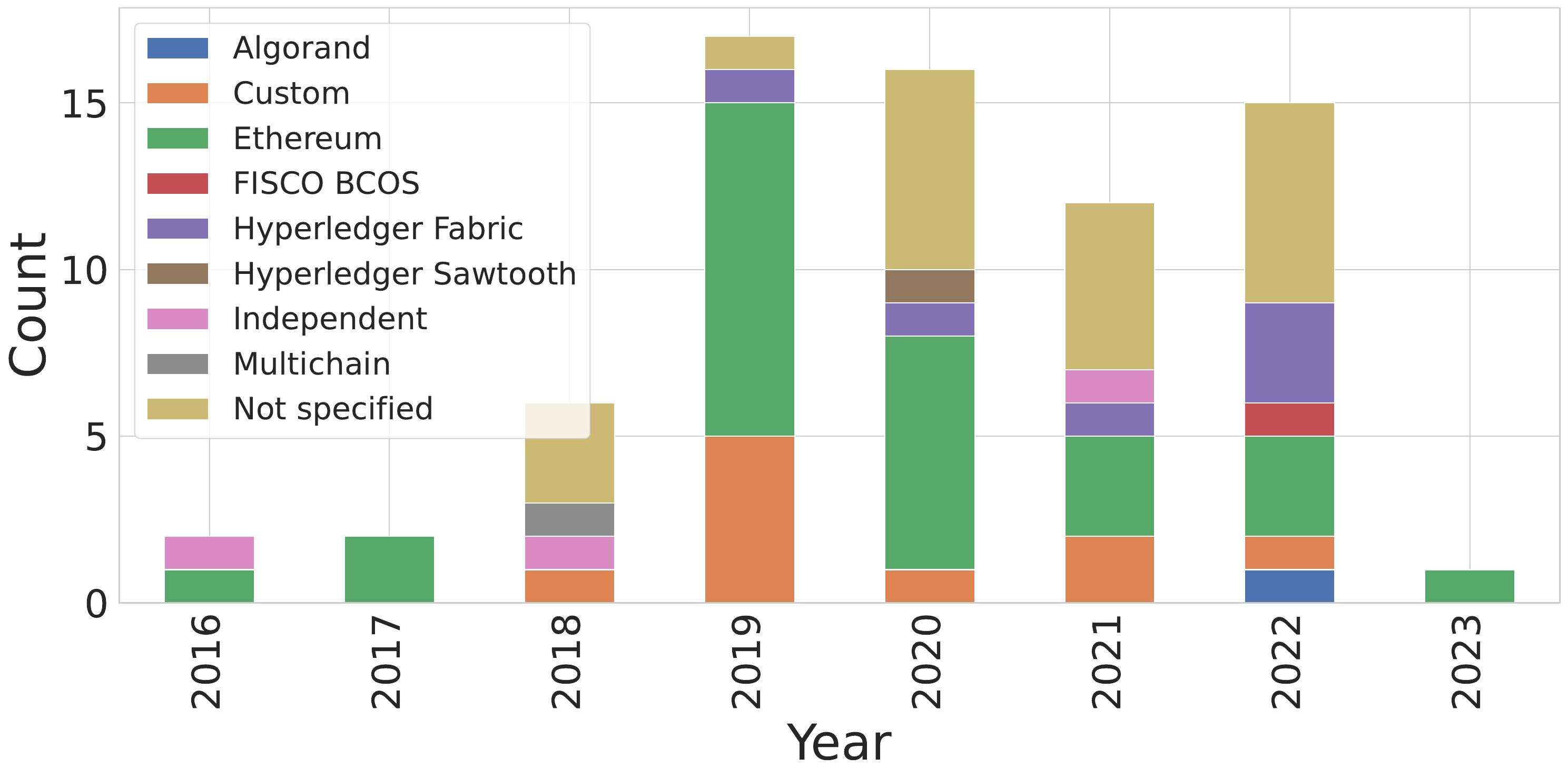}
    \caption{Number of approaches per technology per year. Ethereum usage is decreasing while Hyperledger Fabric usage is increasing.}\label{fig:technology}
\end{figure}

The use of blockchain in different technologies has evolved over the years.
\Cref{fig:technology} shows that the use of Ethereum is dwindling in favor of other technologies like Hyperledger Fabric.
Note the presence in 2022 of FISCO BCOS, which is the largest and most active consortium blockchain ecosystem in China~\cite{fisco-bcos}.

As one can see, a lot of technologies have been used over the years.
While Ethereum was more prevalent in earlier years, the field has fragmented, with competing solutions being used in its stead.

\paragraph*{Consensus}
On the side of consensus, most of the papers once again do not specify any particular consensus method.
33 out of \npapers{} papers do not specify a consensus method.
This ties in to the same remark as for technology; a specific kind of consensus might not be needed for the purposes of the paper.
Note that when a paper does not specify a blockchain technology, it will also most of the time not specify a consensus method as well, solidifying that their use of blockchain is very conceptual.

\begin{figure}
    \includegraphics[width=0.8\linewidth]{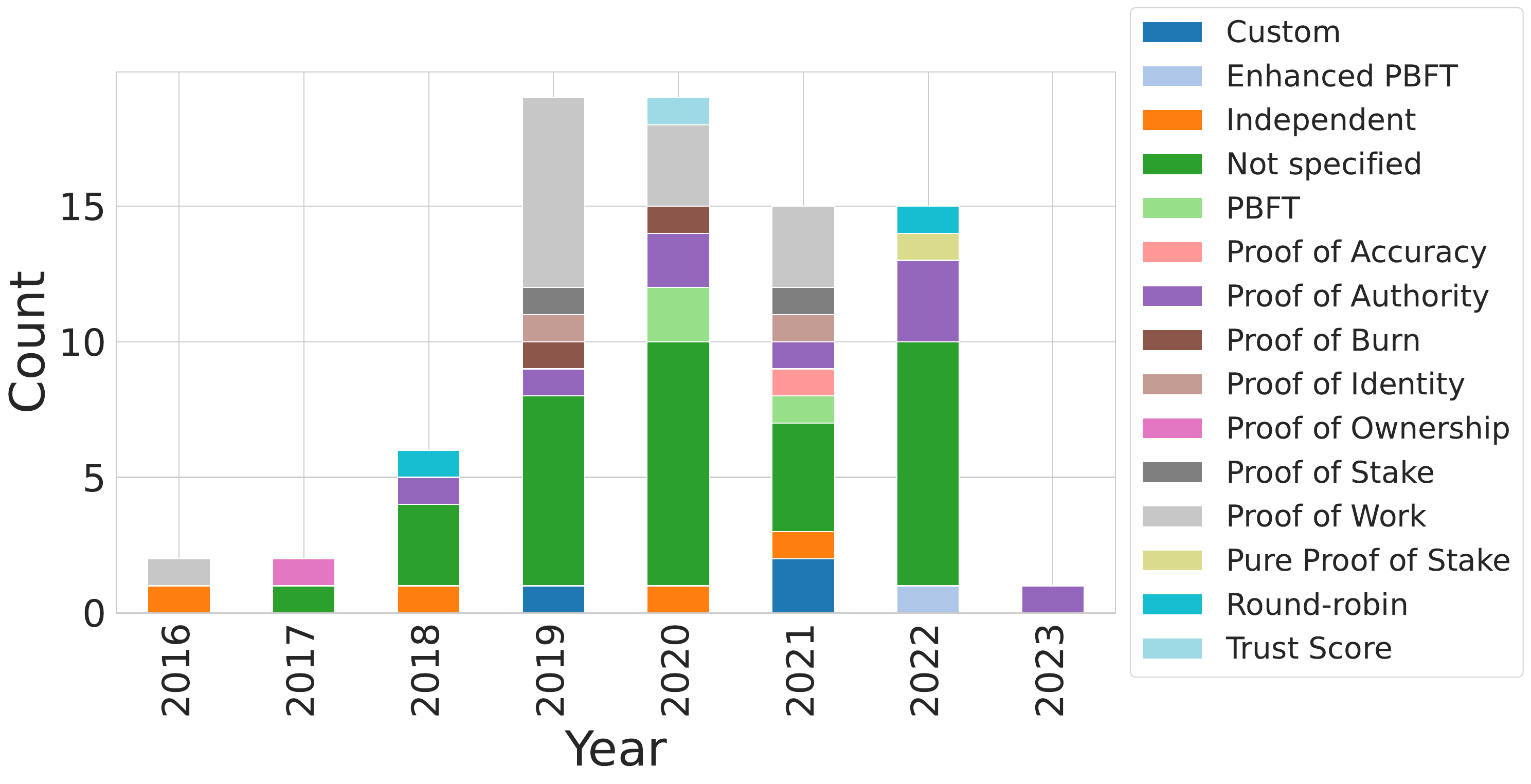}
    \caption{Number of approaches per type of consensus per year.}\label{fig:consensus}
\end{figure}

\Cref{fig:consensus} shows the number of approaches per type of consensus per year.
A lot of different consensus methods have been used over the years, with no clear solution rising above the others.
Even more so than with technology, the research community does not seem to agree on what the best consensus algorithm is for collaborative cybersecurity applications.

Here, we need to make a distinction between public and private blockchains.
Since the assumption of trust in the participants in public blockchains is lower, public blockchains should use a more robust consensus algorithm, like Proof of Work or Proof of Stake.
Private blockchains on the other hand have a better idea of who is participating in the network, since access is restricted by definition.
This means private blockchains can afford a lower trust assumption than public blockchains.
As such, applications using private blockchains should use a consensus protocol primarily as a fault-tolerant means of achieving global consistency by inducing a tamper-proof record~\cite{ramanan2021blockchain}.

\begin{figure}
    \includegraphics[width=0.8\linewidth]{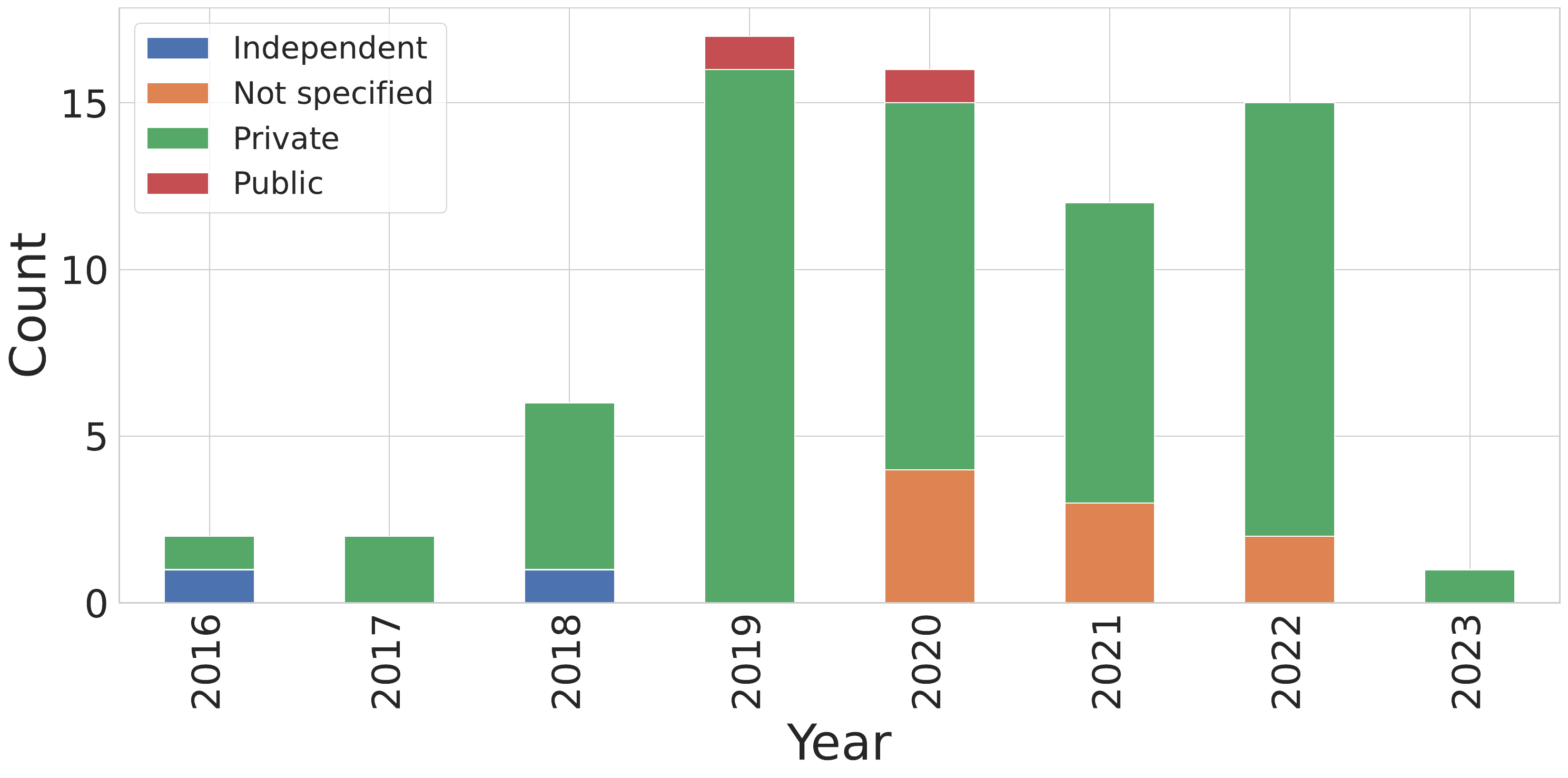}
    \caption{Approaches per access control mode per year.}\label{fig:access-control}
\end{figure}

\paragraph*{Access control}
When looking at access control, we can see that most papers use a private blockchain.
Out of \npapers{} papers, 58 use a private blockchain, which is understandable since collaborative cybersecurity systems usually look to involve a reduce, private set of participants, and where not just anyone can join.
\Cref{fig:access-control} shows just that.

Some attempts at using public blockchains have been made, but overall they are way less prevalent.
This is understandable, because it presents itself as a risk to collaborative cybersecurity systems.
On the one hand, they can gain access to more data, or participants to increase the security of the overall system.
On the other, they risk introducing malicious actors into the system, or even participants which will decrease the overall quality of the data shared on the blockchain.
On top of that, there is always the risk of giving sensitive insights to unwanted parties.
This is relevant in the case of DDoS defense, where attackers could see the blocked IP addresses.
It is also relevant for intrusion detection systems, since an attacker could learn about the systems used to keep him out by learning about the model.

Once again for this characteristic, nine papers do not specify any kind of access control method.
Out of those, some papers just use blockchain as a concept, while others simply omit the fact they use a private blockchain, which we can gather from the application and context of the papers.

\begin{figure}
    \includegraphics[width=0.8\linewidth]{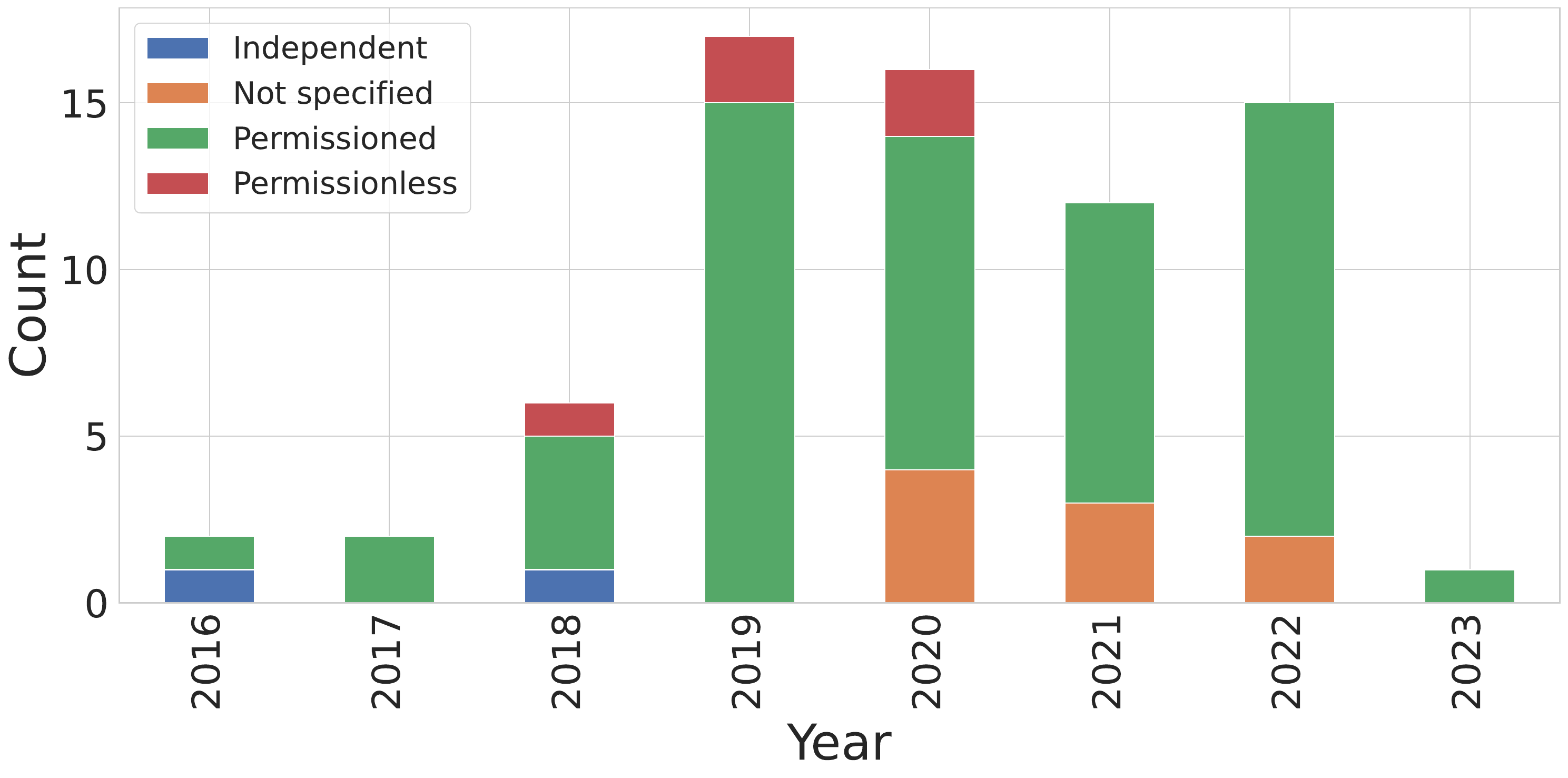}
    \caption{Approaches per permission mode per year.}\label{fig:permission}
\end{figure}

\paragraph*{Data validation policy}
On the side of data validation policy, we can see that most papers prefer using a permissioned blockchain.
55 out of \npapers{} papers use permissioned blockchains.
This is illustrated in \Cref{fig:permission}.
For the same reasons as the prevalent use of private blockchains over public blockchains, most papers prefer to limit the data validation process to just a few actors using the blockchain.
Unsurprisingly, most private blockchains are also permissioned, whereas most public blockchains are permissionless.

Like with access control, nine papers do not specify any kind of data validation policy.
For the same reasons as before, some papers just use blockchain as a concept, while others simply omit the fact they use a permissioned blockchain.

\subsection{Field fragmentation}\label{survey:frag}

The last sections show that the field of blockchain for collaborative cybersecurity is still very fragmented.
This is probably due at least partly to the freshness of blockchain.
Even though the first use of a decentralized blockchain dates back to 2008, it was not until 2015 that smart contracts were introduced.
Moreover, it was not until 2019 that blockchain started really picking up pace, and that the research community started to use this technology for collaborative cybersecurity applications.

The following paragraphs explore additional aspects of the surveyed papers in order to demonstrate the extent of the field's fragmentation.
Namely, we look at the domains the applications were written for, the relevant venues papers were published in, as well as the research groups and the countries the authors came from.

\paragraph*{Domains}

\begin{figure}
    \includegraphics[width=0.8\linewidth]{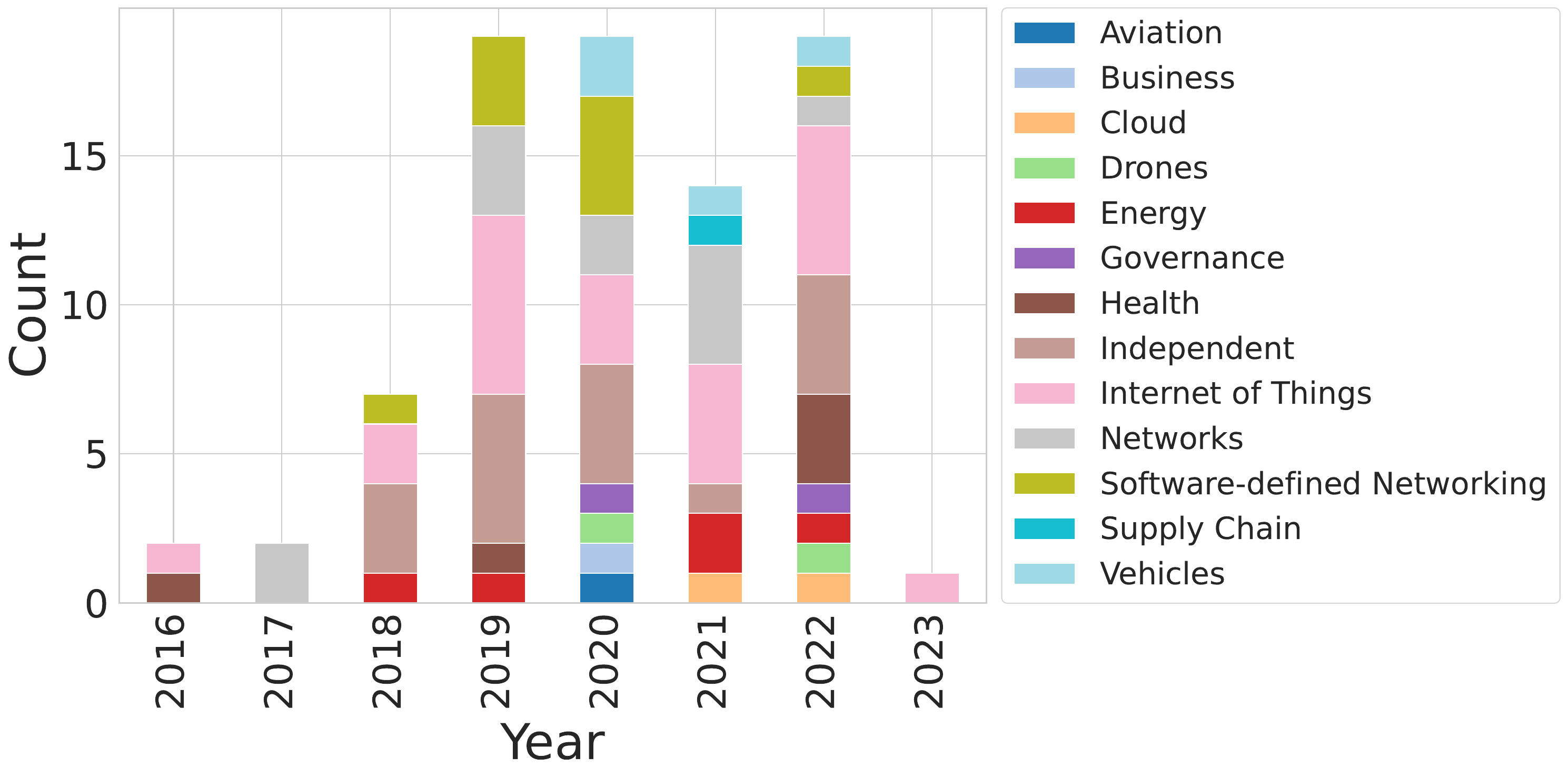}
    \caption{Number of approaches per domain per year. A paper matching multiple domains is counted for each one of those domains.}\label{fig:domain}
\end{figure}

\Cref{fig:domain} depicts the number of approaches per domain per year.
Note that if a paper matched multiple domains, it was counted for each one of those domains.
The figure shows that the Internet of Things is the most popular domain, with 22 papers in total.
Seventeen papers were published independently of any particular domain.
Following domains include Networks and Software-defined Networking, with twelve and nine papers respectively.

One can see with \Cref{fig:domain} that trends seem to remain stable over this time period.
This lack of convergence as well as the high number of different domains are more proof that this domain is still very fragmented, even today.

\paragraph*{Countries}
We also consider in this study the country of origin of the papers.
Note that like for the domains, we also count a paper for multiple countries if the authors were from multiple countries.
Analyzing the papers, we found that China was most represented, with eighteen papers having at least one author from China.
China is closely followed by the United States of America, with twelve papers having at least one author working in America.
This observation does not necessarily mean that China or the United States of America publish more than other countries on the topic because they have a particular interest in it.
Indeed, what we measure here might just be a side effect of population size, with China and the United States of America publishing more papers than other countries overall anyways~\cite{scimagojrpapercount}.
We argue that this finding is still interesting to point out, since it indicates a global tendency on the topic regardless of the reasons for why such a tendency would emerge in the first place.

\begin{figure}
    \includegraphics[width=\linewidth]{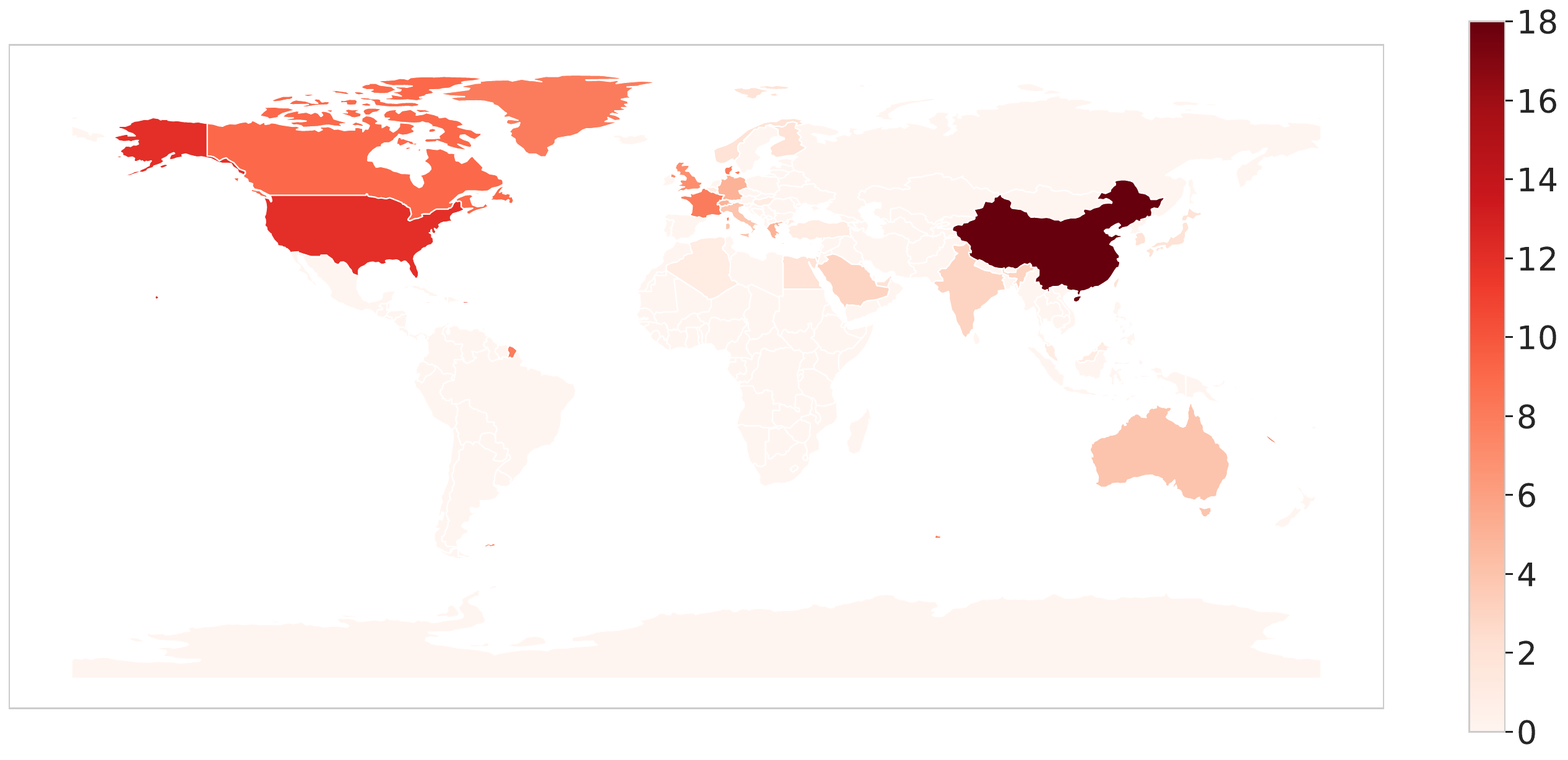}
    \caption{Number of papers per country of origin. A paper with authors from multiple countries counts for each one of those countries.}\label{fig:map}
\end{figure}

\Cref{fig:map} represents the number of papers originating from a specific country.
This map can be referenced for further inspection about which countries published on the topic.
As before, this finding highlights great diversity and fragmentation when it comes to blockchain for collaborative cybersecurity.

\paragraph*{Research groups}
We analyzed the research groups that authored the papers considered in the study.
Like for the domains and countries, we count a paper for multiple research groups if its authors belong to more than one group.
The research groups that contributed the most, with five papers each, were the University of Technology of Troyes, the University of Montréal and the University of Zürich.
Overall, there does not seem to be a prevalent research group, as other groups are not far behind those universities in terms of number of papers.
This is yet another confirmation that research on the topic is still very fragmented, with no group or groups contributing a significant share of the research on the topic.
\Cref{fig:groups} in \Cref{appendix} shows the number of papers per research group per year.
There does not seem to be any particular trend when it comes to the research groups, and the figure indicates a shift in research groups over the years.

\paragraph*{Relevant venues}
We can reach the same conclusion as with research groups when looking at relevant venues.
The top venues to publish in are IEEE Access and the IEEE Internet of Things Journal, with five and four papers respectively.
\Cref{fig:venues} in \Cref{appendix} represents the number of papers per venue per year.
As with the research groups, there is a shift in venues those works are published in over the years.
Yet another bit of evidence that the field is still new and fragmented.
\section{Open Issues and Lessons Learned}\label{lessons}

\subsection{Lessons Learned}

Blockchain technology has rapidly gained traction in the cybersecurity landscape due to its decentralized nature and robust security features.
In the following section, we detail some of the lessons learned from examining the literature around blockchain for collaborative cybersecurity.

\textbf{Explore minor applications and new use cases} --- As we have seen, blockchain has been used in various cybersecurity applications.
Prominent applications include DDoS defense, intrusion detection and data sharing, while minor cybersecurity applications include forensics, distributed honeypots and trust management.
It is crucial to explore those minor applications further to uncover their potential benefits and discover new use cases in the cybersecurity domain.

\textbf{Choose your blockchain meticulously} --- The current literature on the application of blockchain technology in cybersecurity reveals certain shortcomings, particularly in the areas of blockchain selection and consensus mechanisms.
Many papers fail to utilize the most appropriate blockchain for their specific use cases, and some even neglect to specify the type of blockchain employed in their research.
This lack of clarity can hinder the effectiveness of proposed solutions and impede the overall progress of the field.
To address these issues, it is crucial for researchers to meticulously and transparently select the nature of the blockchain they intend to use for their applications.
This process should involve a thorough analysis of the specific requirements and constraints of the use case, as well as a comprehensive understanding of the advantages and limitations of various blockchains.

Similarly, the choice of consensus algorithm is equally important for the success and effectiveness of a blockchain-based solution.
Many researchers have been found to overlook the significance of choosing an appropriate consensus mechanism that aligns with the nature of their selected blockchain.
This oversight can result in suboptimal performance and reduced scalability.
For instance, in the case of public permissionless blockchains, researchers should consider consensus mechanisms like Proof of Work (PoW) or Proof of Stake (PoS), which are designed to maintain security and decentralization in an open environment.
On the other hand, for private permissioned blockchains, algorithms like PBFT are more suitable, as they can ensure quick and secure consensus among a predetermined set of trusted nodes.

By carefully selecting the most appropriate blockchain type and consensus mechanism, researchers can develop more effective and efficient solutions for their cybersecurity applications.
This attention to detail will not only enhance the robustness of the proposed solutions but also foster greater confidence and understanding within the research community, ultimately driving the continued evolution and maturation of the field.

\textbf{Blockchain technology is still emerging} --- The nascent nature of blockchain technology in the realm of cybersecurity is apparent, with the first academic papers and discussions surfacing around 2016 and 2017.
The current state of fragmentation and the absence of a unified approach, as illustrated in \Cref{survey:frag}, further confirms this observation.

Nevertheless, as the field continues to mature and evolve, we can anticipate significant advancements and innovations in leveraging blockchain technology to address various cybersecurity challenges.
As researchers and industry experts collaborate and share insights, we can expect the development of standardized frameworks and methodologies for implementing blockchain-based solutions in cybersecurity.
These standardized approaches will not only facilitate broader adoption but also provide a foundation for evaluating the effectiveness and efficiency of blockchain applications in this context.

Moreover, as more use cases and real-world applications emerge, the field will gain a deeper understanding of the potential limitations and drawbacks of blockchain technology in cybersecurity.
This knowledge will prove instrumental in refining existing solutions and addressing any technological gaps or vulnerabilities.

\textbf{Lack of collaboration incentives} --- Incentives for collaboration remain a critical question in the literature~\cite{han2022can}.
In previous works, there has been a notable absence of incentive mechanisms within the realm of blockchain technology for collaborative cybersecurity.
This oversight is particularly striking given that incentive mechanisms play a crucial role in fostering active participation and collaboration between parties.

Despite the importance of incentive mechanisms, the literature has largely glossed over this aspect, focusing instead on other technological aspects like specific machine learning methods.
This neglect has left a significant gap in our understanding of how to design and implement effective incentive structures within blockchain networks, which is essential for their long-term sustainability and widespread adoption.
In the context of DDoS defense for example, an Autonomous System has incentives beyond rewards to mitigate these attacks, as it saves them bandwidth.
Researchers should explore incentives further to understand the motivation for collaboration and their role in addressing cybersecurity challenges.

\subsection{Blockchain Design Choices}\label{survey:choice}

The current literature highlights certain shortcomings in the design choices made when applying blockchain technology to cybersecurity.
These shortcomings primarily pertain to the selection of the appropriate blockchain type and consensus mechanism for specific use cases.
Insufficient attention is given to these crucial decisions, leading to less effective and optimized solutions.
To address these issues, researchers and practitioners must meticulously evaluate and select the most suitable blockchain for their applications.

Greenspan~\cite{greenspan2015} has proposed eight requirements to ensure justified use of blockchain technology.
These requirements include the need for a shared ledger, involvement of multiple writers, absence of trust between writers, avoidance of centralized third parties, transaction interaction, transaction verifiability, existence of validators, and the storage of value.

\begin{figure}[ht]
    \centering
    \includegraphics[width=\linewidth]{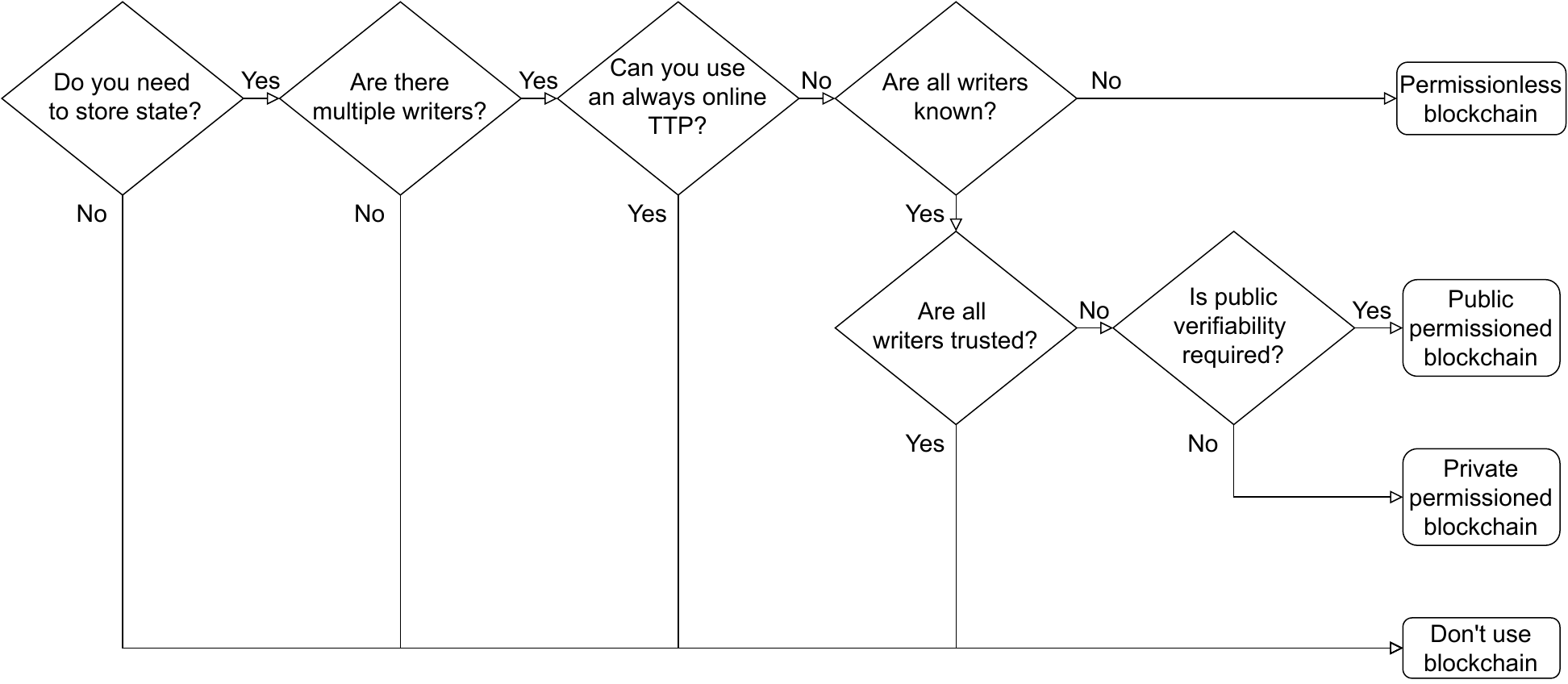}
    \caption{Process to determine if blockchain is an appropriate solution (Following Wüst and Gervais~\cite{wust2018you}).}\label{fig:use-flowchart}
\end{figure}

Additionally, Wüst and Gervais~\cite{wust2018you} have devised a process for determining the suitable type of blockchain, or whether a blockchain should be used at all.
This process is depicted in Figure~\ref{fig:use-flowchart}.

Based on the insights gained from our study, we can offer guidance on the selection of blockchain type and consensus mechanisms for different scenarios.
Considering Figure~\ref{fig:use-flowchart}, in the case of private permissionless blockchains, we recommend to follow the same decision process as for private permissioned blockchains.
The decision of making the private blockchain permissioned or permissionless depends on whether all participants need to validate blocks, or just a subset.

Regarding the choice of underlying technology, we advise to rely on well-established projects such as Ethereum or Hyperledger.
These platforms have a proven track record and extensive community support, which can contribute to the stability and reliability of the cybersecurity system.

In public blockchains where the trust assumption among participants is lower, it becomes crucial to employ a robust consensus algorithm. Consensus mechanisms like Proof of Work (PoW) or Proof of Stake (PoS) are designed to ensure security and decentralization in an open environment.

On the other hand, private blockchains have more control over participant access since access is restricted by design.
This allows for a lower trust assumption compared to public blockchains.
In such cases, the consensus protocol should primarily serve as a fault-tolerant mechanism to achieve global consistency and maintain a tamper-proof record~\cite{ramanan2021blockchain}, like PBFT.

By adhering to a meticulous selection process guided by the decision flowchart provided by Wüst and Gervais as well as our added insights on technology and consensus algorithm choice, researchers and practitioners can make informed decisions regarding their choice of blockchain.
Careful consideration will lead to more effective solutions for cybersecurity applications, improving the overall robustness of those systems.
Ultimately, it will contribute to the continued evolution and maturation of the field.

\subsection{Open Issues}\label{survey:open}

While significant progress has been made in the field of collaborative cybersecurity using blockchain technology, there remain several open issues that warrant further exploration and research.
Some of the open issues have to do with certain applications that have been less explored.

\textbf{Collaborative intrusion response} --- As we saw in this study, a lot of papers focus on intrusion detection, but very few focus on the response to this intrusion.
Intrusion response systems are essential to develop effective countermeasures, minimize the damage caused by unauthorized access to computer networks, limit the impact of any breaches that do occur, and ultimately reduce the risk of future attacks.
More attention should be given to this area, particularly to the automation of response to cyber threats.
Like with collaborative intrusion detection systems, blockchain should be a part of collaborative intrusion response systems.

\textbf{Public collaborative cybersecurity}  --- Collaborative cybersecurity systems using public blockchains remain relatively scarce.
For example, out of all the intrusion detection papers, only one considered using a public permissionless blockchain.
While many organizations prefer to keep their blockchain networks private and permissioned for security reasons, a functioning public system could potentially provide more data to train intrusion detection systems better, leading to improved cybersecurity measures.

\textbf{Self-Sovereign Identities} --- The concept of self-sovereign identities is an area that requires more investigation.
Identity is the foundation of security.
Self-Sovereign Identities (SSIs) are an innovative approach to digital identity management that empowers users with increased control over their personal information, leading to enhanced privacy and security.
In time, those SSIs could constitute the basis for many collaborative cybersecurity applications, as they facilitate the process of trust establishment in those systems.

Several open issues have more to do with intrisic properties of the blockchain itself.

\textbf{Scalability} --- Scalability remains an important challenge for blockchain technology.
Distributed ledgers have scalability limitations when it comes to handling large amounts of data. This can create a bottleneck when trying to share and collaborate on cybersecurity data in real-time.
Developing efficient compression schemes and sharding techniques will be necessary to ensure that the blockchain can handle increasing transaction volumes and data storage demands, especially in the case of public blockchains and certain environments like the Internet of Things.

\textbf{Interoperability} --- Interoperability is another crucial aspect that needs to be addressed.
Different blockchain platforms may not be able to communicate with each other, creating silos of data and limiting collaboration.
Interoperability standards need to be established to allow different blockchain networks to interoperate seamlessly.
In the case of collaborative cybersecurity applications, such interoperability could facilitate collaboration between various collaborative applications with different but converging purposes.
For example, one could imagine a forensics application using blockchain collaborating with a blockchain-based CIDS by sharing its data.

\textbf{Legal and regulatory compliance} --- Legal and regulatory compliance is a significant concern when implementing collaborative cybersecurity solutions using distributed ledgers.
Participants need to ensure that their activities are compliant with relevant laws and regulations, such as data protection and privacy laws.
GDPR compliance in particular poses challenges for blockchain implementation, especially regarding the right to erasure.
The permanence of data on the blockchain conflicts with the GDPR's prohibition on storing Personally Identifiable Information in Europe.
As such, there is a need for blockchain-based collaborative cybersecurity systems that can handle this right to erasure.
Researchers and practitioners should carefully consider these compliance issues when designing and implementing blockchain-based cybersecurity solutions.

\section{Conclusion}\label{conclusion}

This paper presented a comprehensive literature review on the use of blockchain technology in collaborative cybersecurity systems.
After giving an overview on the domain, we identified key applications and trends in blockchain usage in these systems. We also established open research issues and lessons learned to guide future work in the area.

In \Cref{introduction}, we introduced the concept of collaborative cybersecurity and the benefits of using decentralized approaches.
\Cref{related} delved into related work.
We detailed our research methodology in \Cref{methodology}.
\Cref{background} gave a detailed background on blockchain technology, covering nature, technology, and consensus.
We detailed our findings on the identified papers in \Cref{survey}.
It contains a description of the archetypes of collaborative cybersecurity applications using blockchain.
We analyzed the papers in terms of several criteria, e.g.\ access control method, over time.
In \Cref{lessons}, we gave guidelines on an informed choice of the right blockchain technology based on the desired applications.
We discussed open issues and lessons learned, identifying challenges and potential areas for further research.

Regarding our research questions, the main use cases for collaborative cybersecurity applications using blockchain currently are DDoS defense, intrusion detection systems, and secure data sharing (RQ1).
Blockchain technology has been used in more than ten domains, as shown in \Cref{fig:domain} (RQ2).
The blockchains used in collaborative cybersecurity systems (RQ3) are mostly private (\Cref{fig:access-control}), permissioned (\Cref{fig:permission}), but are not characterized by a particular technology (\Cref{fig:technology}) or consensus method (\Cref{fig:consensus}).
Over time (RQ4), collaborative cybersecurity using blockchain has attracted more and more contributions (\Cref{fig:count}), with an increasing number of contributions targeting intrusion detection systems (\Cref{fig:application}).
Nevertheless, the field is still fragmented (\Cref{fig:technology}, \Cref{fig:consensus}, \Cref{fig:domain}), with no prevalent research group (\Cref{fig:groups}) or venue (\Cref{fig:venues}).

Overall, our work shows that blockchain technology holds great potential in the field of collaborative cybersecurity due to its decentralized nature (RQ5).
However, blockchain selection and consensus mechanisms are often chosen poorly or without sufficient justification, negatively impacting the effectiveness of the proposed solutions.

After guiding the reader towards a more-informed choice of blockchain technology,
we showed in \Cref{lessons} the literature has generally overlooked the importance of collaboration incentives, leaving a critical gap in our understanding of how to foster active participation in blockchain networks.

Our study highlighted numerous open challenges that warrant further research.
Key concerns involve collaborative efforts in intrusion response systems, the application of public blockchains to collaborative cybersecurity systems and self-sovereign identities.
Focusing on the blockchain aspect, open issues include enhancing scalability, fostering interoperability, and ensuring compliance with legal and regulatory standards for the blockchains that underpin these cooperative cybersecurity initiatives.

As the field continues to mature, we expect significant advancements and innovations in leveraging blockchain technology to address various cybersecurity challenges.
By addressing these open issues and learning from the lessons identified in this review, researchers and practitioners can develop more effective and efficient blockchain-based solutions for collaborative cybersecurity applications, ultimately enhancing the overall security of organizations and their ability to protect against cyber attacks.

\begin{acks}
  Supported by the industrial chair Cybersecurity for Critical Networked Infrastructures (cyberCNI.fr) with support of the FEDER development fund of the Brittany region.
\end{acks}

\bibliographystyle{ACM-Reference-Format}
\bibliography{bibliography}


\begin{thebibliography}{113}


\ifx \showCODEN    \undefined \def \showCODEN     #1{\unskip}     \fi
\ifx \showDOI      \undefined \def \showDOI       #1{#1}\fi
\ifx \showISBNx    \undefined \def \showISBNx     #1{\unskip}     \fi
\ifx \showISBNxiii \undefined \def \showISBNxiii  #1{\unskip}     \fi
\ifx \showISSN     \undefined \def \showISSN      #1{\unskip}     \fi
\ifx \showLCCN     \undefined \def \showLCCN      #1{\unskip}     \fi
\ifx \shownote     \undefined \def \shownote      #1{#1}          \fi
\ifx \showarticletitle \undefined \def \showarticletitle #1{#1}   \fi
\ifx \showURL      \undefined \def \showURL       {\relax}        \fi
\providecommand\bibfield[2]{#2}
\providecommand\bibinfo[2]{#2}
\providecommand\natexlab[1]{#1}
\providecommand\showeprint[2][]{arXiv:#2}

\bibitem[Abdel-Basset et~al\mbox{.}(2022)]%
        {abdel2022privacy}
\bibfield{author}{\bibinfo{person}{Mohamed Abdel-Basset}, \bibinfo{person}{Nour
  Moustafa}, {and} \bibinfo{person}{Hossam Hawash}.}
  \bibinfo{year}{2022}\natexlab{}.
\newblock \showarticletitle{Privacy-Preserved Cyberattack Detection in
  Industrial Edge of Things (IEoT): A Blockchain-Orchestrated Federated
  Learning Approach}.
\newblock \bibinfo{journal}{\emph{IEEE Transactions on Industrial Informatics}}
  \bibinfo{volume}{18}, \bibinfo{number}{11} (\bibinfo{year}{2022}),
  \bibinfo{pages}{7920--7934}.
\newblock


\bibitem[Abdulkader et~al\mbox{.}(2019)]%
        {abdulkader2019lightweight}
\bibfield{author}{\bibinfo{person}{Omar Abdulkader}, \bibinfo{person}{Alwi~M
  Bamhdi}, \bibinfo{person}{Vijey Thayananthan}, \bibinfo{person}{Fathy
  Elbouraey}, {and} \bibinfo{person}{Bandar Al-Ghamdi}.}
  \bibinfo{year}{2019}\natexlab{}.
\newblock \showarticletitle{A lightweight blockchain based cybersecurity for
  IoT environments}. In \bibinfo{booktitle}{\emph{2019 6th IEEE International
  Conference on Cyber Security and Cloud Computing (CSCloud)/2019 5th IEEE
  International Conference on Edge Computing and Scalable Cloud (EdgeCom)}}.
  IEEE, \bibinfo{pages}{139--144}.
\newblock


\bibitem[Abou El~Houda et~al\mbox{.}(2019a)]%
        {abou2019co}
\bibfield{author}{\bibinfo{person}{Zakaria Abou El~Houda},
  \bibinfo{person}{Abdelhakim Hafid}, {and} \bibinfo{person}{Lyes Khoukhi}.}
  \bibinfo{year}{2019}\natexlab{a}.
\newblock \showarticletitle{Co-IoT: A collaborative DDoS mitigation scheme in
  IoT environment based on blockchain using SDN}. In
  \bibinfo{booktitle}{\emph{2019 IEEE Global Communications Conference
  (GLOBECOM)}}. IEEE, \bibinfo{pages}{1--6}.
\newblock


\bibitem[Abou El~Houda et~al\mbox{.}(2020a)]%
        {abou2020brainchain}
\bibfield{author}{\bibinfo{person}{Zakaria Abou El~Houda},
  \bibinfo{person}{Abdelhakim Hafid}, {and} \bibinfo{person}{Lyes Khoukhi}.}
  \bibinfo{year}{2020}\natexlab{a}.
\newblock \showarticletitle{Brainchain-a machine learning approach for
  protecting blockchain applications using sdn}. In
  \bibinfo{booktitle}{\emph{ICC 2020-2020 IEEE International Conference on
  Communications (ICC)}}. IEEE, \bibinfo{pages}{1--6}.
\newblock


\bibitem[Abou El~Houda et~al\mbox{.}(2019b)]%
        {abou2019cochain}
\bibfield{author}{\bibinfo{person}{Zakaria Abou El~Houda},
  \bibinfo{person}{Abdelhakim~Senhaji Hafid}, {and} \bibinfo{person}{Lyes
  Khoukhi}.} \bibinfo{year}{2019}\natexlab{b}.
\newblock \showarticletitle{Cochain-SC: An intra-and inter-domain DDoS
  mitigation scheme based on blockchain using SDN and smart contract}.
\newblock \bibinfo{journal}{\emph{IEEE Access}}  \bibinfo{volume}{7}
  (\bibinfo{year}{2019}), \bibinfo{pages}{98893--98907}.
\newblock


\bibitem[Abou El~Houda et~al\mbox{.}(2018)]%
        {abou2018chainsecure}
\bibfield{author}{\bibinfo{person}{Zakaria Abou El~Houda},
  \bibinfo{person}{Lyes Khoukhi}, {and} \bibinfo{person}{Abdelhakim Hafid}.}
  \bibinfo{year}{2018}\natexlab{}.
\newblock \showarticletitle{Chainsecure-a scalable and proactive solution for
  protecting blockchain applications using sdn}. In
  \bibinfo{booktitle}{\emph{2018 IEEE global communications conference
  (GLOBECOM)}}. IEEE, \bibinfo{pages}{1--6}.
\newblock


\bibitem[Abou El~Houda et~al\mbox{.}(2020b)]%
        {abou2020bringing}
\bibfield{author}{\bibinfo{person}{Zakaria Abou El~Houda},
  \bibinfo{person}{Lyes Khoukhi}, {and} \bibinfo{person}{Abdelhakim~Senhaji
  Hafid}.} \bibinfo{year}{2020}\natexlab{b}.
\newblock \showarticletitle{Bringing intelligence to software defined networks:
  Mitigating DDoS attacks}.
\newblock \bibinfo{journal}{\emph{IEEE Transactions on Network and Service
  Management}} \bibinfo{volume}{17}, \bibinfo{number}{4}
  (\bibinfo{year}{2020}), \bibinfo{pages}{2523--2535}.
\newblock


\bibitem[Adhikari and Davis(2020)]%
        {adhikari2020application}
\bibfield{author}{\bibinfo{person}{Sam Adhikari} {and} \bibinfo{person}{Chris
  Davis}.} \bibinfo{year}{2020}\natexlab{}.
\newblock \showarticletitle{Application of blockchain within aviation
  cybersecurity framework}. In \bibinfo{booktitle}{\emph{Aiaa Aviation 2020
  Forum}}. \bibinfo{pages}{2931}.
\newblock


\bibitem[Akarca et~al\mbox{.}(2019)]%
        {akarca2019blockchain}
\bibfield{author}{\bibinfo{person}{D Akarca}, \bibinfo{person}{PY Xiu},
  \bibinfo{person}{D Ebbitt}, \bibinfo{person}{B Mustafa}, \bibinfo{person}{H
  Al-Ramadhani}, {and} \bibinfo{person}{A Albeyatti}.}
  \bibinfo{year}{2019}\natexlab{}.
\newblock \showarticletitle{Blockchain secured electronic health records:
  patient rights, privacy and cybersecurity}. In \bibinfo{booktitle}{\emph{2019
  10th International Conference on Dependable Systems, Services and
  Technologies (DESSERT)}}. IEEE, \bibinfo{pages}{108--111}.
\newblock


\bibitem[Alevizos et~al\mbox{.}(2022)]%
        {alevizos2022blockchain}
\bibfield{author}{\bibinfo{person}{Lampis Alevizos},
  \bibinfo{person}{Max~Hashem Eiza}, \bibinfo{person}{Vinh~Thong Ta},
  \bibinfo{person}{Qi Shi}, {and} \bibinfo{person}{Janet Read}.}
  \bibinfo{year}{2022}\natexlab{}.
\newblock \showarticletitle{Blockchain-Enabled Intrusion Detection and
  Prevention System of APTs Within Zero Trust Architecture}.
\newblock \bibinfo{journal}{\emph{IEEE Access}}  \bibinfo{volume}{10}
  (\bibinfo{year}{2022}), \bibinfo{pages}{89270--89288}.
\newblock


\bibitem[Alexopoulos et~al\mbox{.}(2018)]%
        {alexopoulos2018towards}
\bibfield{author}{\bibinfo{person}{Nikolaos Alexopoulos},
  \bibinfo{person}{Emmanouil Vasilomanolakis},
  \bibinfo{person}{Nat{\'a}lia~R{\'e}ka Iv{\'a}nk{\'o}}, {and}
  \bibinfo{person}{Max M{\"u}hlh{\"a}user}.} \bibinfo{year}{2018}\natexlab{}.
\newblock \showarticletitle{Towards blockchain-based collaborative intrusion
  detection systems}. In \bibinfo{booktitle}{\emph{Critical Information
  Infrastructures Security: 12th International Conference, CRITIS 2017, Lucca,
  Italy, October 8-13, 2017, Revised Selected Papers 12}}. Springer,
  \bibinfo{pages}{107--118}.
\newblock


\bibitem[Alexopoulos et~al\mbox{.}(2019)]%
        {alexopoulos2019trident}
\bibfield{author}{\bibinfo{person}{Nikolaos Alexopoulos},
  \bibinfo{person}{Emmanouil Vasilomanolakis}, \bibinfo{person}{Stephane~Le
  Roux}, \bibinfo{person}{Steven Rowe}, {and} \bibinfo{person}{Max
  M{\"u}hlh{\"a}user}.} \bibinfo{year}{2019}\natexlab{}.
\newblock \showarticletitle{TRIDEnT: building decentralized incentives for
  collaborative security}.
\newblock \bibinfo{journal}{\emph{arXiv preprint arXiv:1905.03571}}
  (\bibinfo{year}{2019}).
\newblock


\bibitem[Alexopoulos et~al\mbox{.}(2020)]%
        {alexopoulos2020trident}
\bibfield{author}{\bibinfo{person}{Nikolaos Alexopoulos},
  \bibinfo{person}{Emmanouil Vasilomanolakis}, \bibinfo{person}{Stephane~Le
  Roux}, \bibinfo{person}{Steven Rowe}, {and} \bibinfo{person}{Max
  M{\"u}hlh{\"a}user}.} \bibinfo{year}{2020}\natexlab{}.
\newblock \showarticletitle{TRIDEnT: towards a decentralized threat indicator
  marketplace}. In \bibinfo{booktitle}{\emph{Proceedings of the 35th Annual ACM
  Symposium on Applied Computing}}. \bibinfo{pages}{332--341}.
\newblock


\bibitem[Aljuhani(2022)]%
        {aljuhani2022ids}
\bibfield{author}{\bibinfo{person}{Ahamed Aljuhani}.}
  \bibinfo{year}{2022}\natexlab{}.
\newblock \showarticletitle{IDS-Chain: A Collaborative Intrusion Detection
  Framework Empowered Blockchain for Internet of Medical Things}. In
  \bibinfo{booktitle}{\emph{2022 IEEE Cloud Summit}}. IEEE,
  \bibinfo{pages}{57--62}.
\newblock


\bibitem[Alkadi et~al\mbox{.}(2020)]%
        {alkadi2020deep}
\bibfield{author}{\bibinfo{person}{Osama Alkadi}, \bibinfo{person}{Nour
  Moustafa}, \bibinfo{person}{Benjamin Turnbull}, {and}
  \bibinfo{person}{Kim-Kwang~Raymond Choo}.} \bibinfo{year}{2020}\natexlab{}.
\newblock \showarticletitle{A deep blockchain framework-enabled collaborative
  intrusion detection for protecting IoT and cloud networks}.
\newblock \bibinfo{journal}{\emph{IEEE Internet of Things Journal}}
  \bibinfo{volume}{8}, \bibinfo{number}{12} (\bibinfo{year}{2020}),
  \bibinfo{pages}{9463--9472}.
\newblock


\bibitem[Ashraf et~al\mbox{.}(2022)]%
        {ashraf2022fidchain}
\bibfield{author}{\bibinfo{person}{Eman Ashraf}, \bibinfo{person}{Nihal~FF
  Areed}, \bibinfo{person}{Hanaa Salem}, \bibinfo{person}{Ehab~H Abdelhay},
  {and} \bibinfo{person}{Ahmed Farouk}.} \bibinfo{year}{2022}\natexlab{}.
\newblock \showarticletitle{Fidchain: Federated intrusion detection system for
  blockchain-enabled iot healthcare applications}. In
  \bibinfo{booktitle}{\emph{Healthcare}}, Vol.~\bibinfo{volume}{10}. MDPI,
  \bibinfo{pages}{1110}.
\newblock


\bibitem[Azaria et~al\mbox{.}(2016)]%
        {azaria2016medrec}
\bibfield{author}{\bibinfo{person}{Asaph Azaria}, \bibinfo{person}{Ariel
  Ekblaw}, \bibinfo{person}{Thiago Vieira}, {and} \bibinfo{person}{Andrew
  Lippman}.} \bibinfo{year}{2016}\natexlab{}.
\newblock \showarticletitle{Medrec: Using blockchain for medical data access
  and permission management}. In \bibinfo{booktitle}{\emph{2016 2nd
  international conference on open and big data (OBD)}}. IEEE,
  \bibinfo{pages}{25--30}.
\newblock


\bibitem[Baiod et~al\mbox{.}(2021)]%
        {baiod2021blockchain}
\bibfield{author}{\bibinfo{person}{Wajde Baiod}, \bibinfo{person}{Janet Light},
  {and} \bibinfo{person}{Aniket Mahanti}.} \bibinfo{year}{2021}\natexlab{}.
\newblock \showarticletitle{Blockchain technology and its applications across
  multiple domains: a survey}.
\newblock \bibinfo{journal}{\emph{Journal of International Technology and
  Information Management}} \bibinfo{volume}{29}, \bibinfo{number}{4}
  (\bibinfo{year}{2021}), \bibinfo{pages}{78--119}.
\newblock


\bibitem[Belchior et~al\mbox{.}(2022)]%
        {belchior2022survey}
\bibfield{author}{\bibinfo{person}{Rafael Belchior},
  \bibinfo{person}{S{\'e}rgio Guerreiro}, \bibinfo{person}{Andr{\'e}
  Vasconcelos}, {and} \bibinfo{person}{Miguel Correia}.}
  \bibinfo{year}{2022}\natexlab{}.
\newblock \showarticletitle{A survey on business process view integration:
  past, present and future applications to blockchain}.
\newblock \bibinfo{journal}{\emph{Business Process Management Journal}}
  (\bibinfo{year}{2022}).
\newblock


\bibitem[Briner and Denyer(2012)]%
        {briner2012systematic}
\bibfield{author}{\bibinfo{person}{Rob~B Briner} {and} \bibinfo{person}{David
  Denyer}.} \bibinfo{year}{2012}\natexlab{}.
\newblock \showarticletitle{Systematic review and evidence synthesis as a
  practice and scholarship tool}.
\newblock \bibinfo{journal}{\emph{Oxford Handbook of Evidence-Based
  Management}} (\bibinfo{year}{2012}).
\newblock


\bibitem[Brotsis et~al\mbox{.}(2019)]%
        {brotsis2019blockchain}
\bibfield{author}{\bibinfo{person}{Sotirios Brotsis}, \bibinfo{person}{Nicholas
  Kolokotronis}, \bibinfo{person}{Konstantinos Limniotis},
  \bibinfo{person}{Stavros Shiaeles}, \bibinfo{person}{Dimitris Kavallieros},
  \bibinfo{person}{Emanuele Bellini}, {and} \bibinfo{person}{Cl{\'e}ment
  Pavu{\'e}}.} \bibinfo{year}{2019}\natexlab{}.
\newblock \showarticletitle{Blockchain solutions for forensic evidence
  preservation in IoT environments}. In \bibinfo{booktitle}{\emph{2019 IEEE
  Conference on Network Softwarization (NetSoft)}}. IEEE,
  \bibinfo{pages}{110--114}.
\newblock


\bibitem[Buterin et~al\mbox{.}(2014)]%
        {buterin2014next}
\bibfield{author}{\bibinfo{person}{Vitalik Buterin} {et~al\mbox{.}}}
  \bibinfo{year}{2014}\natexlab{}.
\newblock \showarticletitle{A next-generation smart contract and decentralized
  application platform}.
\newblock \bibinfo{journal}{\emph{white paper}} \bibinfo{volume}{3},
  \bibinfo{number}{37} (\bibinfo{year}{2014}), \bibinfo{pages}{2--1}.
\newblock


\bibitem[Cali et~al\mbox{.}(2021)]%
        {cali2021cybersecurity}
\bibfield{author}{\bibinfo{person}{Umit Cali}, \bibinfo{person}{Murat Kuzlu},
  \bibinfo{person}{Manisa Pipattanasomporn}, \bibinfo{person}{Onur Elma}, {and}
  \bibinfo{person}{Ramesh Reddi}.} \bibinfo{year}{2021}\natexlab{}.
\newblock \showarticletitle{Cybersecurity of Renewable Energy Data and
  Applications Using Distributed Ledger Technology}.
\newblock \bibinfo{journal}{\emph{arXiv preprint arXiv:2110.11354}}
  (\bibinfo{year}{2021}).
\newblock


\bibitem[Chen et~al\mbox{.}(2017)]%
        {chen2017security}
\bibfield{author}{\bibinfo{person}{Lin Chen}, \bibinfo{person}{Lei Xu},
  \bibinfo{person}{Nolan Shah}, \bibinfo{person}{Zhimin Gao},
  \bibinfo{person}{Yang Lu}, {and} \bibinfo{person}{Weidong Shi}.}
  \bibinfo{year}{2017}\natexlab{}.
\newblock \showarticletitle{On security analysis of proof-of-elapsed-time
  (poet)}. In \bibinfo{booktitle}{\emph{Stabilization, Safety, and Security of
  Distributed Systems: 19th International Symposium, SSS 2017, Boston, MA, USA,
  November 5--8, 2017, Proceedings 19}}. Springer, \bibinfo{pages}{282--297}.
\newblock


\bibitem[Chiu et~al\mbox{.}(2022)]%
        {chiu2022folketid}
\bibfield{author}{\bibinfo{person}{Wei-Yang Chiu}, \bibinfo{person}{Weizhi
  Meng}, \bibinfo{person}{Wenjuan Li}, {and} \bibinfo{person}{Liming Fang}.}
  \bibinfo{year}{2022}\natexlab{}.
\newblock \showarticletitle{FolketID: A Decentralized Blockchain-Based NemID
  Alternative Against DDoS Attacks}. In \bibinfo{booktitle}{\emph{Provable and
  Practical Security: 16th International Conference, ProvSec 2022, Nanjing,
  China, November 11--12, 2022, Proceedings}}. Springer,
  \bibinfo{pages}{210--227}.
\newblock


\bibitem[Christidis and Devetsikiotis(2016)]%
        {christidis2016blockchains}
\bibfield{author}{\bibinfo{person}{Konstantinos Christidis} {and}
  \bibinfo{person}{Michael Devetsikiotis}.} \bibinfo{year}{2016}\natexlab{}.
\newblock \showarticletitle{Blockchains and smart contracts for the internet of
  things}.
\newblock \bibinfo{journal}{\emph{Ieee Access}}  \bibinfo{volume}{4}
  (\bibinfo{year}{2016}), \bibinfo{pages}{2292--2303}.
\newblock


\bibitem[Dawit et~al\mbox{.}(2020)]%
        {dawit2020suitability}
\bibfield{author}{\bibinfo{person}{Nahom~Aron Dawit},
  \bibinfo{person}{Sujith~Samuel Mathew}, {and} \bibinfo{person}{Kadhim
  Hayawi}.} \bibinfo{year}{2020}\natexlab{}.
\newblock \showarticletitle{Suitability of blockchain for collaborative
  intrusion detection systems}. In \bibinfo{booktitle}{\emph{2020 12th Annual
  Undergraduate Research Conference on Applied Computing (URC)}}. IEEE,
  \bibinfo{pages}{1--6}.
\newblock


\bibitem[De~Angelis et~al\mbox{.}(2018)]%
        {de2018pbft}
\bibfield{author}{\bibinfo{person}{Stefano De~Angelis},
  \bibinfo{person}{Leonardo Aniello}, \bibinfo{person}{Roberto Baldoni},
  \bibinfo{person}{Federico Lombardi}, \bibinfo{person}{Andrea Margheri},
  \bibinfo{person}{Vladimiro Sassone}, {et~al\mbox{.}}}
  \bibinfo{year}{2018}\natexlab{}.
\newblock \showarticletitle{PBFT vs proof-of-authority: Applying the CAP
  theorem to permissioned blockchain}. In \bibinfo{booktitle}{\emph{CEUR
  workshop proceedings}}, Vol.~\bibinfo{volume}{2058}. CEUR-WS,
  \bibinfo{pages}{1--11}.
\newblock


\bibitem[Demertzis(2021)]%
        {demertzis2021blockchained}
\bibfield{author}{\bibinfo{person}{Konstantinos Demertzis}.}
  \bibinfo{year}{2021}\natexlab{}.
\newblock \showarticletitle{Blockchained federated learning for threat
  defense}.
\newblock \bibinfo{journal}{\emph{arXiv preprint arXiv:2102.12746}}
  (\bibinfo{year}{2021}).
\newblock


\bibitem[Demirkan et~al\mbox{.}(2020)]%
        {demirkan2020blockchain}
\bibfield{author}{\bibinfo{person}{Sebahattin Demirkan}, \bibinfo{person}{Irem
  Demirkan}, {and} \bibinfo{person}{Andrew McKee}.}
  \bibinfo{year}{2020}\natexlab{}.
\newblock \showarticletitle{Blockchain technology in the future of business
  cyber security and accounting}.
\newblock \bibinfo{journal}{\emph{Journal of Management Analytics}}
  \bibinfo{volume}{7}, \bibinfo{number}{2} (\bibinfo{year}{2020}),
  \bibinfo{pages}{189--208}.
\newblock


\bibitem[Dibaei et~al\mbox{.}(2021)]%
        {dibaei2021investigating}
\bibfield{author}{\bibinfo{person}{Mahdi Dibaei}, \bibinfo{person}{Xi Zheng},
  \bibinfo{person}{Youhua Xia}, \bibinfo{person}{Xiwei Xu},
  \bibinfo{person}{Alireza Jolfaei}, \bibinfo{person}{Ali~Kashif Bashir},
  \bibinfo{person}{Usman Tariq}, \bibinfo{person}{Dongjin Yu}, {and}
  \bibinfo{person}{Athanasios~V Vasilakos}.} \bibinfo{year}{2021}\natexlab{}.
\newblock \showarticletitle{Investigating the prospect of leveraging blockchain
  and machine learning to secure vehicular networks: A survey}.
\newblock \bibinfo{journal}{\emph{IEEE Transactions on Intelligent
  Transportation Systems}} \bibinfo{volume}{23}, \bibinfo{number}{2}
  (\bibinfo{year}{2021}), \bibinfo{pages}{683--700}.
\newblock


\bibitem[Diro et~al\mbox{.}(2021)]%
        {diro2021comprehensive}
\bibfield{author}{\bibinfo{person}{Abebe Diro}, \bibinfo{person}{Naveen
  Chilamkurti}, \bibinfo{person}{Van-Doan Nguyen}, {and} \bibinfo{person}{Will
  Heyne}.} \bibinfo{year}{2021}\natexlab{}.
\newblock \showarticletitle{A comprehensive study of anomaly detection schemes
  in IoT networks using machine learning algorithms}.
\newblock \bibinfo{journal}{\emph{Sensors}} \bibinfo{volume}{21},
  \bibinfo{number}{24} (\bibinfo{year}{2021}), \bibinfo{pages}{8320}.
\newblock


\bibitem[Etemadi et~al\mbox{.}(2020)]%
        {etemadi2020blockchain}
\bibfield{author}{\bibinfo{person}{Niloofar Etemadi}, \bibinfo{person}{YG
  Borbon}, {and} \bibinfo{person}{F Strozzi}.} \bibinfo{year}{2020}\natexlab{}.
\newblock \showarticletitle{Blockchain technology for cybersecurity
  applications in the food supply chain: A systematic literature review}.
\newblock \bibinfo{journal}{\emph{Proceedings of the XXIV Summer School
  “Francesco Turco”—Industrial Systems Engineering, Bergamo, Italy}}
  (\bibinfo{year}{2020}), \bibinfo{pages}{9--11}.
\newblock


\bibitem[Fan et~al\mbox{.}(2020)]%
        {fan2020blockchain}
\bibfield{author}{\bibinfo{person}{Wenjun Fan}, \bibinfo{person}{Younghee
  Park}, \bibinfo{person}{Shubham Kumar}, \bibinfo{person}{Priyatham Ganta},
  \bibinfo{person}{Xiaobo Zhou}, {and} \bibinfo{person}{Sang-Yoon Chang}.}
  \bibinfo{year}{2020}\natexlab{}.
\newblock \showarticletitle{Blockchain-enabled collaborative intrusion
  detection in software defined networks}. In \bibinfo{booktitle}{\emph{2020
  IEEE 19th International Conference on Trust, Security and Privacy in
  Computing and Communications (TrustCom)}}. IEEE, \bibinfo{pages}{967--974}.
\newblock


\bibitem[{FISCO BCOS}(2023)]%
        {fisco-bcos}
\bibfield{author}{\bibinfo{person}{{FISCO BCOS}}.}
  \bibinfo{year}{2023}\natexlab{}.
\newblock \bibinfo{title}{FISCO BCOS}.
\newblock
\newblock
\urldef\tempurl%
\url{http://www.fisco-bcos.org/}
\showURL{%
\tempurl}
\newblock
\shownote{[Online; accessed 22-March-2023]}.


\bibitem[Gimenez-Aguilar et~al\mbox{.}(2021)]%
        {gimenez2021achieving}
\bibfield{author}{\bibinfo{person}{Mar Gimenez-Aguilar},
  \bibinfo{person}{Jos{\'e}~Mar{\'\i}a De~Fuentes}, \bibinfo{person}{Lorena
  Gonzalez-Manzano}, {and} \bibinfo{person}{David Arroyo}.}
  \bibinfo{year}{2021}\natexlab{}.
\newblock \showarticletitle{Achieving cybersecurity in blockchain-based
  systems: A survey}.
\newblock \bibinfo{journal}{\emph{Future Generation Computer Systems}}
  \bibinfo{volume}{124} (\bibinfo{year}{2021}), \bibinfo{pages}{91--118}.
\newblock


\bibitem[Golomb et~al\mbox{.}(2018)]%
        {golomb1803ciota}
\bibfield{author}{\bibinfo{person}{T Golomb}, \bibinfo{person}{Y Mirsky}, {and}
  \bibinfo{person}{Y Elovici}.} \bibinfo{year}{2018}\natexlab{}.
\newblock \showarticletitle{CIoTA: Collaborative Anomaly Detection via
  Blockchain}.
\newblock \bibinfo{journal}{\emph{arXiv preprint arXiv:1803.03807}}
  (\bibinfo{year}{2018}).
\newblock


\bibitem[Greenspan(2015)]%
        {greenspan2015}
\bibfield{author}{\bibinfo{person}{Gideon Greenspan}.}
  \bibinfo{year}{2015}\natexlab{}.
\newblock \bibinfo{title}{Avoiding the pointless blockchain project}.
\newblock
\newblock
\urldef\tempurl%
\url{https://www.multichain.com/blog/2015/11/avoiding-pointless-blockchain-project/}
\showURL{%
\tempurl}


\bibitem[Guo et~al\mbox{.}(2021)]%
        {guo2021ldbt}
\bibfield{author}{\bibinfo{person}{Wei Guo}, \bibinfo{person}{Jin Xu},
  \bibinfo{person}{Yukui Pei}, \bibinfo{person}{Liuguo Yin}, {and}
  \bibinfo{person}{Chunxiao Jiang}.} \bibinfo{year}{2021}\natexlab{}.
\newblock \showarticletitle{LDBT: a lightweight DDoS attack tracing scheme
  based on blockchain}. In \bibinfo{booktitle}{\emph{2021 IEEE International
  Conference on Communications Workshops (ICC Workshops)}}. IEEE,
  \bibinfo{pages}{1--6}.
\newblock


\bibitem[Guo et~al\mbox{.}(2022)]%
        {guo2022distributed}
\bibfield{author}{\bibinfo{person}{Wei Guo}, \bibinfo{person}{Jin Xu},
  \bibinfo{person}{Yukui Pei}, \bibinfo{person}{Liuguo Yin},
  \bibinfo{person}{Chunxiao Jiang}, {and} \bibinfo{person}{Ning Ge}.}
  \bibinfo{year}{2022}\natexlab{}.
\newblock \showarticletitle{A distributed collaborative entrance Defense
  framework against DDoS attacks on satellite internet}.
\newblock \bibinfo{journal}{\emph{IEEE Internet of Things Journal}}
  \bibinfo{volume}{9}, \bibinfo{number}{17} (\bibinfo{year}{2022}),
  \bibinfo{pages}{15497--15510}.
\newblock


\bibitem[G{\"u}rcan et~al\mbox{.}(2018)]%
        {gurcan2018industrial}
\bibfield{author}{\bibinfo{person}{{\"O}nder G{\"u}rcan}, \bibinfo{person}{Marc
  Agenis-Nevers}, \bibinfo{person}{Yves-Marie Batany}, \bibinfo{person}{Mohamed
  Elmtiri}, \bibinfo{person}{Fran{\c{c}}ois Le~Fevre}, {and}
  \bibinfo{person}{Sara Tucci-Piergiovanni}.} \bibinfo{year}{2018}\natexlab{}.
\newblock \showarticletitle{An industrial prototype of trusted energy
  performance contracts using blockchain technologies}. In
  \bibinfo{booktitle}{\emph{2018 IEEE 20th International Conference on High
  Performance Computing and Communications; IEEE 16th International Conference
  on Smart City; IEEE 4th International Conference on Data Science and Systems
  (HPCC/SmartCity/DSS)}}. IEEE, \bibinfo{pages}{1336--1343}.
\newblock


\bibitem[Gurung et~al\mbox{.}(2022)]%
        {gurung2022cids}
\bibfield{author}{\bibinfo{person}{Gopal Gurung}, \bibinfo{person}{Gueltoum
  Bendiab}, \bibinfo{person}{Maria Shiaele}, {and} \bibinfo{person}{Stavros
  Shiaeles}.} \bibinfo{year}{2022}\natexlab{}.
\newblock \showarticletitle{CIDS: Collaborative Intrusion Detection System
  using Blockchain Technology}. In \bibinfo{booktitle}{\emph{2022 IEEE
  International Conference on Cyber Security and Resilience (CSR)}}. IEEE,
  \bibinfo{pages}{125--130}.
\newblock


\bibitem[Hakak et~al\mbox{.}(2021)]%
        {hakak2021recent}
\bibfield{author}{\bibinfo{person}{Saqib Hakak}, \bibinfo{person}{Wazir~Zada
  Khan}, \bibinfo{person}{Gulshan~Amin Gilkar}, \bibinfo{person}{Basem Assiri},
  \bibinfo{person}{Mamoun Alazab}, \bibinfo{person}{Sweta Bhattacharya}, {and}
  \bibinfo{person}{G~Thippa Reddy}.} \bibinfo{year}{2021}\natexlab{}.
\newblock \showarticletitle{Recent advances in blockchain technology: A survey
  on applications and challenges}.
\newblock \bibinfo{journal}{\emph{International Journal of Ad Hoc and
  Ubiquitous Computing}} \bibinfo{volume}{38}, \bibinfo{number}{1-3}
  (\bibinfo{year}{2021}), \bibinfo{pages}{82--100}.
\newblock


\bibitem[Han et~al\mbox{.}(2022)]%
        {han2022can}
\bibfield{author}{\bibinfo{person}{Rong Han}, \bibinfo{person}{Zheng Yan},
  \bibinfo{person}{Xueqin Liang}, {and} \bibinfo{person}{Laurence~T Yang}.}
  \bibinfo{year}{2022}\natexlab{}.
\newblock \showarticletitle{How can incentive mechanisms and blockchain benefit
  with each other? a survey}.
\newblock \bibinfo{journal}{\emph{Comput. Surveys}} \bibinfo{volume}{55},
  \bibinfo{number}{7} (\bibinfo{year}{2022}), \bibinfo{pages}{1--38}.
\newblock


\bibitem[Hasanova et~al\mbox{.}(2019)]%
        {hasanova2019survey}
\bibfield{author}{\bibinfo{person}{Huru Hasanova}, \bibinfo{person}{Ui-jun
  Baek}, \bibinfo{person}{Mu-gon Shin}, \bibinfo{person}{Kyunghee Cho}, {and}
  \bibinfo{person}{Myung-Sup Kim}.} \bibinfo{year}{2019}\natexlab{}.
\newblock \showarticletitle{A survey on blockchain cybersecurity
  vulnerabilities and possible countermeasures}.
\newblock \bibinfo{journal}{\emph{International Journal of Network Management}}
  \bibinfo{volume}{29}, \bibinfo{number}{2} (\bibinfo{year}{2019}),
  \bibinfo{pages}{e2060}.
\newblock


\bibitem[He et~al\mbox{.}(2022)]%
        {he2022cgan}
\bibfield{author}{\bibinfo{person}{Xiaoqiang He}, \bibinfo{person}{Qianbin
  Chen}, \bibinfo{person}{Lun Tang}, \bibinfo{person}{Weili Wang}, {and}
  \bibinfo{person}{Tong Liu}.} \bibinfo{year}{2022}\natexlab{}.
\newblock \showarticletitle{CGAN-Based Collaborative Intrusion Detection for
  UAV Networks: A Blockchain-Empowered Distributed Federated Learning
  Approach}.
\newblock \bibinfo{journal}{\emph{IEEE Internet of Things Journal}}
  \bibinfo{volume}{10}, \bibinfo{number}{1} (\bibinfo{year}{2022}),
  \bibinfo{pages}{120--132}.
\newblock


\bibitem[Hu et~al\mbox{.}(2020)]%
        {hu2020contract}
\bibfield{author}{\bibinfo{person}{Bin Hu}, \bibinfo{person}{Xiaofang Zhao},
  \bibinfo{person}{Cheng Zhang}, \bibinfo{person}{Yan Jin}, {and}
  \bibinfo{person}{Bo Wei}.} \bibinfo{year}{2020}\natexlab{}.
\newblock \showarticletitle{A Contract Based User-Centric Computational Trust
  Towards E-Governance}. In \bibinfo{booktitle}{\emph{Web Services--ICWS 2020:
  27th International Conference, Held as Part of the Services Conference
  Federation, SCF 2020, Honolulu, HI, USA, September 18--20, 2020, Proceedings
  27}}. Springer, \bibinfo{pages}{133--149}.
\newblock


\bibitem[Ivanov et~al\mbox{.}(2023)]%
        {ivanov2023security}
\bibfield{author}{\bibinfo{person}{Nikolay Ivanov}, \bibinfo{person}{Chenning
  Li}, \bibinfo{person}{Qiben Yan}, \bibinfo{person}{Zhiyuan Sun},
  \bibinfo{person}{Zhichao Cao}, {and} \bibinfo{person}{Xiapu Luo}.}
  \bibinfo{year}{2023}\natexlab{}.
\newblock \showarticletitle{Security Threat Mitigation For Smart Contracts: A
  Comprehensive Survey}.
\newblock \bibinfo{journal}{\emph{Comput. Surveys}} (\bibinfo{year}{2023}).
\newblock


\bibitem[Keele et~al\mbox{.}(2007)]%
        {keele2007guidelines}
\bibfield{author}{\bibinfo{person}{Staffs Keele} {et~al\mbox{.}}}
  \bibinfo{year}{2007}\natexlab{}.
\newblock \bibinfo{title}{Guidelines for performing systematic literature
  reviews in software engineering}.
\newblock
\newblock


\bibitem[Kolokotronis et~al\mbox{.}(2019)]%
        {kolokotronis2019blockchain}
\bibfield{author}{\bibinfo{person}{Nicholas Kolokotronis},
  \bibinfo{person}{Sotirios Brotsis}, \bibinfo{person}{Georgios Germanos},
  \bibinfo{person}{Costas Vassilakis}, {and} \bibinfo{person}{Stavros
  Shiaeles}.} \bibinfo{year}{2019}\natexlab{}.
\newblock \showarticletitle{On blockchain architectures for trust-based
  collaborative intrusion detection}. In \bibinfo{booktitle}{\emph{2019 IEEE
  world congress on services (SERVICES)}}, Vol.~\bibinfo{volume}{2642}. IEEE,
  \bibinfo{pages}{21--28}.
\newblock


\bibitem[Kolokotronis et~al\mbox{.}(2022)]%
        {kolokotronis2022intelligent}
\bibfield{author}{\bibinfo{person}{Nicholas Kolokotronis},
  \bibinfo{person}{Maria Dareioti}, \bibinfo{person}{Stavros Shiaeles}, {and}
  \bibinfo{person}{Emanuele Bellini}.} \bibinfo{year}{2022}\natexlab{}.
\newblock \showarticletitle{An Intelligent Platform for Threat Assessment and
  Cyber-Attack Mitigation in IoMT Ecosystems}. In
  \bibinfo{booktitle}{\emph{2022 IEEE Globecom Workshops (GC Wkshps)}}. IEEE,
  \bibinfo{pages}{541--546}.
\newblock


\bibitem[Krichen et~al\mbox{.}(2022)]%
        {krichen2022blockchain}
\bibfield{author}{\bibinfo{person}{Moez Krichen}, \bibinfo{person}{Meryem
  Ammi}, \bibinfo{person}{Alaeddine Mihoub}, {and} \bibinfo{person}{Mutiq
  Almutiq}.} \bibinfo{year}{2022}\natexlab{}.
\newblock \showarticletitle{Blockchain for modern applications: A survey}.
\newblock \bibinfo{journal}{\emph{Sensors}} \bibinfo{volume}{22},
  \bibinfo{number}{14} (\bibinfo{year}{2022}), \bibinfo{pages}{5274}.
\newblock


\bibitem[Kumar and Singh(2020)]%
        {kumar2020distributed}
\bibfield{author}{\bibinfo{person}{Manish Kumar} {and}
  \bibinfo{person}{Ashish~Kumar Singh}.} \bibinfo{year}{2020}\natexlab{}.
\newblock \showarticletitle{Distributed intrusion detection system using
  blockchain and cloud computing infrastructure}. In
  \bibinfo{booktitle}{\emph{2020 4th international conference on trends in
  electronics and informatics (ICOEI)(48184)}}. IEEE,
  \bibinfo{pages}{248--252}.
\newblock


\bibitem[Kumar et~al\mbox{.}(2022)]%
        {kumar2022distributed}
\bibfield{author}{\bibinfo{person}{Randhir Kumar}, \bibinfo{person}{Prabhat
  Kumar}, \bibinfo{person}{Rakesh Tripathi}, \bibinfo{person}{Govind~P Gupta},
  \bibinfo{person}{Sahil Garg}, {and} \bibinfo{person}{Mohammad~Mehedi
  Hassan}.} \bibinfo{year}{2022}\natexlab{}.
\newblock \showarticletitle{A distributed intrusion detection system to detect
  DDoS attacks in blockchain-enabled IoT network}.
\newblock \bibinfo{journal}{\emph{J. Parallel and Distrib. Comput.}}
  \bibinfo{volume}{164} (\bibinfo{year}{2022}), \bibinfo{pages}{55--68}.
\newblock


\bibitem[Kumar et~al\mbox{.}(2021)]%
        {kumar2021survey}
\bibfield{author}{\bibinfo{person}{Sathish Kumar},
  \bibinfo{person}{Sarveshwaran Velliangiri}, \bibinfo{person}{Periyasami
  Karthikeyan}, \bibinfo{person}{Saru Kumari}, \bibinfo{person}{Sachin Kumar},
  {and} \bibinfo{person}{Muhammad~Khurram Khan}.}
  \bibinfo{year}{2021}\natexlab{}.
\newblock \showarticletitle{A survey on the blockchain techniques for the
  Internet of Vehicles security}.
\newblock \bibinfo{journal}{\emph{Transactions on Emerging Telecommunications
  Technologies}} (\bibinfo{year}{2021}), \bibinfo{pages}{e4317}.
\newblock


\bibitem[Kumari and McPherson(2009)]%
        {RFC5635}
\bibfield{author}{\bibinfo{person}{W. Kumari} {and} \bibinfo{person}{D.
  McPherson}.} \bibinfo{year}{2009}\natexlab{}.
\newblock \bibinfo{booktitle}{\emph{Remote Triggered Black Hole Filtering with
  Unicast Reverse Path Forwarding (uRPF)}}.
\newblock \bibinfo{type}{RFC} 5635. \bibinfo{institution}{RFC Editor}.
\newblock
\showISSN{2070-1721}


\bibitem[Lavaur et~al\mbox{.}(2022)]%
        {lavaur2022evolution}
\bibfield{author}{\bibinfo{person}{L{\'e}o Lavaur},
  \bibinfo{person}{Marc-Oliver Pahl}, \bibinfo{person}{Yann Busnel}, {and}
  \bibinfo{person}{Fabien Autrel}.} \bibinfo{year}{2022}\natexlab{}.
\newblock \showarticletitle{The Evolution of Federated Learning-based Intrusion
  Detection and Mitigation: a Survey}.
\newblock \bibinfo{journal}{\emph{IEEE Transactions on Network and Service
  Management}} (\bibinfo{year}{2022}).
\newblock


\bibitem[Lee and Kwon(2021)]%
        {lee2021distributed}
\bibfield{author}{\bibinfo{person}{JongHyup Lee} {and}
  \bibinfo{person}{Taekyoung Kwon}.} \bibinfo{year}{2021}\natexlab{}.
\newblock \showarticletitle{Distributed watchdogs based on blockchain for
  securing industrial internet of things}.
\newblock \bibinfo{journal}{\emph{Sensors}} \bibinfo{volume}{21},
  \bibinfo{number}{13} (\bibinfo{year}{2021}), \bibinfo{pages}{4393}.
\newblock


\bibitem[Leng et~al\mbox{.}(2020)]%
        {leng2020blockchain}
\bibfield{author}{\bibinfo{person}{Jiewu Leng}, \bibinfo{person}{Shide Ye},
  \bibinfo{person}{Man Zhou}, \bibinfo{person}{J~Leon Zhao},
  \bibinfo{person}{Qiang Liu}, \bibinfo{person}{Wei Guo}, \bibinfo{person}{Wei
  Cao}, {and} \bibinfo{person}{Leijie Fu}.} \bibinfo{year}{2020}\natexlab{}.
\newblock \showarticletitle{Blockchain-secured smart manufacturing in industry
  4.0: A survey}.
\newblock \bibinfo{journal}{\emph{IEEE Transactions on Systems, Man, and
  Cybernetics: Systems}} \bibinfo{volume}{51}, \bibinfo{number}{1}
  (\bibinfo{year}{2020}), \bibinfo{pages}{237--252}.
\newblock


\bibitem[Li et~al\mbox{.}(2020a)]%
        {li2020scalable}
\bibfield{author}{\bibinfo{person}{Wenyu Li}, \bibinfo{person}{Chenglin Feng},
  \bibinfo{person}{Lei Zhang}, \bibinfo{person}{Hao Xu}, \bibinfo{person}{Bin
  Cao}, {and} \bibinfo{person}{Muhammad~Ali Imran}.}
  \bibinfo{year}{2020}\natexlab{a}.
\newblock \showarticletitle{A scalable multi-layer PBFT consensus for
  blockchain}.
\newblock \bibinfo{journal}{\emph{IEEE Transactions on Parallel and Distributed
  Systems}} \bibinfo{volume}{32}, \bibinfo{number}{5} (\bibinfo{year}{2020}),
  \bibinfo{pages}{1146--1160}.
\newblock


\bibitem[Li et~al\mbox{.}(2020b)]%
        {li2020framework}
\bibfield{author}{\bibinfo{person}{Wenjuan Li}, \bibinfo{person}{Jiao Tan},
  {and} \bibinfo{person}{Yu Wang}.} \bibinfo{year}{2020}\natexlab{b}.
\newblock \showarticletitle{A framework of blockchain-based collaborative
  intrusion detection in software defined networking}. In
  \bibinfo{booktitle}{\emph{Network and System Security: 14th International
  Conference, NSS 2020, Melbourne, VIC, Australia, November 25--27, 2020,
  Proceedings 14}}. Springer, \bibinfo{pages}{261--276}.
\newblock


\bibitem[Li et~al\mbox{.}(2019a)]%
        {li2019designing}
\bibfield{author}{\bibinfo{person}{Wenjuan Li}, \bibinfo{person}{Steven Tug},
  \bibinfo{person}{Weizhi Meng}, {and} \bibinfo{person}{Yu Wang}.}
  \bibinfo{year}{2019}\natexlab{a}.
\newblock \showarticletitle{Designing collaborative blockchained
  signature-based intrusion detection in IoT environments}.
\newblock \bibinfo{journal}{\emph{Future Generation Computer Systems}}
  \bibinfo{volume}{96} (\bibinfo{year}{2019}), \bibinfo{pages}{481--489}.
\newblock


\bibitem[Li et~al\mbox{.}(2019b)]%
        {li2019towards}
\bibfield{author}{\bibinfo{person}{Wenjuan Li}, \bibinfo{person}{Yu Wang},
  \bibinfo{person}{Jin Li}, {and} \bibinfo{person}{Man~Ho Au}.}
  \bibinfo{year}{2019}\natexlab{b}.
\newblock \showarticletitle{Towards blockchained challenge-based collaborative
  intrusion detection}. In \bibinfo{booktitle}{\emph{Applied Cryptography and
  Network Security Workshops: ACNS 2019 Satellite Workshops, SiMLA, Cloud S\&P,
  AIBlock, and AIoTS, Bogota, Colombia, June 5--7, 2019, Proceedings 17}}.
  Springer, \bibinfo{pages}{122--139}.
\newblock


\bibitem[Li et~al\mbox{.}(2021)]%
        {li2021toward}
\bibfield{author}{\bibinfo{person}{Wenjuan Li}, \bibinfo{person}{Yu Wang},
  \bibinfo{person}{Jin Li}, {and} \bibinfo{person}{Man~Ho Au}.}
  \bibinfo{year}{2021}\natexlab{}.
\newblock \showarticletitle{Toward a blockchain-based framework for
  challenge-based collaborative intrusion detection}.
\newblock \bibinfo{journal}{\emph{International Journal of Information
  Security}}  \bibinfo{volume}{20} (\bibinfo{year}{2021}),
  \bibinfo{pages}{127--139}.
\newblock


\bibitem[Li et~al\mbox{.}(2022)]%
        {li2022blockcsdn}
\bibfield{author}{\bibinfo{person}{Wenjuan Li}, \bibinfo{person}{Yu Wang},
  \bibinfo{person}{Weizhi Meng}, \bibinfo{person}{Jin Li}, {and}
  \bibinfo{person}{Chunhua Su}.} \bibinfo{year}{2022}\natexlab{}.
\newblock \showarticletitle{BlockCSDN: towards blockchain-based collaborative
  intrusion detection in software defined networking}.
\newblock \bibinfo{journal}{\emph{IEICE TRANSACTIONS on Information and
  Systems}} \bibinfo{volume}{105}, \bibinfo{number}{2} (\bibinfo{year}{2022}),
  \bibinfo{pages}{272--279}.
\newblock


\bibitem[Lin and Liao(2017)]%
        {lin2017survey}
\bibfield{author}{\bibinfo{person}{Iuon-Chang Lin} {and}
  \bibinfo{person}{Tzu-Chun Liao}.} \bibinfo{year}{2017}\natexlab{}.
\newblock \showarticletitle{A survey of blockchain security issues and
  challenges}.
\newblock \bibinfo{journal}{\emph{Int. J. Netw. Secur.}} \bibinfo{volume}{19},
  \bibinfo{number}{5} (\bibinfo{year}{2017}), \bibinfo{pages}{653--659}.
\newblock


\bibitem[Liu et~al\mbox{.}(2021a)]%
        {liu2021sedid}
\bibfield{author}{\bibinfo{person}{Gao Liu}, \bibinfo{person}{Zheng Yan},
  \bibinfo{person}{Wei Feng}, \bibinfo{person}{Xuyang Jing},
  \bibinfo{person}{Yaxing Chen}, {and} \bibinfo{person}{Mohammed Atiquzzaman}.}
  \bibinfo{year}{2021}\natexlab{a}.
\newblock \showarticletitle{SeDID: An SGX-enabled decentralized intrusion
  detection framework for network trust evaluation}.
\newblock \bibinfo{journal}{\emph{Information Fusion}}  \bibinfo{volume}{70}
  (\bibinfo{year}{2021}), \bibinfo{pages}{100--114}.
\newblock


\bibitem[Liu et~al\mbox{.}(2021b)]%
        {liu2021blockchain}
\bibfield{author}{\bibinfo{person}{Hong Liu}, \bibinfo{person}{Shuaipeng
  Zhang}, \bibinfo{person}{Pengfei Zhang}, \bibinfo{person}{Xinqiang Zhou},
  \bibinfo{person}{Xuebin Shao}, \bibinfo{person}{Geguang Pu}, {and}
  \bibinfo{person}{Yan Zhang}.} \bibinfo{year}{2021}\natexlab{b}.
\newblock \showarticletitle{Blockchain and federated learning for collaborative
  intrusion detection in vehicular edge computing}.
\newblock \bibinfo{journal}{\emph{IEEE Transactions on Vehicular Technology}}
  \bibinfo{volume}{70}, \bibinfo{number}{6} (\bibinfo{year}{2021}),
  \bibinfo{pages}{6073--6084}.
\newblock


\bibitem[Liu et~al\mbox{.}(2020)]%
        {liu2020blockchain}
\bibfield{author}{\bibinfo{person}{Yang Liu}, \bibinfo{person}{Debiao He},
  \bibinfo{person}{Mohammad~S Obaidat}, \bibinfo{person}{Neeraj Kumar},
  \bibinfo{person}{Muhammad~Khurram Khan}, {and}
  \bibinfo{person}{Kim-Kwang~Raymond Choo}.} \bibinfo{year}{2020}\natexlab{}.
\newblock \showarticletitle{Blockchain-based identity management systems: A
  review}.
\newblock \bibinfo{journal}{\emph{Journal of network and computer
  applications}}  \bibinfo{volume}{166} (\bibinfo{year}{2020}),
  \bibinfo{pages}{102731}.
\newblock


\bibitem[Mathew et~al\mbox{.}(2022)]%
        {mathew2022integration}
\bibfield{author}{\bibinfo{person}{Sujith~Samuel Mathew},
  \bibinfo{person}{Kadhim Hayawi}, \bibinfo{person}{Nahom~Aron Dawit},
  \bibinfo{person}{Ikbal Taleb}, {and} \bibinfo{person}{Zouheir Trabelsi}.}
  \bibinfo{year}{2022}\natexlab{}.
\newblock \showarticletitle{Integration of blockchain and collaborative
  intrusion detection for secure data transactions in industrial IoT: a
  survey}.
\newblock \bibinfo{journal}{\emph{Cluster Computing}} (\bibinfo{year}{2022}),
  \bibinfo{pages}{1--21}.
\newblock


\bibitem[Meng et~al\mbox{.}(2020)]%
        {meng2020enhancing}
\bibfield{author}{\bibinfo{person}{Weizhi Meng}, \bibinfo{person}{Wenjuan Li},
  \bibinfo{person}{Laurence~T Yang}, {and} \bibinfo{person}{Peng Li}.}
  \bibinfo{year}{2020}\natexlab{}.
\newblock \showarticletitle{Enhancing challenge-based collaborative intrusion
  detection networks against insider attacks using blockchain}.
\newblock \bibinfo{journal}{\emph{International Journal of Information
  Security}}  \bibinfo{volume}{19} (\bibinfo{year}{2020}),
  \bibinfo{pages}{279--290}.
\newblock


\bibitem[Meng et~al\mbox{.}(2018)]%
        {meng2018intrusion}
\bibfield{author}{\bibinfo{person}{Weizhi Meng},
  \bibinfo{person}{Elmar~Wolfgang Tischhauser}, \bibinfo{person}{Qingju Wang},
  \bibinfo{person}{Yu Wang}, {and} \bibinfo{person}{Jinguang Han}.}
  \bibinfo{year}{2018}\natexlab{}.
\newblock \showarticletitle{When intrusion detection meets blockchain
  technology: a review}.
\newblock \bibinfo{journal}{\emph{Ieee Access}}  \bibinfo{volume}{6}
  (\bibinfo{year}{2018}), \bibinfo{pages}{10179--10188}.
\newblock


\bibitem[Miller and Pelsser(2019)]%
        {miller2019taxonomy}
\bibfield{author}{\bibinfo{person}{Lo{\"\i}c Miller} {and}
  \bibinfo{person}{Cristel Pelsser}.} \bibinfo{year}{2019}\natexlab{}.
\newblock \showarticletitle{A taxonomy of attacks using bgp blackholing}. In
  \bibinfo{booktitle}{\emph{Computer Security--ESORICS 2019: 24th European
  Symposium on Research in Computer Security, Luxembourg, September 23--27,
  2019, Proceedings, Part I 24}}. Springer, \bibinfo{pages}{107--127}.
\newblock


\bibitem[Mylrea and Gourisetti(2018)]%
        {mylrea2018blockchain}
\bibfield{author}{\bibinfo{person}{Michael Mylrea} {and} \bibinfo{person}{Sri
  Nikhil~Gupta Gourisetti}.} \bibinfo{year}{2018}\natexlab{}.
\newblock \showarticletitle{Blockchain for supply chain cybersecurity,
  optimization and compliance}. In \bibinfo{booktitle}{\emph{2018 resilience
  week (RWS)}}. IEEE, \bibinfo{pages}{70--76}.
\newblock


\bibitem[Nakamoto(2008)]%
        {nakamoto2008bitcoin}
\bibfield{author}{\bibinfo{person}{Satoshi Nakamoto}.}
  \bibinfo{year}{2008}\natexlab{}.
\newblock \showarticletitle{Bitcoin: A peer-to-peer electronic cash system}.
\newblock \bibinfo{journal}{\emph{Decentralized Business Review}}
  (\bibinfo{year}{2008}), \bibinfo{pages}{21260}.
\newblock


\bibitem[Neureither et~al\mbox{.}(2020)]%
        {neureither2020legiot}
\bibfield{author}{\bibinfo{person}{Jens Neureither}, \bibinfo{person}{Alexandra
  Dmitrienko}, \bibinfo{person}{David Koisser}, \bibinfo{person}{Ferdinand
  Brasser}, {and} \bibinfo{person}{Ahmad-Reza Sadeghi}.}
  \bibinfo{year}{2020}\natexlab{}.
\newblock \showarticletitle{LegIoT: Ledgered trust management platform for
  IoT}. In \bibinfo{booktitle}{\emph{Computer Security--ESORICS 2020: 25th
  European Symposium on Research in Computer Security, ESORICS 2020, Guildford,
  UK, September 14--18, 2020, Proceedings, Part I 25}}. Springer,
  \bibinfo{pages}{377--396}.
\newblock


\bibitem[Ossamah(2020)]%
        {ossamah2020blockchain}
\bibfield{author}{\bibinfo{person}{Almotery Ossamah}.}
  \bibinfo{year}{2020}\natexlab{}.
\newblock \showarticletitle{Blockchain as a solution to Drone Cybersecurity}.
  In \bibinfo{booktitle}{\emph{2020 IEEE 6th World Forum on Internet of Things
  (WF-IoT)}}. IEEE, \bibinfo{pages}{1--9}.
\newblock


\bibitem[Pahlevan et~al\mbox{.}(2021)]%
        {pahlevan2021secure}
\bibfield{author}{\bibinfo{person}{Maryam Pahlevan}, \bibinfo{person}{Artemis
  Voulkidis}, {and} \bibinfo{person}{Terpsichori-Helen Velivassaki}.}
  \bibinfo{year}{2021}\natexlab{}.
\newblock \showarticletitle{Secure exchange of cyber threat intelligence using
  TAXII and distributed ledger technologies-application for electrical power
  and energy system}. In \bibinfo{booktitle}{\emph{Proceedings of the 16th
  International Conference on Availability, Reliability and Security}}.
  \bibinfo{pages}{1--8}.
\newblock


\bibitem[Peng et~al\mbox{.}(2007)]%
        {peng2007survey}
\bibfield{author}{\bibinfo{person}{Tao Peng}, \bibinfo{person}{Christopher
  Leckie}, {and} \bibinfo{person}{Kotagiri Ramamohanarao}.}
  \bibinfo{year}{2007}\natexlab{}.
\newblock \showarticletitle{Survey of network-based defense mechanisms
  countering the DoS and DDoS problems}.
\newblock \bibinfo{journal}{\emph{ACM Computing Surveys (CSUR)}}
  \bibinfo{volume}{39}, \bibinfo{number}{1} (\bibinfo{year}{2007}),
  \bibinfo{pages}{3--es}.
\newblock


\bibitem[Preuveneers et~al\mbox{.}(2018)]%
        {preuveneers2018chained}
\bibfield{author}{\bibinfo{person}{Davy Preuveneers}, \bibinfo{person}{Vera
  Rimmer}, \bibinfo{person}{Ilias Tsingenopoulos}, \bibinfo{person}{Jan
  Spooren}, \bibinfo{person}{Wouter Joosen}, {and} \bibinfo{person}{Elisabeth
  Ilie-Zudor}.} \bibinfo{year}{2018}\natexlab{}.
\newblock \showarticletitle{Chained anomaly detection models for federated
  learning: An intrusion detection case study}.
\newblock \bibinfo{journal}{\emph{Applied Sciences}} \bibinfo{volume}{8},
  \bibinfo{number}{12} (\bibinfo{year}{2018}), \bibinfo{pages}{2663}.
\newblock


\bibitem[Ramanan et~al\mbox{.}(2021)]%
        {ramanan2021blockchain}
\bibfield{author}{\bibinfo{person}{Paritosh Ramanan}, \bibinfo{person}{Dan Li},
  {and} \bibinfo{person}{Nagi Gebraeel}.} \bibinfo{year}{2021}\natexlab{}.
\newblock \showarticletitle{Blockchain-based decentralized replay attack
  detection for large-scale power systems}.
\newblock \bibinfo{journal}{\emph{IEEE Transactions on Systems, Man, and
  Cybernetics: Systems}} \bibinfo{volume}{52}, \bibinfo{number}{8}
  (\bibinfo{year}{2021}), \bibinfo{pages}{4727--4739}.
\newblock


\bibitem[Rathore et~al\mbox{.}(2019)]%
        {rathore2019blockseciotnet}
\bibfield{author}{\bibinfo{person}{Shailendra Rathore},
  \bibinfo{person}{Byung~Wook Kwon}, {and} \bibinfo{person}{Jong~Hyuk Park}.}
  \bibinfo{year}{2019}\natexlab{}.
\newblock \showarticletitle{BlockSecIoTNet: Blockchain-based decentralized
  security architecture for IoT network}.
\newblock \bibinfo{journal}{\emph{Journal of Network and Computer
  Applications}}  \bibinfo{volume}{143} (\bibinfo{year}{2019}),
  \bibinfo{pages}{167--177}.
\newblock


\bibitem[Rodrigues et~al\mbox{.}(2017b)]%
        {rodrigues2017blockchain}
\bibfield{author}{\bibinfo{person}{Bruno Rodrigues}, \bibinfo{person}{Thomas
  Bocek}, \bibinfo{person}{Andri Lareida}, \bibinfo{person}{David Hausheer},
  \bibinfo{person}{Sina Rafati}, {and} \bibinfo{person}{Burkhard Stiller}.}
  \bibinfo{year}{2017}\natexlab{b}.
\newblock \showarticletitle{A blockchain-based architecture for collaborative
  DDoS mitigation with smart contracts}. In \bibinfo{booktitle}{\emph{Security
  of Networks and Services in an All-Connected World: 11th IFIP WG 6.6
  International Conference on Autonomous Infrastructure, Management, and
  Security, AIMS 2017, Zurich, Switzerland, July 10-13, 2017, Proceedings 11}}.
  Springer International Publishing, \bibinfo{pages}{16--29}.
\newblock


\bibitem[Rodrigues et~al\mbox{.}(2017a)]%
        {rodrigues2017enabling}
\bibfield{author}{\bibinfo{person}{Bruno Rodrigues}, \bibinfo{person}{Thomas
  Bocek}, {and} \bibinfo{person}{Burkhard Stiller}.}
  \bibinfo{year}{2017}\natexlab{a}.
\newblock \showarticletitle{Enabling a cooperative, multi-domain DDoS defense
  by a blockchain signaling system (BloSS)}.
\newblock \bibinfo{journal}{\emph{Semantic Scholar}} (\bibinfo{year}{2017}).
\newblock


\bibitem[Rodrigues et~al\mbox{.}(2019)]%
        {rodrigues2019evaluating}
\bibfield{author}{\bibinfo{person}{Bruno Rodrigues}, \bibinfo{person}{Lukas
  Eisenring}, \bibinfo{person}{Eder Scheid}, \bibinfo{person}{Thomas Bocek},
  {and} \bibinfo{person}{Burkhard Stiller}.} \bibinfo{year}{2019}\natexlab{}.
\newblock \showarticletitle{Evaluating a blockchain-based cooperative defense}.
  In \bibinfo{booktitle}{\emph{2019 IFIP/IEEE Symposium on Integrated Network
  and Service Management (IM)}}. IEEE, \bibinfo{pages}{533--538}.
\newblock


\bibitem[Rodrigues and Stiller(2019)]%
        {rodrigues2019cooperative}
\bibfield{author}{\bibinfo{person}{Bruno Rodrigues} {and}
  \bibinfo{person}{Burkhard Stiller}.} \bibinfo{year}{2019}\natexlab{}.
\newblock \showarticletitle{Cooperative signaling of DDoS attacks in a
  blockchain-based network}. In \bibinfo{booktitle}{\emph{Proceedings of the
  ACM SIGCOMM 2019 Conference Posters and Demos}}. \bibinfo{pages}{39--41}.
\newblock


\bibitem[Rodrigues et~al\mbox{.}(2020)]%
        {rodrigues2020sc}
\bibfield{author}{\bibinfo{person}{Bruno Rodrigues}, \bibinfo{person}{Spasen
  Trendafilov}, \bibinfo{person}{Eder Scheid}, {and} \bibinfo{person}{Burkhard
  Stiller}.} \bibinfo{year}{2020}\natexlab{}.
\newblock \showarticletitle{SC-FLARE: Cooperative DDoS signaling based on smart
  contracts}. In \bibinfo{booktitle}{\emph{2020 IEEE International Conference
  on Blockchain and Cryptocurrency (ICBC)}}. IEEE, \bibinfo{pages}{1--3}.
\newblock


\bibitem[Santin et~al\mbox{.}(2022)]%
        {santin2022framework}
\bibfield{author}{\bibinfo{person}{Gabriele Santin}, \bibinfo{person}{Inna
  Skarbovsky}, \bibinfo{person}{Fabiana Fournier}, {and} \bibinfo{person}{Bruno
  Lepri}.} \bibinfo{year}{2022}\natexlab{}.
\newblock \showarticletitle{A Framework for Verifiable and Auditable
  Collaborative Anomaly Detection}.
\newblock \bibinfo{journal}{\emph{IEEE Access}}  \bibinfo{volume}{10}
  (\bibinfo{year}{2022}), \bibinfo{pages}{82896--82909}.
\newblock


\bibitem[Saveetha and Maragatham(2022)]%
        {saveetha2022design}
\bibfield{author}{\bibinfo{person}{D Saveetha} {and} \bibinfo{person}{G
  Maragatham}.} \bibinfo{year}{2022}\natexlab{}.
\newblock \showarticletitle{Design of Blockchain enabled intrusion detection
  model for detecting security attacks using deep learning}.
\newblock \bibinfo{journal}{\emph{Pattern Recognition Letters}}
  \bibinfo{volume}{153} (\bibinfo{year}{2022}), \bibinfo{pages}{24--28}.
\newblock


\bibitem[{Scimago}(2022)]%
        {scimagojrpapercount}
\bibfield{author}{\bibinfo{person}{{Scimago}}.}
  \bibinfo{year}{2022}\natexlab{}.
\newblock \bibinfo{title}{Scimago Country Rank}.
\newblock
\newblock
\urldef\tempurl%
\url{https://www.scimagojr.com/countryrank.php}
\showURL{%
\tempurl}
\newblock
\shownote{[Online; accessed 22-March-2023]}.


\bibitem[Shi et~al\mbox{.}(2019)]%
        {shi2019dynamic}
\bibfield{author}{\bibinfo{person}{Leyi Shi}, \bibinfo{person}{Yang Li},
  \bibinfo{person}{Tianxu Liu}, \bibinfo{person}{Jia Liu},
  \bibinfo{person}{Baoying Shan}, {and} \bibinfo{person}{Honglong Chen}.}
  \bibinfo{year}{2019}\natexlab{}.
\newblock \showarticletitle{Dynamic distributed honeypot based on blockchain}.
\newblock \bibinfo{journal}{\emph{IEEE Access}}  \bibinfo{volume}{7}
  (\bibinfo{year}{2019}), \bibinfo{pages}{72234--72246}.
\newblock


\bibitem[Singh et~al\mbox{.}(2023)]%
        {singh2023fusionfedblock}
\bibfield{author}{\bibinfo{person}{Sushil~Kumar Singh},
  \bibinfo{person}{Laurence~T Yang}, {and} \bibinfo{person}{Jong~Hyuk Park}.}
  \bibinfo{year}{2023}\natexlab{}.
\newblock \showarticletitle{FusionFedBlock: Fusion of blockchain and federated
  learning to preserve privacy in industry 5.0}.
\newblock \bibinfo{journal}{\emph{Information Fusion}}  \bibinfo{volume}{90}
  (\bibinfo{year}{2023}), \bibinfo{pages}{233--240}.
\newblock


\bibitem[Snider et~al\mbox{.}(2018)]%
        {snider2018delegated}
\bibfield{author}{\bibinfo{person}{Myles Snider}, \bibinfo{person}{Kyle
  Samani}, {and} \bibinfo{person}{Tushar Jain}.}
  \bibinfo{year}{2018}\natexlab{}.
\newblock \showarticletitle{Delegated proof of stake: features \& tradeoffs}.
\newblock \bibinfo{journal}{\emph{Multicoin Cap}}  \bibinfo{volume}{19}
  (\bibinfo{year}{2018}), \bibinfo{pages}{1--19}.
\newblock


\bibitem[Tang et~al\mbox{.}(2022)]%
        {tang2022blockchain}
\bibfield{author}{\bibinfo{person}{Fengxiao Tang}, \bibinfo{person}{Cong Wen},
  \bibinfo{person}{Linfeng Luo}, \bibinfo{person}{Ming Zhao}, {and}
  \bibinfo{person}{Nei Kato}.} \bibinfo{year}{2022}\natexlab{}.
\newblock \showarticletitle{Blockchain-Based Trusted Traffic Offloading in
  Space-Air-Ground Integrated Networks (SAGIN): A Federated Reinforcement
  Learning Approach}.
\newblock \bibinfo{journal}{\emph{IEEE Journal on Selected Areas in
  Communications}} \bibinfo{volume}{40}, \bibinfo{number}{12}
  (\bibinfo{year}{2022}), \bibinfo{pages}{3501--3516}.
\newblock


\bibitem[Taylor et~al\mbox{.}(2020)]%
        {taylor2020systematic}
\bibfield{author}{\bibinfo{person}{Paul~J Taylor}, \bibinfo{person}{Tooska
  Dargahi}, \bibinfo{person}{Ali Dehghantanha}, \bibinfo{person}{Reza~M
  Parizi}, {and} \bibinfo{person}{Kim-Kwang~Raymond Choo}.}
  \bibinfo{year}{2020}\natexlab{}.
\newblock \showarticletitle{A systematic literature review of blockchain cyber
  security}.
\newblock \bibinfo{journal}{\emph{Digital Communications and Networks}}
  \bibinfo{volume}{6}, \bibinfo{number}{2} (\bibinfo{year}{2020}),
  \bibinfo{pages}{147--156}.
\newblock


\bibitem[Tug et~al\mbox{.}(2018)]%
        {tug2018cbsigids}
\bibfield{author}{\bibinfo{person}{Steven Tug}, \bibinfo{person}{Weizhi Meng},
  {and} \bibinfo{person}{Yu Wang}.} \bibinfo{year}{2018}\natexlab{}.
\newblock \showarticletitle{CBSigIDS: towards collaborative blockchained
  signature-based intrusion detection}. In \bibinfo{booktitle}{\emph{2018 IEEE
  International Conference on Internet of Things (iThings) and IEEE Green
  Computing and Communications (GreenCom) and IEEE Cyber, Physical and Social
  Computing (CPSCom) and IEEE Smart Data (SmartData)}}. IEEE,
  \bibinfo{pages}{1228--1235}.
\newblock


\bibitem[Ujjan et~al\mbox{.}(2019)]%
        {ujjan2019snort}
\bibfield{author}{\bibinfo{person}{Raja Majid~Ali Ujjan},
  \bibinfo{person}{Zeeshan Pervez}, {and} \bibinfo{person}{Keshav Dahal}.}
  \bibinfo{year}{2019}\natexlab{}.
\newblock \showarticletitle{Snort based collaborative intrusion detection
  system using blockchain in SDN}. In \bibinfo{booktitle}{\emph{2019 13th
  International Conference on Software, Knowledge, Information Management and
  Applications (SKIMA)}}. IEEE, \bibinfo{pages}{1--8}.
\newblock


\bibitem[Vance and Vance(2019)]%
        {vance2019cybersecurity}
\bibfield{author}{\bibinfo{person}{Taylor~Rodriguez Vance} {and}
  \bibinfo{person}{Andrew Vance}.} \bibinfo{year}{2019}\natexlab{}.
\newblock \showarticletitle{Cybersecurity in the blockchain era: a survey on
  examining critical infrastructure protection with blockchain-based
  technology}. In \bibinfo{booktitle}{\emph{2019 IEEE International
  Scientific-Practical Conference Problems of Infocommunications, Science and
  Technology (PIC S\&T)}}. IEEE, \bibinfo{pages}{107--112}.
\newblock


\bibitem[Vasilomanolakis et~al\mbox{.}(2015)]%
        {vasilomanolakis2015taxonomy}
\bibfield{author}{\bibinfo{person}{Emmanouil Vasilomanolakis},
  \bibinfo{person}{Shankar Karuppayah}, \bibinfo{person}{Max
  M{\"u}hlh{\"a}user}, {and} \bibinfo{person}{Mathias Fischer}.}
  \bibinfo{year}{2015}\natexlab{}.
\newblock \showarticletitle{Taxonomy and survey of collaborative intrusion
  detection}.
\newblock \bibinfo{journal}{\emph{ACM Computing Surveys (CSUR)}}
  \bibinfo{volume}{47}, \bibinfo{number}{4} (\bibinfo{year}{2015}),
  \bibinfo{pages}{1--33}.
\newblock


\bibitem[Vasin(2014)]%
        {vasin2014blackcoin}
\bibfield{author}{\bibinfo{person}{Pavel Vasin}.}
  \bibinfo{year}{2014}\natexlab{}.
\newblock \showarticletitle{Blackcoin’s proof-of-stake protocol v2}.
\newblock \bibinfo{journal}{\emph{URL: https://blackcoin.
  co/blackcoin-pos-protocol-v2-whitepaper. pdf}}  \bibinfo{volume}{71}
  (\bibinfo{year}{2014}).
\newblock


\bibitem[Wang et~al\mbox{.}(2023)]%
        {wang2023sok}
\bibfield{author}{\bibinfo{person}{Qin Wang}, \bibinfo{person}{Jiangshan Yu},
  \bibinfo{person}{Shiping Chen}, {and} \bibinfo{person}{Yang Xiang}.}
  \bibinfo{year}{2023}\natexlab{}.
\newblock \showarticletitle{Sok: Dag-based blockchain systems}.
\newblock \bibinfo{journal}{\emph{Comput. Surveys}} \bibinfo{volume}{55},
  \bibinfo{number}{12} (\bibinfo{year}{2023}), \bibinfo{pages}{1--38}.
\newblock


\bibitem[Wang et~al\mbox{.}(2020)]%
        {wang2020survey}
\bibfield{author}{\bibinfo{person}{Xifeng Wang}, \bibinfo{person}{Changqiao
  Xu}, \bibinfo{person}{Zan Zhou}, \bibinfo{person}{Shujie Yang}, {and}
  \bibinfo{person}{Limin Sun}.} \bibinfo{year}{2020}\natexlab{}.
\newblock \showarticletitle{A survey of blockchain-based cybersecurity for
  vehicular networks}.
\newblock \bibinfo{journal}{\emph{2020 International Wireless Communications
  and Mobile Computing (IWCMC)}} (\bibinfo{year}{2020}),
  \bibinfo{pages}{740--745}.
\newblock


\bibitem[Wang et~al\mbox{.}(2022)]%
        {wang2022infedge}
\bibfield{author}{\bibinfo{person}{Xiaofei Wang}, \bibinfo{person}{Yunfeng
  Zhao}, \bibinfo{person}{Chao Qiu}, \bibinfo{person}{Zhicheng Liu},
  \bibinfo{person}{Jiangtian Nie}, {and} \bibinfo{person}{Victor~CM Leung}.}
  \bibinfo{year}{2022}\natexlab{}.
\newblock \showarticletitle{Infedge: A blockchain-based incentive mechanism in
  hierarchical federated learning for end-edge-cloud communications}.
\newblock \bibinfo{journal}{\emph{IEEE Journal on Selected Areas in
  Communications}} \bibinfo{volume}{40}, \bibinfo{number}{12}
  (\bibinfo{year}{2022}), \bibinfo{pages}{3325--3342}.
\newblock


\bibitem[Winanto et~al\mbox{.}(2021)]%
        {winanto2021designing}
\bibfield{author}{\bibinfo{person}{Eko~Arip Winanto},
  \bibinfo{person}{Mohd~Yazid Idris}, \bibinfo{person}{Deris Stiawan}, {and}
  \bibinfo{person}{Mohammad~Sulkhan Nurfatih}.}
  \bibinfo{year}{2021}\natexlab{}.
\newblock \showarticletitle{Designing consensus algorithm for collaborative
  signature-based intrusion detection system}.
\newblock \bibinfo{journal}{\emph{Indones J. Electron. Eng. Comput. Sci.}}
  \bibinfo{volume}{22}, \bibinfo{number}{1} (\bibinfo{year}{2021}),
  \bibinfo{pages}{485--496}.
\newblock


\bibitem[W{\"u}st and Gervais(2018)]%
        {wust2018you}
\bibfield{author}{\bibinfo{person}{Karl W{\"u}st} {and} \bibinfo{person}{Arthur
  Gervais}.} \bibinfo{year}{2018}\natexlab{}.
\newblock \showarticletitle{Do you need a blockchain?}. In
  \bibinfo{booktitle}{\emph{2018 Crypto Valley Conference on Blockchain
  Technology (CVCBT)}}. IEEE, \bibinfo{pages}{45--54}.
\newblock


\bibitem[Wylde et~al\mbox{.}(2022)]%
        {wylde2022cybersecurity}
\bibfield{author}{\bibinfo{person}{Vinden Wylde}, \bibinfo{person}{Nisha
  Rawindaran}, \bibinfo{person}{John Lawrence}, \bibinfo{person}{Rushil
  Balasubramanian}, \bibinfo{person}{Edmond Prakash}, \bibinfo{person}{Ambikesh
  Jayal}, \bibinfo{person}{Imtiaz Khan}, \bibinfo{person}{Chaminda Hewage},
  {and} \bibinfo{person}{Jon Platts}.} \bibinfo{year}{2022}\natexlab{}.
\newblock \showarticletitle{Cybersecurity, data privacy and blockchain: a
  review}.
\newblock \bibinfo{journal}{\emph{SN Computer Science}} \bibinfo{volume}{3},
  \bibinfo{number}{2} (\bibinfo{year}{2022}), \bibinfo{pages}{1--12}.
\newblock


\bibitem[Xu et~al\mbox{.}(2023)]%
        {xu2023survey}
\bibfield{author}{\bibinfo{person}{Jie Xu}, \bibinfo{person}{Cong Wang}, {and}
  \bibinfo{person}{Xiaohua Jia}.} \bibinfo{year}{2023}\natexlab{}.
\newblock \showarticletitle{A survey of blockchain consensus protocols}.
\newblock \bibinfo{journal}{\emph{Comput. Surveys}} (\bibinfo{year}{2023}).
\newblock


\bibitem[Yeh et~al\mbox{.}(2019)]%
        {yeh2019collaborative}
\bibfield{author}{\bibinfo{person}{Lo-Yao Yeh}, \bibinfo{person}{Jiun-Long
  Huang}, \bibinfo{person}{Ting-Yin Yen}, {and} \bibinfo{person}{Jen-Wei Hu}.}
  \bibinfo{year}{2019}\natexlab{}.
\newblock \showarticletitle{A collaborative DDoS defense platform based on
  blockchain technology}. In \bibinfo{booktitle}{\emph{2019 Twelfth
  International Conference on Ubi-Media Computing (Ubi-Media)}}. IEEE,
  \bibinfo{pages}{1--6}.
\newblock


\bibitem[Yeh et~al\mbox{.}(2020)]%
        {yeh2020sochain}
\bibfield{author}{\bibinfo{person}{Lo-Yao Yeh}, \bibinfo{person}{Peggy~Joy Lu},
  \bibinfo{person}{Szu-Hao Huang}, {and} \bibinfo{person}{Jiun-Long Huang}.}
  \bibinfo{year}{2020}\natexlab{}.
\newblock \showarticletitle{SOChain: A privacy-preserving DDoS data exchange
  service over soc consortium blockchain}.
\newblock \bibinfo{journal}{\emph{IEEE Transactions on Engineering Management}}
  \bibinfo{volume}{67}, \bibinfo{number}{4} (\bibinfo{year}{2020}),
  \bibinfo{pages}{1487--1500}.
\newblock


\bibitem[Yenugunti and Yau(2020)]%
        {yenugunti2020blockchain}
\bibfield{author}{\bibinfo{person}{Chandralekha Yenugunti} {and}
  \bibinfo{person}{Stephen~S Yau}.} \bibinfo{year}{2020}\natexlab{}.
\newblock \showarticletitle{A blockchain approach to identifying compromised
  nodes in collaborative intrusion detection systems}. In
  \bibinfo{booktitle}{\emph{2020 IEEE Intl Conf on Dependable, Autonomic and
  Secure Computing, Intl Conf on Pervasive Intelligence and Computing, Intl
  Conf on Cloud and Big Data Computing, Intl Conf on Cyber Science and
  Technology Congress (DASC/PiCom/CBDCom/CyberSciTech)}}. IEEE,
  \bibinfo{pages}{87--93}.
\newblock


\bibitem[Zhang et~al\mbox{.}(2020)]%
        {zhang2020blockchain}
\bibfield{author}{\bibinfo{person}{Weishan Zhang}, \bibinfo{person}{Qinghua
  Lu}, \bibinfo{person}{Qiuyu Yu}, \bibinfo{person}{Zhaotong Li},
  \bibinfo{person}{Yue Liu}, \bibinfo{person}{Sin~Kit Lo},
  \bibinfo{person}{Shiping Chen}, \bibinfo{person}{Xiwei Xu}, {and}
  \bibinfo{person}{Liming Zhu}.} \bibinfo{year}{2020}\natexlab{}.
\newblock \showarticletitle{Blockchain-based federated learning for device
  failure detection in industrial IoT}.
\newblock \bibinfo{journal}{\emph{IEEE Internet of Things Journal}}
  \bibinfo{volume}{8}, \bibinfo{number}{7} (\bibinfo{year}{2020}),
  \bibinfo{pages}{5926--5937}.
\newblock


\bibitem[Zheng et~al\mbox{.}(2018)]%
        {zheng2018blockchain}
\bibfield{author}{\bibinfo{person}{Zibin Zheng}, \bibinfo{person}{Shaoan Xie},
  \bibinfo{person}{Hong-Ning Dai}, \bibinfo{person}{Xiangping Chen}, {and}
  \bibinfo{person}{Huaimin Wang}.} \bibinfo{year}{2018}\natexlab{}.
\newblock \showarticletitle{Blockchain challenges and opportunities: A survey}.
\newblock \bibinfo{journal}{\emph{International journal of web and grid
  services}} \bibinfo{volume}{14}, \bibinfo{number}{4} (\bibinfo{year}{2018}),
  \bibinfo{pages}{352--375}.
\newblock


\bibitem[Zhuang et~al\mbox{.}(2020)]%
        {zhuang2020blockchain}
\bibfield{author}{\bibinfo{person}{Peng Zhuang}, \bibinfo{person}{Talha Zamir},
  {and} \bibinfo{person}{Hao Liang}.} \bibinfo{year}{2020}\natexlab{}.
\newblock \showarticletitle{Blockchain for cybersecurity in smart grid: A
  comprehensive survey}.
\newblock \bibinfo{journal}{\emph{IEEE Transactions on Industrial Informatics}}
  \bibinfo{volume}{17}, \bibinfo{number}{1} (\bibinfo{year}{2020}),
  \bibinfo{pages}{3--19}.
\newblock


\end{thebibliography}

\appendix

\section{Additional Material}
\label{appendix}

\subsection{Detailed data on the papers of this study}

Tables~\ref{tab:papers} and~\ref{tab:papers_2} detail all the papers we treated in this survey.
\Cref{tab:papers} contains information on the application, the domain and the number of citations.
\Cref{tab:papers_2} on the other hand contains information about the nature of the blockchain, namely technology, consensus, access control and data validation policy.
A reader looking for an collaborative application using a specific blockchain nature may find great use of those tables, to read the papers using this nature in particular.

\begin{table*}[ht]
    \centering
    \caption{Papers on blockchain and collaborative cybersecurity --- Application, domain and citations.}
    \begin{adjustbox}{max width=\textwidth}
        \begin{tabular}{clllr}
            \toprule
            Year & Name                                                                                    & Application                              & Domain                                         & Cited \\
            \midrule
            2016 & MedRec: Using Blockchain for Medical Data Acce\ldots~\cite{azaria2016medrec}            & Health                                   & Health                                         & 2100  \\
            2016 & Blockchains and Smart Contracts for the Intern\ldots~\cite{christidis2016blockchains}   & Internet of Things                       & Internet of Things                             & 4334  \\
            2017 & A Blockchain-Based Architecture for Collaborat\ldots~\cite{rodrigues2017blockchain}     & DDoS Defense                             & Networks                                       & 167   \\
            2017 & Enabling a Cooperative Multi-domain DDoS Defen\ldots~\cite{rodrigues2017enabling}       & DDoS Defense                             & Networks                                       & 49    \\
            2018 & ChainSecure - A Scalable and Proactive Solutio\ldots~\cite{abou2018chainsecure}         & DDoS Defense                             & Software-defined Networking                    & 32    \\
            2018 & CBSigIDS: Towards Collaborative Blockchained S\ldots~\cite{tug2018cbsigids}             & Intrusion Detection                      & Independent                                    & 35    \\
            2018 & Chained Anomaly Detection Models for Federated\ldots~\cite{preuveneers2018chained}      & Intrusion Detection                      & Independent                                    & 173   \\
            2018 & CIoTA: Collaborative Anomaly Detection via Blo\ldots~\cite{golomb1803ciota}             & Intrusion Detection                      & Internet of Things                             & 3     \\
            2018 & Blockchain for Supply Chain Cybersecurity Opti\ldots~\cite{mylrea2018blockchain}        & Supply Chain                             & Internet of Things;Energy                      & 56    \\
            2018 & Towards Blockchain-Based Collaborative Intrusi\ldots~\cite{alexopoulos2018towards}      & Intrusion Detection                      & Independent                                    & 103   \\
            2019 & A Collaborative DDoS Defense Platform Based on\ldots~\cite{yeh2019collaborative}        & DDoS Defense                             & Networks                                       & 10    \\
            2019 & Blockchain Solutions for Forensic Evidence Pre\ldots~\cite{brotsis2019blockchain}       & Forensics                                & Internet of Things                             & 62    \\
            2019 & On Blockchain Architectures for Trust-based Co\ldots~\cite{kolokotronis2019blockchain}  & Intrusion Detection                      & Independent                                    & 35    \\
            2019 & Snort Based Collaborative Intrusion Detection \ldots~\cite{ujjan2019snort}              & Intrusion Detection                      & Software-defined Networking                    & 19    \\
            2019 & Towards Blockchained Challenge-based Collabora\ldots~\cite{li2019towards}               & Intrusion Detection                      & Independent                                    & 20    \\
            2019 & Evaluating a Blockchain-based Cooperative Defense~\cite{rodrigues2019evaluating}        & DDoS Defense                             & Networks                                       & 12    \\
            2019 & Co-IoT: A Collaborative DDoS mitigation scheme\ldots~\cite{abou2019co}                  & DDoS Defense                             & Internet of Things;Software-defined Networking & 56    \\
            2019 & Cooperative Signaling of DDoS Attacks in a Blo\ldots~\cite{rodrigues2019cooperative}    & DDoS Defense                             & Networks                                       & 12    \\
            2019 & Cochain-SC: An Intra-and Inter-Domain Ddos Mit\ldots~\cite{abou2019cochain}             & DDoS Defense                             & Internet of Things;Software-defined Networking & 102   \\
            2019 & TRIDEnT: Building Decentralized Incentives for\ldots~\cite{alexopoulos2019trident}      & Intrusion Detection                      & Independent                                    & 4     \\
            2019 & Dynamic Distributed Honeypot Based on Blockchain~\cite{shi2019dynamic}                  & Honeypot                                 & Independent                                    & 30    \\
            2019 & Enhancing challenge-based collaborative intrus\ldots~\cite{meng2020enhancing}           & Intrusion Detection                      & Independent                                    & 36    \\
            2019 & BlockSecIoTNet: Blockchain-based decentralized\ldots~\cite{rathore2019blockseciotnet}   & Intrusion Detection                      & Internet of Things                             & 192   \\
            2019 & Designing collaborative blockchained signature\ldots~\cite{li2019designing}             & Intrusion Detection                      & Internet of Things                             & 132   \\
            2019 & An Industrial Prototype of Trusted Energy Perf\ldots~\cite{gurcan2018industrial}        & Energy                                   & Energy                                         & 23    \\
            2019 & Blockchain Secured Electronic Health Records: \ldots~\cite{akarca2019blockchain}        & Health                                   & Health                                         & 14    \\
            2019 & A Lightweight Blockchain Based Cybersecurity f\ldots~\cite{abdulkader2019lightweight}   & Internet of Things                       & Internet of Things                             & 16    \\
            2020 & SC-FLARE: Cooperative DDoS Signaling based on \ldots~\cite{rodrigues2020sc}             & DDoS Defense                             & Networks                                       & 9     \\
            2020 & BrainChain - A Machine learning Approach for p\ldots~\cite{abou2020brainchain}          & DDoS Defense                             & Software-defined Networking                    & 25    \\
            2020 & Bringing Intelligence to Software Defined Netw\ldots~\cite{abou2020bringing}            & DDoS Defense                             & Software-defined Networking                    & 36    \\
            2020 & A Blockchain Approach to Identifying Compromis\ldots~\cite{yenugunti2020blockchain}     & Intrusion Detection                      & Independent                                    & 5     \\
            2020 & A Contract Based User-Centric Computational Tr\ldots~\cite{hu2020contract}              & Governance                               & Governance                                     & 1     \\
            2020 & A Framework of Blockchain-Based Collaborative \ldots~\cite{li2020framework}             & Intrusion Detection                      & Software-defined Networking                    & 10    \\
            2020 & Blockchain-enabled Collaborative Intrusion Det\ldots~\cite{fan2020blockchain}           & Intrusion Detection                      & Software-defined Networking                    & 8     \\
            2020 & Distributed Intrusion Detection System using B\ldots~\cite{kumar2020distributed}        & Intrusion Detection                      & Independent                                    & 31    \\
            2020 & LegIoT: Ledgered Trust Management Platform for\ldots~\cite{neureither2020legiot}        & Trust Management                         & Internet of Things                             & 4     \\
            2020 & SOChain: A Privacy-Preserving DDoS Data Exchan\ldots~\cite{yeh2020sochain}              & DDoS Defense                             & Networks                                       & 32    \\
            2020 & TRIDEnT: Towards a Decentralized Threat Indica\ldots~\cite{alexopoulos2020trident}      & Intrusion Detection                      & Independent                                    & 3     \\
            2020 & Blockchain-based Federated Learning for Device\ldots~\cite{zhang2020blockchain}         & Intrusion Detection                      & Internet of Things                             & 116   \\
            2020 & Application of Blockchain within Aviation Cybe\ldots~\cite{adhikari2020application}     & Intrusion Mitigation                     & Vehicles;Aviation                              & 5     \\
            2020 & Blockchain as a solution to Drone Cybersecurity~\cite{ossamah2020blockchain}            & Drones                                   & Internet of Things;Vehicles;Drones             & 16    \\
            2020 & Blockchain technology in the future of busines\ldots~\cite{demirkan2020blockchain}      & Accounting                               & Business                                       & 130   \\
            2020 & Suitability of Blockchain for Collaborative In\ldots~\cite{dawit2020suitability}        & Intrusion Detection                      & Independent                                    & 12    \\
            2021 & A Distributed Collaborative Entrance Defense F\ldots~\cite{guo2022distributed}          & DDoS Defense                             & Networks                                       & 3     \\
            2021 & A Deep Blockchain Framework-Enabled Collaborat\ldots~\cite{alkadi2020deep}              & Intrusion Detection                      & Internet of Things;Cloud                       & 165   \\
            2021 & A Comprehensive Study of Anomaly Detection Sch\ldots~\cite{diro2021comprehensive}       & Intrusion Detection                      & Internet of Things                             & 16    \\
            2021 & Distributed Watchdogs Based on Blockchain for \ldots~\cite{lee2021distributed}          & Supply Chain                             & Internet of Things;Supply Chain                & 6     \\
            2021 & Toward a blockchain-based framework for challe\ldots~\cite{li2021toward}                & Intrusion Detection                      & Networks                                       & 41    \\
            2021 & SeDID: An SGX-enabled decentralized intrusion \ldots~\cite{liu2021sedid}                & Intrusion Detection                      & Networks                                       & 16    \\
            2021 & Designing consensus algorithm for collaborativ\ldots~\cite{winanto2021designing}        & Intrusion Detection                      & Independent                                    & 4     \\
            2021 & LDBT: A Lightweight DDoS Attack Tracing Scheme\ldots~\cite{guo2021ldbt}                 & DDoS Defense                             & Networks                                       & 5     \\
            2021 & Blockchain and Federated Learning for Collabor\ldots~\cite{liu2021blockchain}           & Intrusion Detection                      & Vehicles                                       & 60    \\
            2021 & Blockchained Federated Learning for Threat Def\ldots~\cite{demertzis2021blockchained}   & Intrusion Detection                      & Internet of Things                             & 7     \\
            2021 & Cybersecurity of Renewable Energy Data and App\ldots~\cite{cali2021cybersecurity}       & Energy                                   & Energy                                         & 1     \\
            2021 & Secure exchange of cyber threat intelligence u\ldots~\cite{pahlevan2021secure}          & Intrusion Detection                      & Energy                                         & 2     \\
            2022 & FolketID: A Decentralized Blockchain-Based Nem\ldots~\cite{chiu2022folketid}            & DDoS Defense                             & Governance                                     & 0     \\
            2022 & Blockchain-Based Decentralized Replay Attack D\ldots~\cite{ramanan2021blockchain}       & Intrusion Detection                      & Energy                                         & 14    \\
            2022 & BlockCSDN: Towards Blockchain-Based Collaborat\ldots~\cite{li2022blockcsdn}             & Intrusion Detection                      & Software-defined Networking                    & 8     \\
            2022 & CIDS: Collaborative Intrusion Detection System\ldots~\cite{gurung2022cids}              & Intrusion Detection                      & Independent                                    & 2     \\
            2022 & IDS-Chain: A Collaborative Intrusion Detection\ldots~\cite{aljuhani2022ids}             & Intrusion Detection                      & Health;Internet of Things                      & 0     \\
            2022 & Design of Blockchain enabled intrusion detecti\ldots~\cite{saveetha2022design}          & Intrusion Detection                      & Independent                                    & 4     \\
            2022 & Blockchain-Enabled Intrusion Detection and Pre\ldots~\cite{alevizos2022blockchain}      & Intrusion Detection;Intrusion Prevention & Independent                                    & 2     \\
            2022 & FIDChain: Federated Intrusion Detection System\ldots~\cite{ashraf2022fidchain}          & Intrusion Detection                      & Health                                         & 5     \\
            2022 & Privacy-preserved Cyberattack Detection in Ind\ldots~\cite{abdel2022privacy}            & Intrusion Detection                      & Internet of Things                             & 1     \\
            2022 & CGAN-Based Collaborative Intrusion Detection f\ldots~\cite{he2022cgan}                  & Intrusion Detection                      & Internet of Things;Vehicles;Drones             & 0     \\
            2022 & A Framework for Verifiable and Auditable Feder\ldots~\cite{santin2022framework}         & Intrusion Detection                      & Independent                                    & 0     \\
            2022 & InFEDge: A Blockchain-based Incentive Mechanis\ldots~\cite{wang2022infedge}             & Federated Learning                       & Cloud                                          & 3     \\
            2022 & An Intelligent Platform for Threat Assessment \ldots~\cite{kolokotronis2022intelligent} & Intrusion Detection;Intrusion Response   & Health;Internet of Things                      & 0     \\
            2022 & Blockchain-based Trusted Traffic Offloading in\ldots~\cite{tang2022blockchain}          & Networks                                 & Networks                                       & 19    \\
            2022 & A Distributed Intrusion Detection System to De\ldots~\cite{kumar2022distributed}        & Intrusion Detection                      & Internet of Things                             & 0     \\
            2023 & FusionFedBlock: Fusion of Blockchain and Feder\ldots~\cite{singh2023fusionfedblock}     & Federated Learning                       & Internet of Things                             & 2     \\
            \bottomrule
        \end{tabular}
    \end{adjustbox}
    \label{tab:papers}
\end{table*}

\begin{table*}[ht]
    \centering
    \caption{Papers on blockchain and collaborative cybersecurity --- Technology, consensus, access control and permission.}
    \begin{adjustbox}{max width=\textwidth}
        \begin{tabular}{clllll}
            \toprule
            Year & Name                                                                                    & Technology           & Consensus                           & Access control & Permission     \\
            \midrule
            2016 & MedRec: Using Blockchain for Medical Data Acce\ldots~\cite{azaria2016medrec}            & Ethereum             & Proof of Work                       & Private        & Permissioned   \\
            2016 & Blockchains and Smart Contracts for the Intern\ldots~\cite{christidis2016blockchains}   & Independent          & Independent                         & Independent    & Independent    \\
            2017 & A Blockchain-Based Architecture for Collaborat\ldots~\cite{rodrigues2017blockchain}     & Ethereum             & Proof of Ownership                  & Private        & Permissioned   \\
            2017 & Enabling a Cooperative Multi-domain DDoS Defen\ldots~\cite{rodrigues2017enabling}       & Ethereum             & Not specified                       & Private        & Permissioned   \\
            2018 & ChainSecure - A Scalable and Proactive Solutio\ldots~\cite{abou2018chainsecure}         & Not specified        & Not specified                       & Private        & Permissioned   \\
            2018 & CBSigIDS: Towards Collaborative Blockchained S\ldots~\cite{tug2018cbsigids}             & Not specified        & Not specified                       & Private        & Permissioned   \\
            2018 & Chained Anomaly Detection Models for Federated\ldots~\cite{preuveneers2018chained}      & Multichain           & Round-robin                         & Private        & Permissioned   \\
            2018 & CIoTA: Collaborative Anomaly Detection via Blo\ldots~\cite{golomb1803ciota}             & Custom               & Not specified                       & Private        & Permissionless \\
            2018 & Blockchain for Supply Chain Cybersecurity Opti\ldots~\cite{mylrea2018blockchain}        & Not specified        & Proof of Authority                  & Private        & Permissioned   \\
            2018 & Towards Blockchain-Based Collaborative Intrusi\ldots~\cite{alexopoulos2018towards}      & Independent          & Independent                         & Independent    & Independent    \\
            2019 & A Collaborative DDoS Defense Platform Based on\ldots~\cite{yeh2019collaborative}        & Ethereum             & Proof of Work                       & Private        & Permissioned   \\
            2019 & Blockchain Solutions for Forensic Evidence Pre\ldots~\cite{brotsis2019blockchain}       & Hyperledger Fabric   & Not specified                       & Private        & Permissioned   \\
            2019 & On Blockchain Architectures for Trust-based Co\ldots~\cite{kolokotronis2019blockchain}  & Custom               & Proof of Work;Proof of Stake        & Private        & Permissioned   \\
            2019 & Snort Based Collaborative Intrusion Detection \ldots~\cite{ujjan2019snort}              & Ethereum             & Proof of Work                       & Private        & Permissioned   \\
            2019 & Towards Blockchained Challenge-based Collabora\ldots~\cite{li2019towards}               & Custom               & Proof of Identity                   & Private        & Permissioned   \\
            2019 & Evaluating a Blockchain-based Cooperative Defense~\cite{rodrigues2019evaluating}        & Ethereum             & Not specified                       & Private        & Permissioned   \\
            2019 & Co-IoT: A Collaborative DDoS mitigation scheme\ldots~\cite{abou2019co}                  & Ethereum             & Not specified                       & Private        & Permissioned   \\
            2019 & Cooperative Signaling of DDoS Attacks in a Blo\ldots~\cite{rodrigues2019cooperative}    & Ethereum             & Not specified                       & Private        & Permissioned   \\
            2019 & Cochain-SC: An Intra-and Inter-Domain Ddos Mit\ldots~\cite{abou2019cochain}             & Ethereum             & Not specified                       & Private        & Permissioned   \\
            2019 & TRIDEnT: Building Decentralized Incentives for\ldots~\cite{alexopoulos2019trident}      & Ethereum             & Proof of Burn;Proof of Work         & Public         & Permissionless \\
            2019 & Dynamic Distributed Honeypot Based on Blockchain~\cite{shi2019dynamic}                  & Ethereum             & Proof of Work                       & Private        & Permissioned   \\
            2019 & Enhancing challenge-based collaborative intrus\ldots~\cite{meng2020enhancing}           & Custom               & Not specified                       & Private        & Permissioned   \\
            2019 & BlockSecIoTNet: Blockchain-based decentralized\ldots~\cite{rathore2019blockseciotnet}   & Ethereum             & Proof of Work                       & Private        & Permissioned   \\
            2019 & Designing collaborative blockchained signature\ldots~\cite{li2019designing}             & Custom               & Custom                              & Private        & Permissionless \\
            2019 & An Industrial Prototype of Trusted Energy Perf\ldots~\cite{gurcan2018industrial}        & Ethereum             & Proof of Work                       & Private        & Permissioned   \\
            2019 & Blockchain Secured Electronic Health Records: \ldots~\cite{akarca2019blockchain}        & Not specified        & Not specified                       & Private        & Permissioned   \\
            2019 & A Lightweight Blockchain Based Cybersecurity f\ldots~\cite{abdulkader2019lightweight}   & Custom               & Proof of Authority                  & Private        & Permissioned   \\
            2020 & SC-FLARE: Cooperative DDoS Signaling based on \ldots~\cite{rodrigues2020sc}             & Ethereum             & Proof of Authority                  & Private        & Permissioned   \\
            2020 & BrainChain - A Machine learning Approach for p\ldots~\cite{abou2020brainchain}          & Ethereum             & Not specified                       & Private        & Permissioned   \\
            2020 & Bringing Intelligence to Software Defined Netw\ldots~\cite{abou2020bringing}            & Ethereum             & Not specified                       & Private        & Permissioned   \\
            2020 & A Blockchain Approach to Identifying Compromis\ldots~\cite{yenugunti2020blockchain}     & Custom               & Trust Score;PBFT                    & Private        & Permissioned   \\
            2020 & A Contract Based User-Centric Computational Tr\ldots~\cite{hu2020contract}              & Hyperledger Fabric   & Not specified                       & Private        & Permissioned   \\
            2020 & A Framework of Blockchain-Based Collaborative \ldots~\cite{li2020framework}             & Not specified        & Not specified                       & Private        & Permissioned   \\
            2020 & Blockchain-enabled Collaborative Intrusion Det\ldots~\cite{fan2020blockchain}           & Ethereum             & Proof of Authority;PBFT             & Private        & Permissioned   \\
            2020 & Distributed Intrusion Detection System using B\ldots~\cite{kumar2020distributed}        & Not specified        & Not specified                       & Not specified  & Not specified  \\
            2020 & LegIoT: Ledgered Trust Management Platform for\ldots~\cite{neureither2020legiot}        & Hyperledger Sawtooth & Not specified                       & Private        & Permissioned   \\
            2020 & SOChain: A Privacy-Preserving DDoS Data Exchan\ldots~\cite{yeh2020sochain}              & Ethereum             & Proof of Work                       & Private        & Permissioned   \\
            2020 & TRIDEnT: Towards a Decentralized Threat Indica\ldots~\cite{alexopoulos2020trident}      & Ethereum             & Proof of Burn;Proof of Work         & Public         & Permissionless \\
            2020 & Blockchain-based Federated Learning for Device\ldots~\cite{zhang2020blockchain}         & Ethereum             & Proof of Work                       & Private        & Permissionless \\
            2020 & Application of Blockchain within Aviation Cybe\ldots~\cite{adhikari2020application}     & Not specified        & Not specified                       & Not specified  & Not specified  \\
            2020 & Blockchain as a solution to Drone Cybersecurity~\cite{ossamah2020blockchain}            & Not specified        & Not specified                       & Private        & Permissioned   \\
            2020 & Blockchain technology in the future of busines\ldots~\cite{demirkan2020blockchain}      & Not specified        & Not specified                       & Not specified  & Not specified  \\
            2020 & Suitability of Blockchain for Collaborative In\ldots~\cite{dawit2020suitability}        & Not specified        & Independent                         & Not specified  & Not specified  \\
            2021 & A Distributed Collaborative Entrance Defense F\ldots~\cite{guo2022distributed}          & Not specified        & Not specified                       & Private        & Permissioned   \\
            2021 & A Deep Blockchain Framework-Enabled Collaborat\ldots~\cite{alkadi2020deep}              & Ethereum             & Proof of Work                       & Private        & Permissioned   \\
            2021 & A Comprehensive Study of Anomaly Detection Sch\ldots~\cite{diro2021comprehensive}       & Independent          & Independent                         & Not specified  & Not specified  \\
            2021 & Distributed Watchdogs Based on Blockchain for \ldots~\cite{lee2021distributed}          & Ethereum             & Proof of Authority                  & Private        & Permissioned   \\
            2021 & Toward a blockchain-based framework for challe\ldots~\cite{li2021toward}                & Custom               & Proof of Identity                   & Private        & Permissioned   \\
            2021 & SeDID: An SGX-enabled decentralized intrusion \ldots~\cite{liu2021sedid}                & Custom               & Custom                              & Private        & Permissioned   \\
            2021 & Designing consensus algorithm for collaborativ\ldots~\cite{winanto2021designing}        & Not specified        & Proof of Work;Proof of Stake;Custom & Not specified  & Not specified  \\
            2021 & LDBT: A Lightweight DDoS Attack Tracing Scheme\ldots~\cite{guo2021ldbt}                 & Hyperledger Fabric   & PBFT                                & Private        & Permissioned   \\
            2021 & Blockchain and Federated Learning for Collabor\ldots~\cite{liu2021blockchain}           & Ethereum             & Proof of Work;Proof of Accuracy     & Private        & Permissioned   \\
            2021 & Blockchained Federated Learning for Threat Def\ldots~\cite{demertzis2021blockchained}   & Not specified        & Not specified                       & Not specified  & Not specified  \\
            2021 & Cybersecurity of Renewable Energy Data and App\ldots~\cite{cali2021cybersecurity}       & Not specified        & Not specified                       & Private        & Permissioned   \\
            2021 & Secure exchange of cyber threat intelligence u\ldots~\cite{pahlevan2021secure}          & Not specified        & Not specified                       & Private        & Permissioned   \\
            2022 & FolketID: A Decentralized Blockchain-Based Nem\ldots~\cite{chiu2022folketid}            & Ethereum             & Proof of Authority                  & Private        & Permissioned   \\
            2022 & Blockchain-Based Decentralized Replay Attack D\ldots~\cite{ramanan2021blockchain}       & Ethereum             & Proof of Authority                  & Private        & Permissioned   \\
            2022 & BlockCSDN: Towards Blockchain-Based Collaborat\ldots~\cite{li2022blockcsdn}             & Not specified        & Not specified                       & Private        & Permissioned   \\
            2022 & CIDS: Collaborative Intrusion Detection System\ldots~\cite{gurung2022cids}              & Hyperledger Fabric   & Not specified                       & Private        & Permissioned   \\
            2022 & IDS-Chain: A Collaborative Intrusion Detection\ldots~\cite{aljuhani2022ids}             & Not specified        & Not specified                       & Private        & Permissioned   \\
            2022 & Design of Blockchain enabled intrusion detecti\ldots~\cite{saveetha2022design}          & Ethereum             & Not specified                       & Private        & Permissioned   \\
            2022 & Blockchain-Enabled Intrusion Detection and Pre\ldots~\cite{alevizos2022blockchain}      & Hyperledger Fabric   & Not specified                       & Private        & Permissioned   \\
            2022 & FIDChain: Federated Intrusion Detection System\ldots~\cite{ashraf2022fidchain}          & Custom               & Round-robin                         & Private        & Permissioned   \\
            2022 & Privacy-preserved Cyberattack Detection in Ind\ldots~\cite{abdel2022privacy}            & Algorand             & Pure Proof of Stake                 & Private        & Permissioned   \\
            2022 & CGAN-Based Collaborative Intrusion Detection f\ldots~\cite{he2022cgan}                  & Not specified        & Proof of Authority                  & Private        & Permissioned   \\
            2022 & A Framework for Verifiable and Auditable Feder\ldots~\cite{santin2022framework}         & Hyperledger Fabric   & Not specified                       & Private        & Permissioned   \\
            2022 & InFEDge: A Blockchain-based Incentive Mechanis\ldots~\cite{wang2022infedge}             & FISCO BCOS           & Not specified                       & Private        & Permissioned   \\
            2022 & An Intelligent Platform for Threat Assessment \ldots~\cite{kolokotronis2022intelligent} & Not specified        & Not specified                       & Not specified  & Not specified  \\
            2022 & Blockchain-based Trusted Traffic Offloading in\ldots~\cite{tang2022blockchain}          & Not specified        & Enhanced PBFT                       & Private        & Permissioned   \\
            2022 & A Distributed Intrusion Detection System to De\ldots~\cite{kumar2022distributed}        & Not specified        & Not specified                       & Not specified  & Not specified  \\
            2023 & FusionFedBlock: Fusion of Blockchain and Feder\ldots~\cite{singh2023fusionfedblock}     & Ethereum             & Proof of Authority                  & Private        & Permissioned   \\
            \bottomrule
        \end{tabular}
    \end{adjustbox}
    \label{tab:papers_2}
\end{table*}

\subsection{Research groups and venues}

\begin{landscape}
    \begin{figure}
        \includegraphics[width=\linewidth]{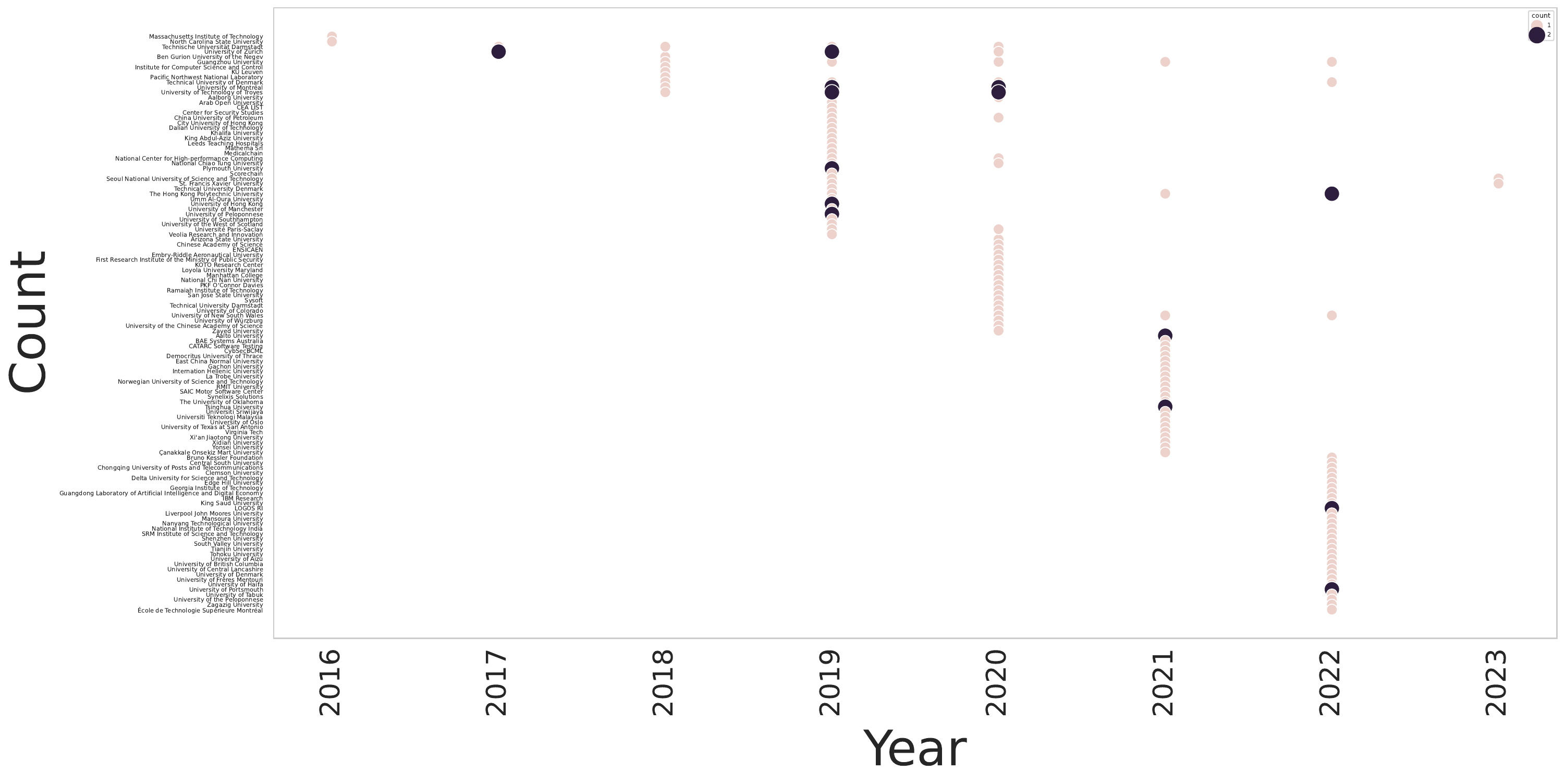}
        \caption{Number of papers per research group per year.}\label{fig:groups}
    \end{figure}
\end{landscape}

\Cref{fig:groups} shows the number of papers per research group per year.
There does not seem to be any particular trend when it comes to the research groups, and the figure indicates a shift in research groups over the years.

\begin{landscape}
    \begin{figure}
        \includegraphics[width=\linewidth]{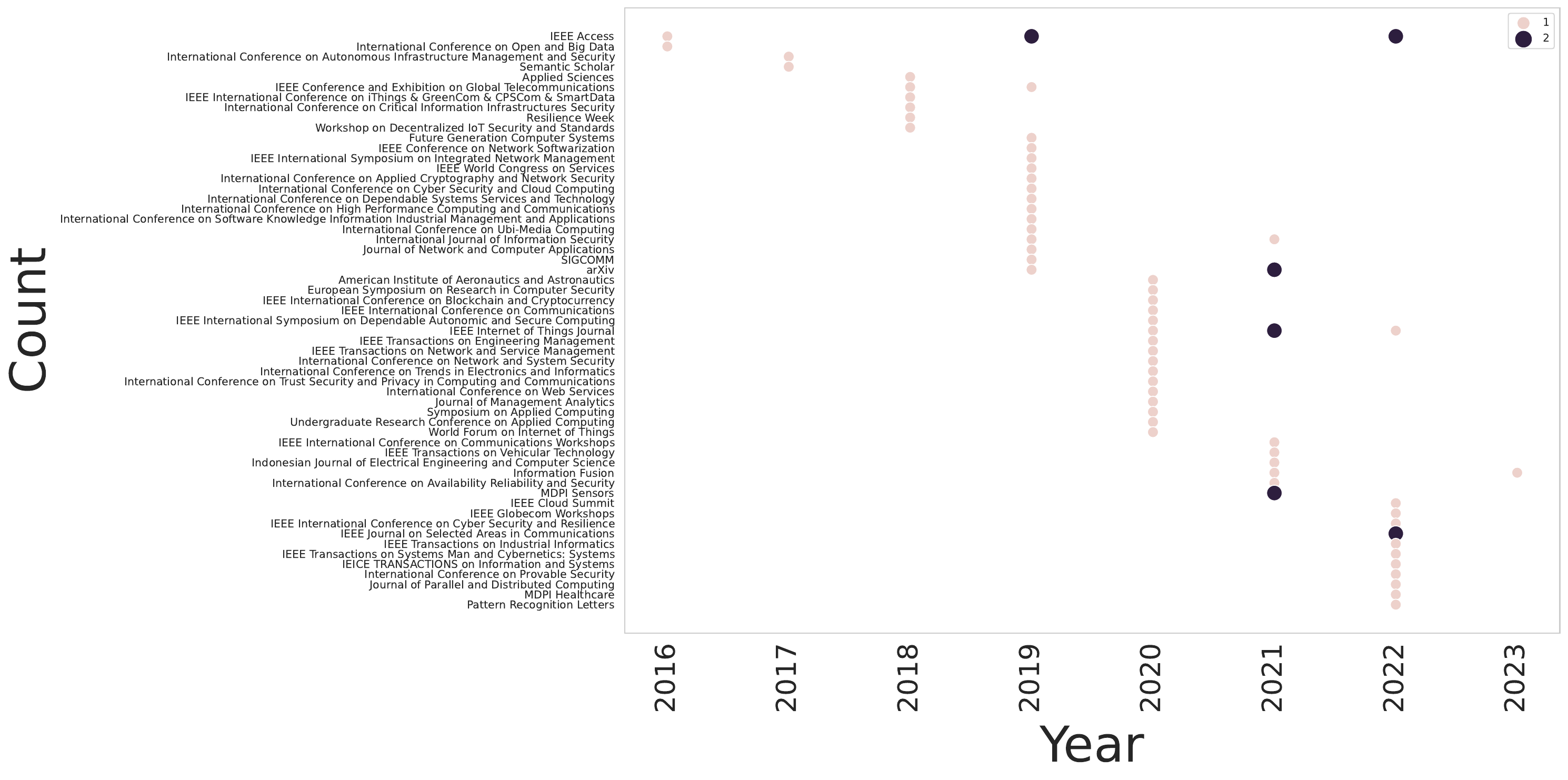}
        \caption{Number of papers per venue per year.}\label{fig:venues}
    \end{figure}
\end{landscape}

\Cref{fig:venues} represents the number of papers per venue per year.
As with the research groups, there is a shift in venues those works are published in over the years.
Yet another bit of evidence that the field is still new and fragmented.

\end{document}